\documentclass[traditabstract]{aa} 
\usepackage{epsfig}
\usepackage{graphicx}
\usepackage{txfonts}
\begin{document}

\authorrunning{Chris Pearson et al.}
\titlerunning{FU-HYU - First Results}
\title{The AKARI FU-HYU galaxy evolution program: First results from the GOODS-N field}

   \subtitle{}

   \author{C.P.~Pearson
          \inst{1,2,3}
          \and
          S. Serjeant\inst{3}
          \and
          M. Negrello\inst{3}
          \and
          T. Takagi\inst{4}
          \and
          W.-S. Jeong\inst{5}
          \and
          H. Matsuhara\inst{4}
          \and
          T. Wada\inst{4}
          \and
          S. Oyabu\inst{4}
          \and
          H.M. Lee\inst{6}
          \and
          M.S. Im\inst{6}
          }

   \institute{Space Science and Technology Department, CCLRC Rutherford Appleton Laboratory, Chilton, Didcot, Oxfordshire OX11 0QX, UK\\
              \email{chris.pearson@stfc.ac.uk}
         \and
             Department of Physics, University of Lethbridge, 4401 University Drive, Lethbridge, Alberta T1J 1B1, Canada\\
          \and
             Astrophysics Group, Department of Physics, The Open University, Milton Keynes, MK7 6AA, UK\\
         \and
             Institute of Space and Astronautical Science, Yoshinodai 3-1-1, Sagamihara, Kanagawa 229 8510, Japan\\
          \and
             KASI, 61-1, Whaam-dong, Yuseong-gu, Deajeon, 305-348, South Korea\\
         \and
             Department of Physics and Astronomy, Seoul National University, Shillim-Dong, Kwanak-Gu, Seoul 151-742, Korea\\
             }
   \date{Received {\today}; accepted {\today}}

\abstract
{The AKARI FU-HYU mission program carried out mid-infrared imaging of several well studied Spitzer fields preferentially selecting fields already rich in multi-wavelength data from radio to X-ray
 wavelengths filling in the wavelength desert between the Spitzer IRAC and MIPS bands. We present the initial results for the FU-HYU survey in the GOODS-N field. We utilize the supreme multiwavelength coverage in the GOODS-N field to produce a multiwavelength catalogue from infrared to ultraviolet wavelengths, containing more than 4393 sources, including photometric redshifts. Using the FU-HYU catalogue we present colour-colour diagrams that map the passage of PAH features through our observation bands. We find that the longer mid-infrared bands from AKARI (IRC-L18W 18 micron band) and Spitzer (MIPS24 24 micron band) provide an accurate measure of the total MIR emission of the sources and therefore their probable total mid-infrared luminosity. We also find that colours incorporating the AKARI IRC-S11 11 micron band produce a bimodal distribution where an excess at 11 microns preferentially selects moderate redshift star-forming galaxies. These powerful colour-colour diagnostics are further used as tools to extract anomalous colour populations, in particular a population of Silicate Break galaxies from the GOODS-N field showing that dusty starbursts can be selected of specific redshift ranges (z=1.2 -- 1.6) by mid-infrared drop-out techniques. The FU-HYU catalogue will be made publically available to the astronomical community. 
 }
   \maketitle

\section{Introduction}\label{sec:introduction}
 
 Studies with the Infrared Space Observatory ({\it ISO}) of the Hubble deep fields, North and South (HDF-N, HDF-S) have revealed star formation rates at least comparable to or higher than those of the optical/UV studies (\cite{mann02}) . At submillimetre wavelengths, surveys in the same HDF-N field with SCUBA on the JCMT has revealed a large ($>$3000 deg$^{-2}$ at $S_{850} > $2 mJy) population of strongly evolving sources with bolometric luminosities  $>10^{12}L_{\sun}$ and star formation rates of $\sim$300--1000M$_{\sun}$yr$^{-1}$ with a median redshift of 2.4 (\cite{chapman05}). The overwhelming conclusion is that the star formation rate at z$\sim$1--2 requires significant evolution in the IR galaxy population from the current epoch. Deep observations with the {\it Spitzer} Space Telescope have confirmed this strong evolution in the galaxy population out to the intermediate redshift range probed by {\it ISO} (z$\sim$0.3--1)  and furthermore provided insight into the higher redshift Universe in the so called redshift desert z$\sim$1-3. To connect the local and intermediate redshift Universe to the higher and high z Universe observed by  {\it Spitzer} and SCUBA, comprehensive multiwavelength imaging and spectroscopy is required throughout the extragalactic population. The Fields surveyed by the  {\it Spitzer} space telescope are some of the most richest fields in multiwavelength data in the entire sky. The SWIRE wide area survey (50 square degrees over 6 fields, \cite{lonsdale04}), the GOODS survey (300 square arcmins over 2 fields,  \cite{giavalisco04}), the other GTO fields and the Spitzer First Look Survey (FLS, e.g. \cite{frayer06}) as well as having coverage over the 3.5-160$\mu$m wavelength range also include a wealth of ancillary data at other wavelengths from radio to X-rays. However, Spitzer has limited imaging capability between 8$\mu$m$<\lambda<$24$\mu$m and no spectroscopic capability shortward of 5$\mu$m leaving a conspicuous gap in these ranges in both wavelength and redshift.
 
The {\it AKARI} satellite, (formerly known as {\it ASTRO-F}, \cite{murakami07}) is the first Japanese space mission dedicated to infrared astrophysics and was launched on board JAXA's M-V Launch Vehicle No. 8 (M-V-8) at 6:28 a.m. on February 22, 2006 Japan Standard Time, JST (February 21st, 9.28 p.m. UT) from the Uchinoura Space Center (USC) on the southern tip of Japan. The satellite is in a Sun-synchronous polar orbit at an altitude of 700 km and a period of 100 minutes.  {\it AKARI} has a 68.5 cm cooled telescope with two focal plane instruments, namely the Far-Infrared Surveyor (FIS) and the Infrared Camera (IRC). The FIS has two 2-dimensional detector arrays and observes in four far-infrared bands between 50 and 180 $\mu$m (\cite{kawada07}). The IRC consists of three cameras covering 1.7-- 26 $\mu$m in 9 bands (see Table \ref{table:irc}) with fields of view of approximately 10$\arcmin$  $\times$ 10$\arcmin$ (\cite{onaka07}). Both instruments have low- to moderate-resolution spectroscopic capability.

\begin{table}
\caption{Photometric Filter band definitions for the three cameras of the IRC.}
\begin{center}
\begin{tabular}{cccccc}
\hline\hline
Camera & Array Area$^*$ & Pixel Scale & Band    &     $\lambda_{\rm ref}$    &  $\Delta \lambda$  \\
               &    (pixels)      &   (arcsec) &            &  ($\mu$m)    & ($\mu$m) \\
\hline
NIR   &  319$\times$412 &   1.46$\times$1.46    & N2       & 2.4 & 1.9--2.8     \\
          &                            &                                   & N3       & 3.2 & 2.7--3.8    \\
          &                            &                                   & N4      & 4.1 & 3.6--5.3    \\
 MIR-S &  233$\times$256&   2.34$\times$2.34  & S7       & 7.0  & 5.9--8.4    \\
          &                              &                                   & S9W      & 9.0  & 6.7--11.6   \\
           &                               &                                   & S11      & 11.0 & 8.5--13.1    \\
MIR-L &  246$\times239$ &   2.51$\times$2.39  & L15      & 15.0 & 12.6--19.4  \\
           &                             &                                    & L18W& 18.0 & 13.9--25.6   \\
           &                              &                                    & L24      & 24.0 & 20.3--26.5  \\
\hline
\multicolumn{6}{l}{* This is the effective imaging area of the array}\\
\end{tabular}
\end{center}
\label{table:irc}
\end{table}

 The  {\it AKARI} mission is primarily a survey mission with an All-Sky Survey in the far-infrared and large mid-infrared legacy survey programs being carried out in the region of the North Ecliptic Pole (\cite{matsuhara06}) and the Large Magellanic Cloud. However in addition to these Large Surveys (LS), a campaign of guaranteed time observations, referred to as Mission Programs (MP) was also undertaken.
 
\begin{figure}
\centering
\centerline{
\psfig{ figure=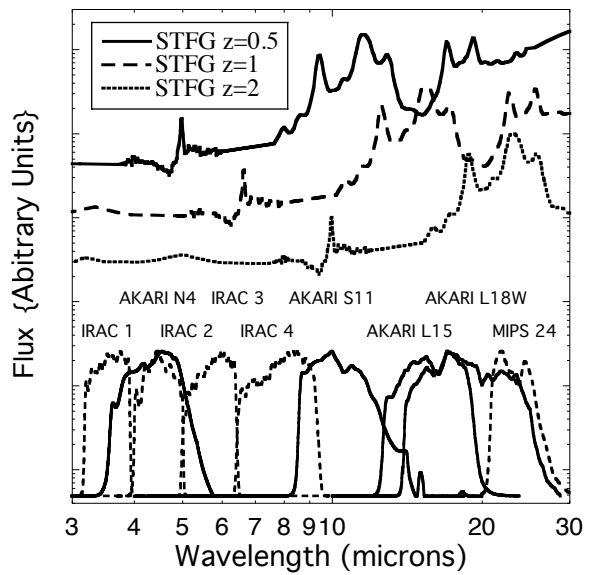,width=7.5cm}
}
\caption{The comprehensive coverage of the {\it AKARI} and {\it Spitzer} bands in the FU-HYU fields. The combination of the {\it AKARI} and {\it Spitzer} bands completely cover the features in the mid-infrared spectra of star-forming galaxies  (STFG, shown in the figure at three different redshifts) locally, out to redshifts of $\sim$2.
\label{fig:bandcoverage}}
\end{figure}

We report the initial results from the  {\it AKARI}-FU-HYU Mission Program (Follow-Up Hayai-Yasui-Umai). The FU-HYU MP strategically targeted well-studied fields to maximise the legacy value of the {\it AKARI} data and in particular to target {\it Spitzer} fields in the above mentioned conspicuous wavelength "gap", at wavelengths of 11, 15 \& 18$\mu$m. The FU-HYU observations fill in vital gaps in the wavelength coverage (see Figure \ref{fig:bandcoverage}) and provide invaluable insight into the connection between {\it ISO} \& {\it Spitzer} populations and linking the far-infrared Universe to the high redshift sub-mm Universe as observed by the SCUBA instrument on the James Clerk Maxell Telescope. 

The FU-HYU program imaged three major fields (GOODS-N, Lockman Hole, ELAIS-N1, see Section \ref{sec:FU-HYU}) and in this work we report on the initial data reduction and results from the GOODS-N observations. Observation in the Lockman Hole will be reported by  \cite{serjeant09}.

Unfortunately, the FU-HYU observations in the GOODS-N field were not dithered during the observation operation resulting in many hot and bad pixels leaving {\it holes} in the final co-added images. Moreover the distortion correction in the standard IRC pipeline toolkit blurs the hot pixels compounding the problem. In Section ~\ref{sec:Reduction} we describe in detail our extensive additional processing utilizing intrinsic {\it jitter} in the spacecraft observations as an effective {\it dithering} and interrupting the standard IRC pipeline toolkit before the distortion correction stage to carry out independent re-gridding of the pixels onto a finer mesh, co-adding and re-binning, resulting in final dithered images without the hot pixels and line artifacts seen in the original processed data.

 The cross associations with other data sets in the GOOD-N field, most notably the  {\it Spitzer} data and the construction of the FU-HYU catalogue is described in Section \ref{sec:Association}. The results and conclusions are given in Sections \ref{sec:results} \& \ref{sec:summary} respectively.

\section{The AKARI FU-HYU program}\label{sec:FU-HYU}

The objective of the FU-HYU Mission Program is to add {\it AKARI} data to contemporary data rich fields. On selection of our target fields, the constraint on the visibility of the  {\it AKARI} satellite had to be seriously considered. The orbit of  {\it AKARI} is Sun-synchronous meaning that the ecliptic poles enjoy many pointing opportunities whilst fields near the ecliptic may only be visible on a few orbits every six months. Taking as an example the two well known fields of {\it ISO}-ELAIS N1  (\cite{oliver00})  at an ecliptic latitude of $\beta \sim$73$^{\circ}$ and the UKIDSS-UDS  field (\cite{lawrence07}) at  $\beta \sim$-17$^{\circ}$, we find that the former is observable of 17 orbits during half a year whilst the latter is observable only 5 times. Given that not every {\it AKARI} orbit is available for observation (calibration, directors time, etc) and further constraints placed on the observations by the necessity of choice of the deep imaging AOT (see below), there exist very clear criteria for our choice of target fields.

For the FU-HYU program we selected three contemporary data rich fields;

\begin{itemize}
\item The GOODS North field (GOODS-N)
\item The ELAIS-N1 field
\item The Lockman Hole
\end{itemize}

The ELAIS-N1 field (R.A.= 16h09m20s; DEC=+54$^{\circ}$57$^{\prime}$00$^{\prime \prime}$, J2000, $\beta$=73$^{\circ}$) is one of the the most richest fields in multiwavelength data on the entire sky. The ELAIS-N1  field was selected as part of the Extragalactic Large Area ISO Survey (ELAIS,  \cite{oliver00} ). The fields were selected on the basis of their high ecliptic latitude  ($\beta >$40$^{\circ}$, to reduce the impact of zodiacal dust emission), low cirrus emission (I$_{100\mu m} <$ 1.5 MJy/sr) and the absence of any bright (S$_{12\mu m} >$ 0.6 Jy) IRAS 12$\mu$m sources. A field of $\sim$2 square degrees was observed to  $\sim$1.0, 0.7, 70, 223mJy at 6.7, 15, 90, 175$\mu$m respectively and has extensive spectroscopic and imaging follow up at optical and near-IR wavelengths in the U, g, r, i  , Z, J, H, K bands  (\cite{vaisanen02}, \cite{gonzalez-solares05}, \cite{serjeant04}). X-ray follow up with Chandra in the central region of N1 and a comparison with ROSAT has been made by  \cite{manners03}, \cite{willott03},  \cite{basilakos02}. Radio observations also exist at 20cm (\cite{ciliegi99}). The ELAIS N1 field was also partially observed with the H$\alpha$ survey of  \cite{pascual01}. Most significantly, ELAIS-N1 was a {\it Spitzer} SWIRE Field and was also selected as part of the GOODS Science Verification observations (\cite{chary04}). This is one of the deepest ever observations of the 24$\mu$m sky down to around 20$\mu$Jy over 130arcmin$^{2}$ centred on R.A.=16h09m00.16s, Dec=+54$^{\circ}$55$^{\prime}$50.9$^{\prime \prime}$ and is the centre of the FU-HYU observation in the ELAIS-N1 field. Of the 3 FU-HYU fields, the ELAIS-N1 field has the best visibility and was observed with a total of 22 pointings over one 10$\arcmin \times$ 10$\arcmin$ field of view (7 in the IRC S11 band, 8 in the IRC L15 band and 7 in the IRC L18W band, see Table \ref{tab:obslog}).

The Lockman Hole field (R.A.= 10h52m43s; DEC=+57$^{\circ}$28$^{\prime}$48$^{\prime \prime}$, J2000, $\beta$=53$^{\circ}$) has one of the lowest HI column densities and background cirrus emission on the sky (I$_{100\mu m} <$ 0.4 MJy/sr).  The field has been covered by {\it Spitzer} by both IRAC and MIPS instruments (\cite{lefloch04}, \cite{huang04}). The Lockman field has also been mapped as part of the SCUBA Half Degree Extragalactic Survey (SHADES, \cite{mortier05}) and recently supplemented by equivalently deep 1.1mm data with the AzTEC camera (\cite{austermann09}). The Lockmann Hole is also the site of the UKIDSS DXS survey JHK data to K=21, H=22, J=22.5 over  8.75 sq deg. (\cite{lawrence07}). At radio and X-ray wavelengths there is VLA data at 1.4GHz to about 15$\mu$Jy (\cite{biggs06}) and XMM data (\cite{hasinger01}) respectively. The objective of the  Lockman field observations was to gather {\it AKARI} data on the submillimetre sources in the field however the visibility was unfortunate in that the opportunities to observe the Lockman Hole fell between observation phases in the AKARI mission planning (i.e. PV phase and the end of Phase I observations) limiting the number of possible 10$\arcmin \times$ 10$\arcmin$ fields of view to only three (see Table \ref{tab:obslog}). The results from the FU-HYU Lockman Hole Observations are presented in \cite{serjeant09}

The GOODS-N field (R.A.= 12h36m50s; DEC=+62$^{\circ}$12$^{\prime}$58$^{\prime \prime}$, J2000, $\beta$=57$^{\circ}$, \cite{dickenson01})  has perhaps the best multi-wavelength coverage of any pencil-beam survey (160$\arcmin ^2$), and is presented here as the first priority for our initial analysis. It is of course co-located with the Hubble deep field North (\cite{williams96}) with optical data from the HST treasury programme (\cite{giavalisco04}), deep ground based UBVRIJHKz$^ \prime$ Subaru/KPNO data (\cite {capak04}) and Keck optical spectroscopy (\cite{reddy06}). The GOODS-N field has also been surveyed in the infrared with {\it ISO} (\cite{serjeant97}) \& {\it Spitzer} (See Section \ref{sec:xcorrelate}) and at submillimetre wavelengths with the SCUBA and AzTEC instruments at 850$\mu$m and 1.1mm respectively (\cite{pope05}, \cite{perera08}). The visibility of the GOODS-N field was marginal for {\it AKARI} and the emphasis was placed on a single 10$\arcmin \times$ 10$\arcmin$ field of view  in the unique  IRC L18W (3 pointings) and IRC N4, S11 (6 pointings each) bands. The full details of the observations are given in Table \ref{tab:obslog}. Combining the {\it AKARI} and {\it Spitzer} data in the GOODS-N field together provides comprehensive coverage of the entire near-mid infrared spectrum of dusty galaxies and their multitude of emission/absorption features from redshifts 0 to 2. The supreme coverage of the combined {\it AKARI} and {\it Spitzer} bands is highlighted in Figure \ref{fig:bandcoverage}.

\begin{figure}
\centering
\centerline{
\psfig{ figure=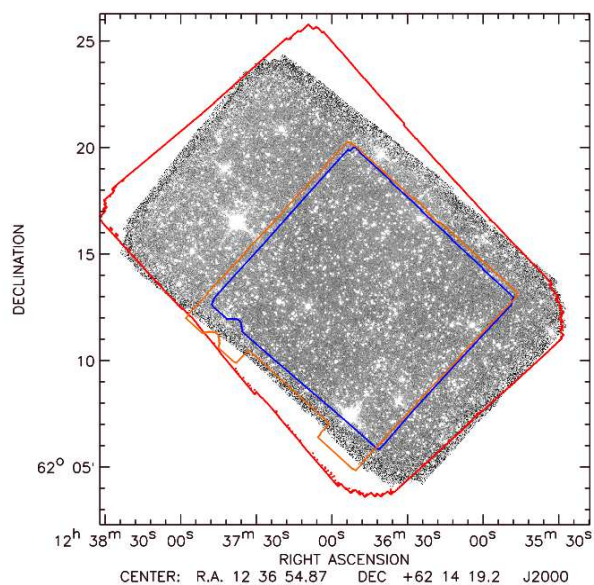,width=8cm}
}
\caption{Coverage of the {\it AKARI} FU-HYU area within the GOODS-N field. The image is the {\it Spitzer}  IRAC 3.6$\mu$m band image. The large {\it round rectangle} depicts the  {\it Spitzer} coverage in the MIPS24$\mu$m band. The larger of the two inner  {\it squares} is the {\it AKARI}-IRC MIR-L L18W band coverage and the smaller inner  {\it square} is the coverage of the  IRC MIR-S S11 \& NIR N4 bands  (i.e. the foot print of the detector arrays)
\label{fig:coverage}}
\end{figure}

In Figure \ref{fig:coverage} we show the FU-HYU imaging field of view ($\sim$10$\times$10 arcmin for both the MIR-L and MIR-S/NIR IRC channels) in the GOODS-N field. These, along with the  {\it Spitzer} MIPS 24$\mu$m coverage are overlaid on the image in the  {\it Spitzer}  IRAC 3.6$\mu$m band.

\begin{table*}
  \caption{Observation Log for the FU-HYU Mission Program}
  \label{tab:obslog}
  \begin{center}
    \begin{tabular}{lllllllll}
      \hline \hline
Field&RA&DEC&Observation ID&Band&AOT&Parameters&Date&Status\\
 \hline
 GOODS-N&189.2079&62.2161&1320028-001&S11&IRC05&c;N&19/11/06&Observed\\
&&&1320029-001&S11&IRC05&c;N&19/11/06&Observed\\
&&&1320030-001&S11&IRC05&c;N&20/11/06&Observed\\
&&&1320031-001&S11&IRC05&c;N&21/11/06&Observed\\
&&&1320032-001&S11&IRC05&c;N&21/11/06&Observed\\
&&&1320033-001&S11&IRC05&c;N&21/11/06&Observed\\
&&&1320025-001&L18W&IRC05&a;L&19/11/06&Observed\\
&&&1320026-001&L18W&IRC05&a;L&19/11/06&Observed\\
&&&1320027-001&L18W&IRC05&a;L&19/11/06&Observed\\
ELAIS-N1&242.2507&54.9308&1320226-001&S11&IRC05&c;N&19/7/07&Observed\\
&&&1320226-003&S11&IRC05&c;N&19/7/07&Observed\\
&&&1320226-004&S11&IRC05&c;N&19/7/07&Observed\\
&&&1320226-005&S11&IRC05&c;N&19/7/07&STT WDT error\\
&&&1320226-002&S11&IRC05&c;N&19/7/07&Data lost\\
&&&1320226-006&S11&IRC05&c;N&20/7/07&Observed\\
&&&1320226-007&S11&IRC05&c;N&20/7/07&Observed\\
&&&1320014-001&L15&IRC05&b;L&15/1/07&Observed\\
&&&1320013-001&L15&IRC05&b;L&16/1/07&Observed\\
&&&1320015-001&L15&IRC05&b;L&17/1/07&Observed\\
&&&1320235-001&L15&IRC05&b;L&21/7/07&Observed\\
&&&1320235-002&L15&IRC05&b;L&21/7/07&Observed\\
&&&1320235-003&L15&IRC05&b;L&21/7/07&Observed\\
&&&1320235-004&L15&IRC05&b;L&22/7/07&Data lost\\
&&&1320235-005&L15&IRC05&b;L&22/7/07&Observed\\
&&&1320232-001&L18W&IRC05&a;L&20/7/07&Observed\\
&&&1320232-002&L18W&IRC05&a;L&20/7/07&Observed\\
&&&1320232-003&L18W&IRC05&a;L&20/7/07&Observed\\
&&&1320232-004&L18W&IRC05&a;L&21/7/07&Observed\\
&&&1320232-005&L18W&IRC05&a;L&22/7/07&Observed\\
&&&1320232-006&L18W&IRC05&a;L&22/7/07&Observed\\
&&&1320232-007&L18W&IRC05&a;L&22/7/07&Observed\\
Lockman Hole&162.7016&57.5555&1320102-001&L15&IRC05&b;L&7/5/07&Observed\\
&&&1320103-001&L15&IRC05&b;L&7/5/07&Observed\\
&&&1320104-001&L15&IRC05&b;L&7/5/07&Observed\\
&162.9408&57.4731&1320099-001&L15&IRC05&b;L&7/5/07&Observed\\
&&&1320300-001&L15&IRC05&b;L&8/5/07&Observed\\
&&&1320301-001&L15&IRC05&b;L&8/5/07&Observed\\
&163.1986&57.3838&1320305-001&L15&IRC05&b;L&8/5/07&Observed\\
&&&1320306-001&L15&IRC05&b;L&9/5/07&Observed\\
&&&1320307-001&L15&IRC05&b;L&9/5/07&Observed\\
    \hline
    \end{tabular}
  \end{center}
\end{table*}

\section{Data reduction}\label{sec:Reduction}

\subsection{Standard processing with the IRC pipeline}\label{sec:pipeline}

In nominal circumstances, $AKARI$-IRC data is reduced by the standard IRC imaging pipeline inside the $IRAF$ environment (\cite{lorente07}). The IRC pipeline consists of three stages;

\begin{itemize}
\item Pre-pipeline (red-box): Takes the packaged raw data as input and then slices these into individual exposure frames and creates an exposure frame observation log.
\item Pipeline (green-box): The output of the green box produces the basic data for all exposure frames, processing each frame individually, correcting for instrumental effects (e.g. dark subtraction, cosmic ray rejection, saturation correction, flat fielding, distortion correction, etc).
\item Pipeline (blue-box): Co-add individual frames together and astrometry solution. The output is the final image and signal-to-noise map
\end{itemize}

 A series of three AOT sets for pointed observations with the IRC have
been prepared. Two of these, referred to as IRC02 and IRC03, are intended for medium deep 
multi-band surveys, in which both changes of filters and dithering in target position
are performed. The remaining set are for deep surveys and are referred to as IRC00 or IRC05,
in which no filter change and no dithering is performed, in order to minimize the dead time and maximize the observation time. For the IRC00 and IRC05 AOTs, at least three pointed observations are
required in order to ensure redundant and reliable observations. For these AOTs dithering is performed by the operations team by a slight offset of each pointing from each other.
 
Note that in addition, there exists a jitter between frames in the IRC images causing frames to become misaligned with each other (note this is not an intentional dither which is a separate procedure). The attitude of these
frames must be matched (for any shift and rotation in position) before stacking to produce the final image.

\subsection{Image correction and stacking}\label{sec:correction}

\begin{figure*}
\centering
\centerline{
\psfig{ figure=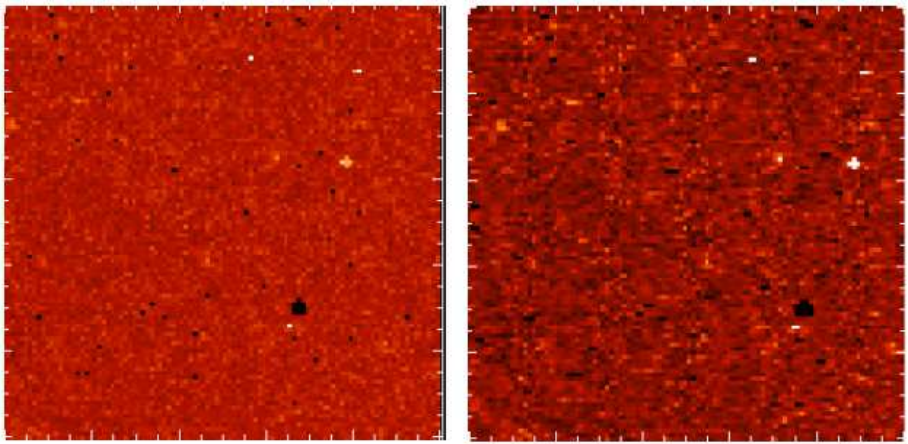,width=12cm}
}
\caption{100$\times$100 pixel segment of an L18W frame before {\it left} and after {\it right} the field distortion
correction in the standard IRC pipeline toolkit. Note the bad pixels which have been spread horizontally over
multiple pixels in the right-hand frame, and the subtle vertical strobing.
\label{fig:distortion}}
\end{figure*}

Unfortunately, the FU-HYU data in GOODS-N was taken without any spacecraft dithering. This means that bad pixels and other image artifacts such as slight differences between columns in the array, etc always fall in the same positions and are therefore not removed from the final images. In order to overcome this we have made use of
the intrinsic spacecraft jitter naturally present during any {\it AKARI} observation to simulate a dithering operation. Furthermore, the image distortion stage of the standard IRC pipeline toolkit involves many interpolations, with unsatisfactory results particularly near bad pixels (see Figure \ref{fig:distortion}). Most perniciously, the pixels are no longer statistically independent after this stage.

To overcome the combined problems of the dither operation and the pixel distortion, we customised the standard IRC pipeline toolkit to output FITS data products before the field distortion stage. At this stage, the data should be dark-subtracted and flat-fielded, however bad columns were still evident in some of the data in both the L18W and S11
channels. Our approach to removing these features exploits the fact that features confined to single columns cannot be real astronomical signals. Initially, after normalising each
frame to its median value, we fit Gaussian profiles to the histograms of the data values for each
pixel, and used the inferred modal values to construct a new sky flat. This technique has been applied in the past to 
{\it ISO}-ISOCAM data in the ELAIS survey (\cite{serjeant00}) but unlike the
ELAIS case there is little spacecraft movement between frames. Therefore, the flat field
constructed in this way will inevitably contain real astronomical features as well as
the bad columns we are seeking to correct.
In order to isolate the bad columns from the real astronomical point
sources we opted to median filter the sky flat using periodic boundary conditions,
with a 20-pixel boxcar window, and subtracted this smoothed image from our sky flat.
Outlier pixels in the sky flat were replaced with global median values for the
purpose of this smoothing. The resulting image represents the fractional excess or
deficit of each pixel over its neighbours, minus one. We then fit each column of
this image with a sixth-order Fourier expansion;

\begin{equation}
y(x) =  \!\!\sum_{i=0}^{6}\!\! a_{i} sin \left[ i \left( 2\pi \frac{x}{x_{max}}+ \phi_{i} \right)  \right]
\end{equation}

where the row number $x$ runs from 1 to $x_{max}$=256, the order $i$ runs from 0 to 6, and
$\phi_{i}$ and $a_{i}$ are the $i^{th}$ order phase and amplitude respectively. There are 13 free
parameters in total, and 256 data points in each column. Data points outside the range
[-0.006,0.006] were excluded which was sufficiently of low-order as to not remove point
sources from the image, while having a sufficiently high order to track the profiles of the
bad columns. The periodic boundary conditions implicit in the Fourier expansion
ensured no over-fitting at the image edges. These fits gave us a model for the
fractional excess flux in each column of the detector which was then used 
to correct the bad columns in the FITS data products output from the standard IRC pipeline toolkit.

The next stage of the processing is the determination of the intrinsic jitter in the IRC pointed observations.  We identified known bright sources in the image field, and calculated their centroids in every frame. We then corrected the pixel positions for field distortion and calculated the mean offset of each frame from the first frame,
which was assumed to have no jitter offset as a reference frame.

Noise images were then created by estimating the noise level in each frame by
fitting a Gaussian to the histogram of the pixel values. This estimator has previously been
used in data reduction for {\it ISO}-ISOCAM and at sub-millitre wavelengths for the SCUBA  instrument (\cite{serjeant00}  \cite{serjeant03}). Pixels greater than10$\sigma$ above the
modal value or less than -5$\sigma$ below it were masked, as were their neighbours and next neighbours. 
Any  regions removed by the standard IRC pipeline toolkit were also masked. Masking
was achieved by assigning an arbitrarily high noise value to the affected pixels.

\begin{figure}
\centering
\centerline{
\psfig{ figure=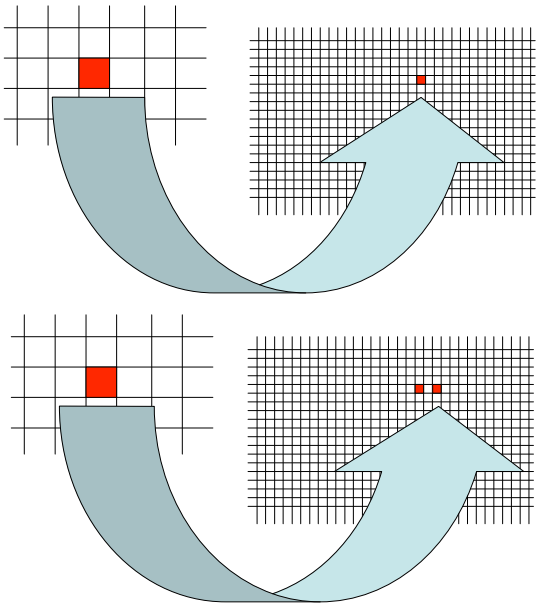,width=7cm}
}
\caption{The methodology for resampling the IRC pixels to a finer grid. In the {\it upper panel}, the flux of an IRC pixel is inserted into a single pixel of a finer grid. Altogether, the 256$\times$256 IRC pixels only sparsely populate the finer grid. The {\it lower panel} shows a later, jittered observation. Due to the jitter, the IRC pixel fluxes are not inserted into the same fine pixels as before. As more jittered observations are added, the fine grid fills. Where a fine grid pixel already contains a measurement, a noise-weighted co-addition is made.
\label{fig:zerofootprint}}
\end{figure}

As eluded to earlier, the field distortion correction carried out by the standard IRC pipeline toolkit involves many interpolations with every stage of interpolation
adding correlated noise to the final products. To avoid such interpolation steps the IRC pixel fluxes were
inserted and coadded directly into a zerofootprint map (drawing an analogy to drizzling, c.f. \cite{fruchter02}), accounting for the field distortion, a procedure again adopted from processing submillimetre data from the SCUBA instrument ((e.g. \cite{serjeant03}, \cite{mortier05},  \cite{coppin06}) - pictorially shown in Figure \ref{fig:zerofootprint}). 
In total 90 individual L18W frames (one frame was discarded due to a moving object in the frame), 180 S11 frames and  31 N4 frames were co-added in this way.

\begin{figure}
\centering
\centerline{
\psfig{ figure=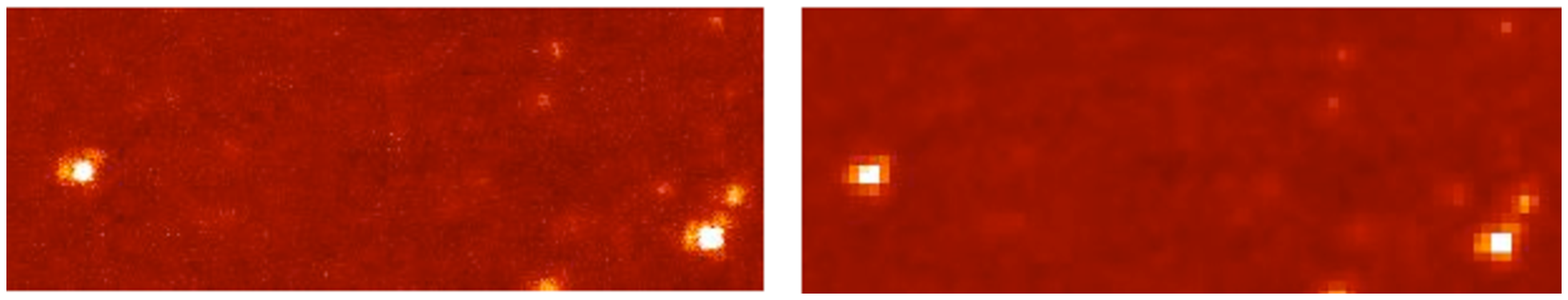,width=7cm}
}
\centerline{
\psfig{ figure=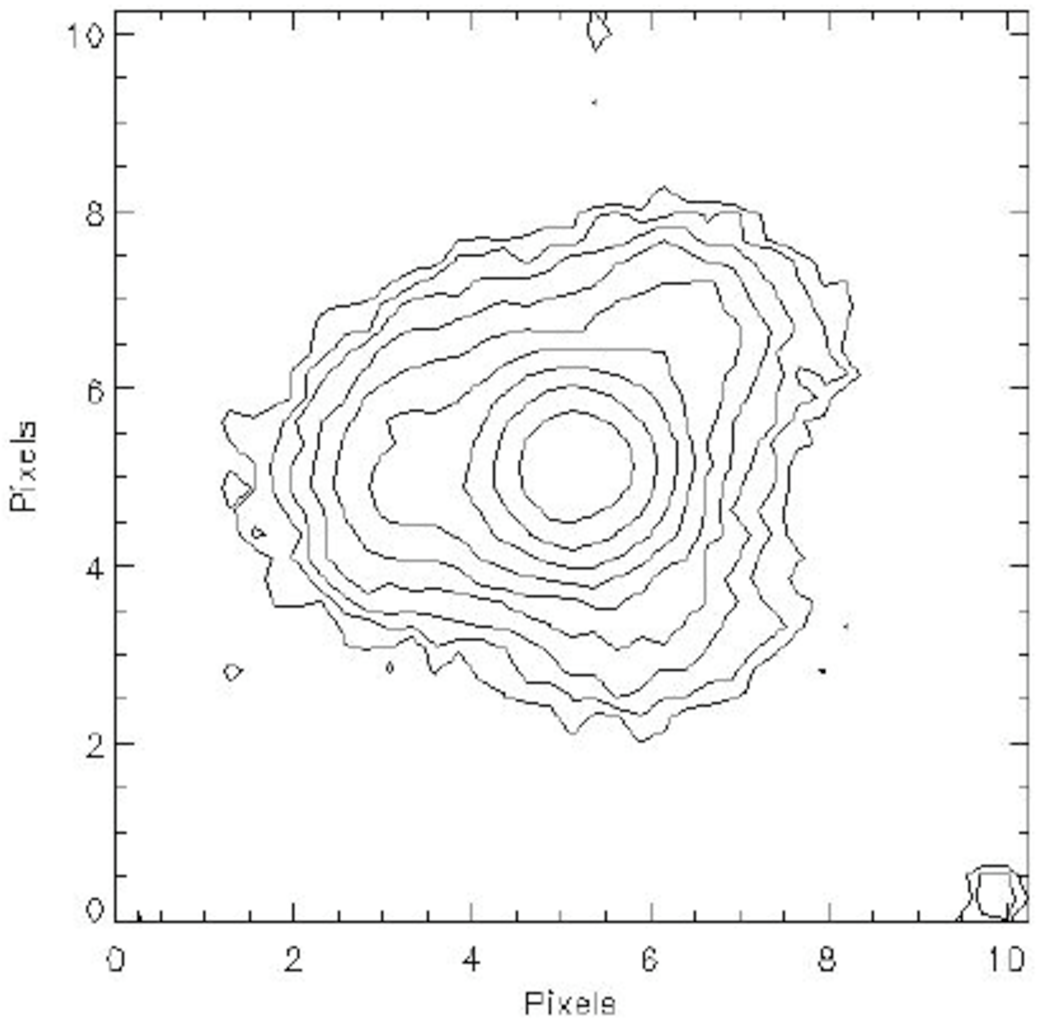,width=5cm}
}
\caption{{\it Top}: A section of the S11 zerofootprint image before ({\it left panel}) and after ({\it right panel}) rebinning to the IRC pixel
scale. Note the features in the  ({\it left panel}) at  about 9 o'clock and 1 o'clock  around the two brightest point sources in the image These features are not apparent in the rebinned image in the {\it right panel}. {\it Bottom}: Stacked point spread function in the S11 band, estimated from 101 bright point sources as discussed in the text. Contours are spaced at logarithmic intervals of $\sqrt{2}$ from a starting value of $1/ \sqrt{2}$. The point spread function is normalised to a peak flux of unity. The axes indicate the sizes of IRC pixels, though the image sampling is 4$\times$4 finer. Note the features at 9 o'clock and 1 o'clock, also visible in the {\it Top left panel} , and another fainter feature at about 5 o'clock. Companion galaxies or noise spikes may cause the weak features in the bottom-right corner and  top-centre of the image. 
\label{fig:s11noise}}
\end{figure}

 It was necessary to perform a further stage of deglitching to remove negative
spikes and we rejected features below -3.5$\sigma$ in the L18W band image and -4$\sigma$ in the  S11 band image. The resulting image still showed some subtle large-scale structure, possibly a result of poor flat
fielding, so we constructed a background model by boxcar median smoothing the
image with a width equivalent to 25 IRC pixels, with periodic boundary conditions,
and subtracted this background model from our images. The  zerofootprint map was then resampled back to the IRC pixel scale. During this process it became evident that the IRC S11 data still exhibited structure on smaller scales as shown in  Figure \ref{fig:s11noise}. A possible reason for this is that  the IRC pixel scale has only 1.68 pixels spanning the expected full-width half maximum (FWHM) of a diffraction-limited 11$\mu$m PSF ($\theta_{FWHM} = 1.22\lambda / D$,  $\lambda$=11$\mu$m and $D$=0.69m), which is just below the Nyquist scale.
The same excess structure was not evident in the L18W image, therefore we decided to use the rebinned to the original IRC pixel scale image at 18$\mu$m for later source extraction  but to retain the unbinned zerofootprint map at 11$\mu$m. 

\begin{figure*}
\centering
\centerline{
\psfig{ figure=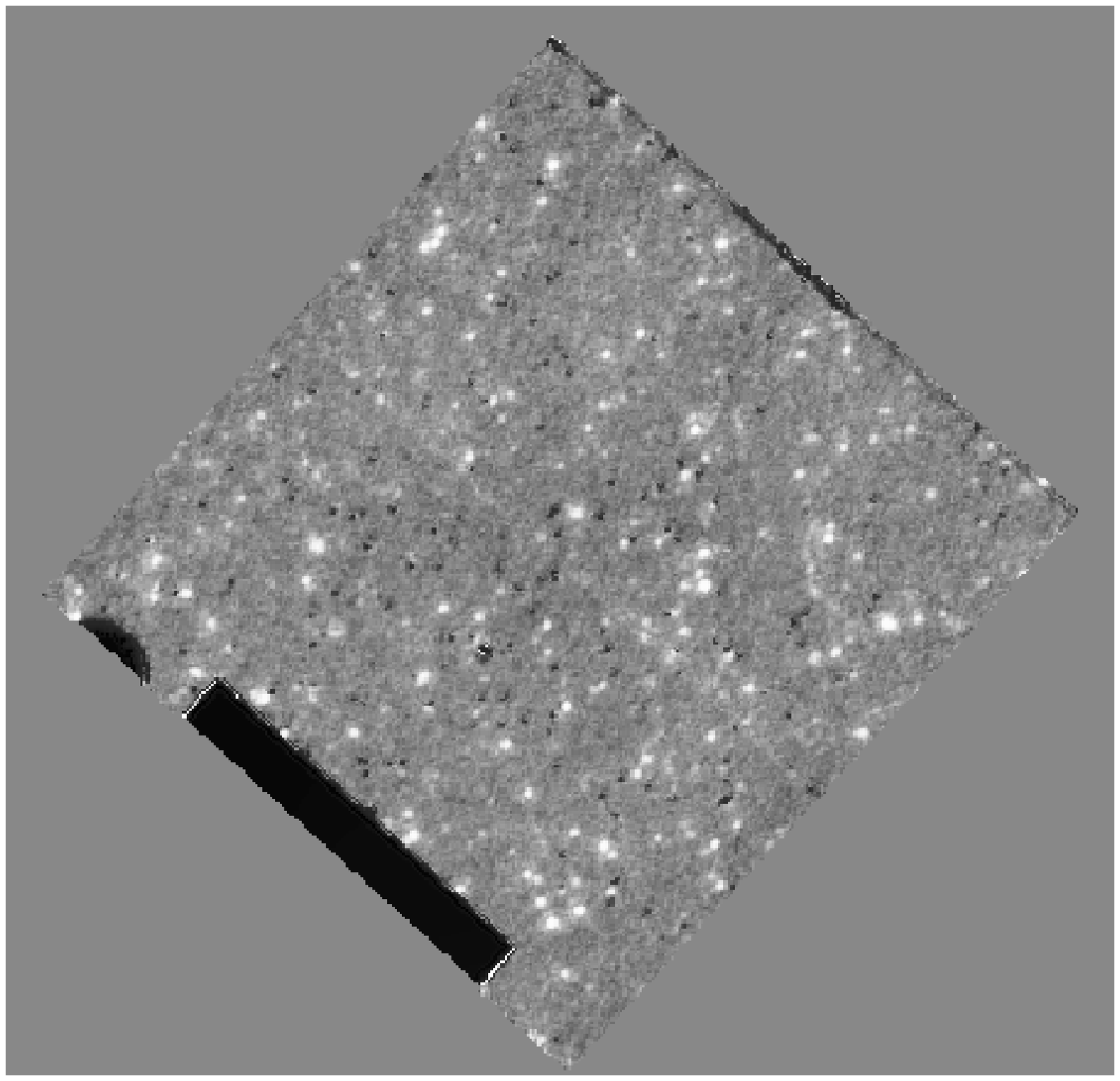,width=6cm}
\psfig{ figure=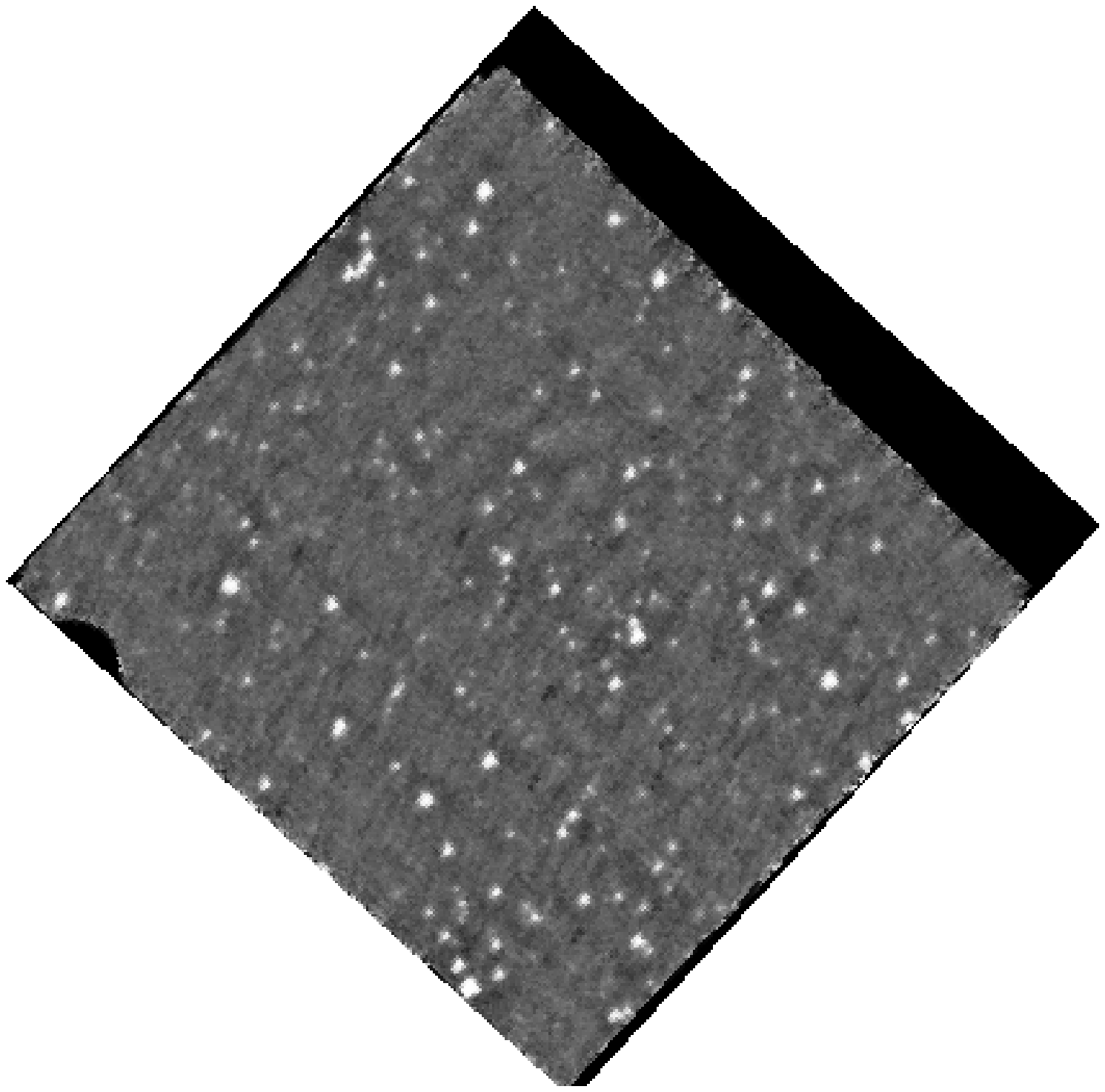,width=6cm}
\psfig{ figure=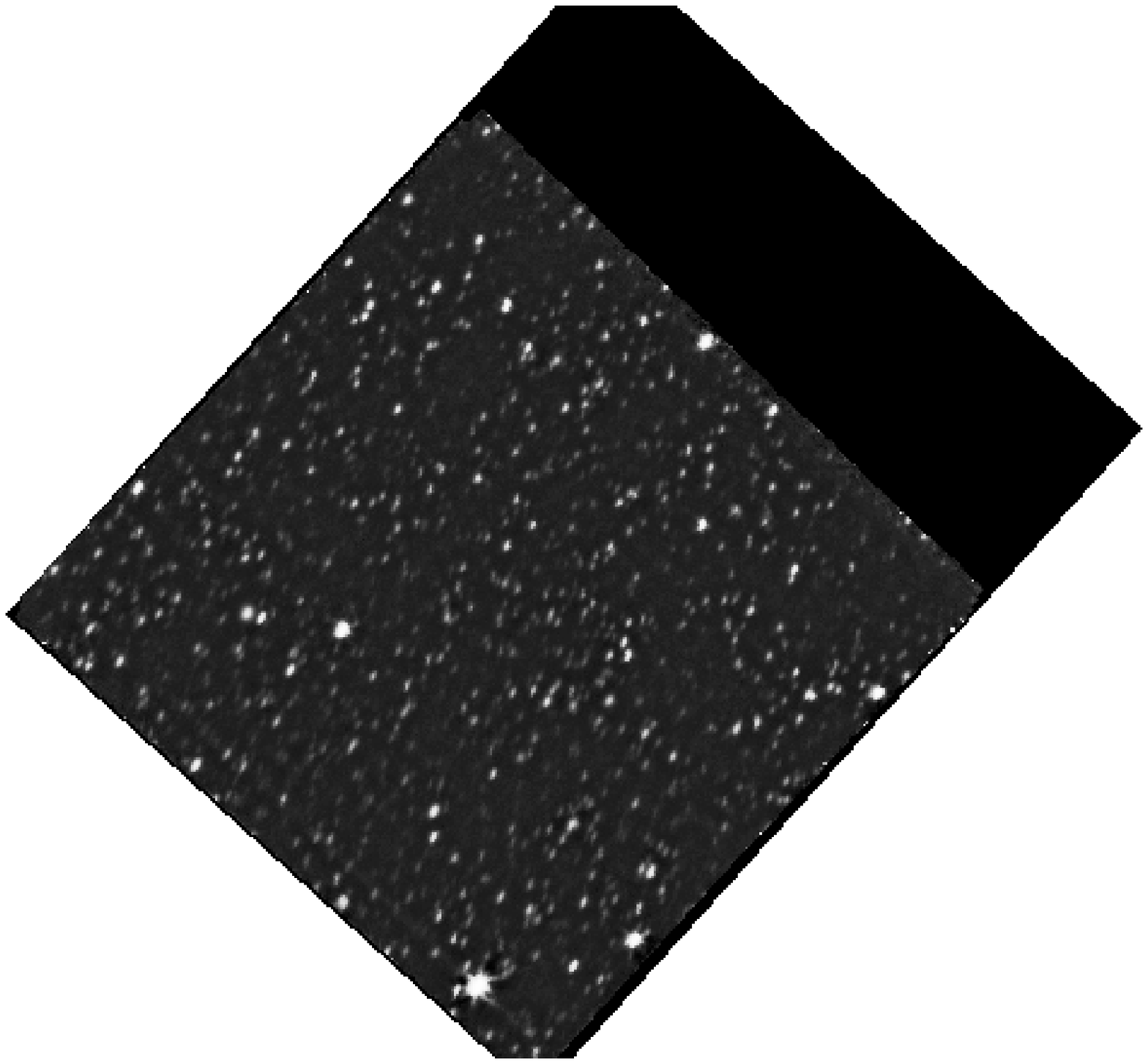,width=6cm}
}
\vspace{0.2cm}
\centerline{
\psfig{ figure=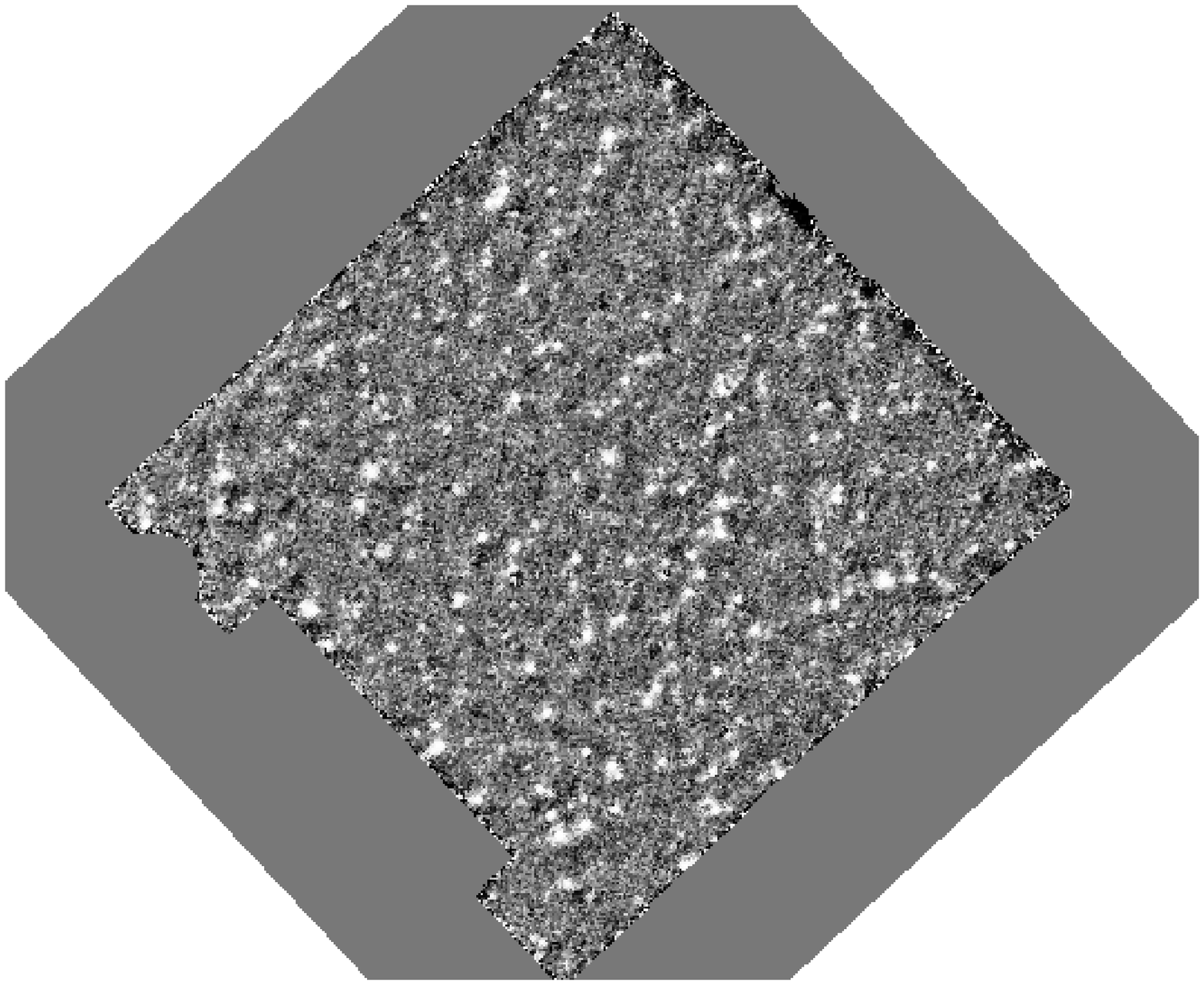,width=6cm}
\psfig{ figure=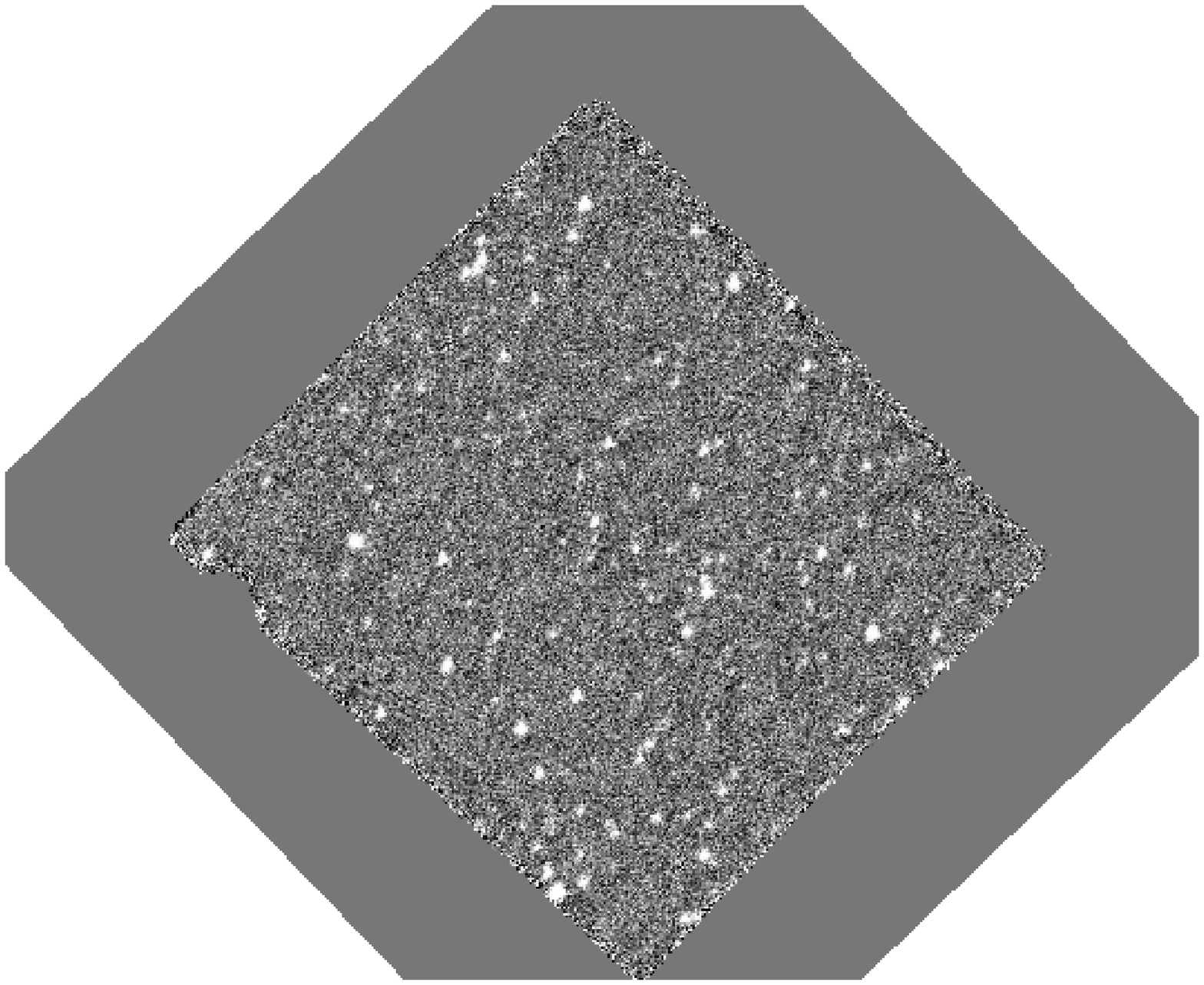,width=6cm}
\psfig{ figure=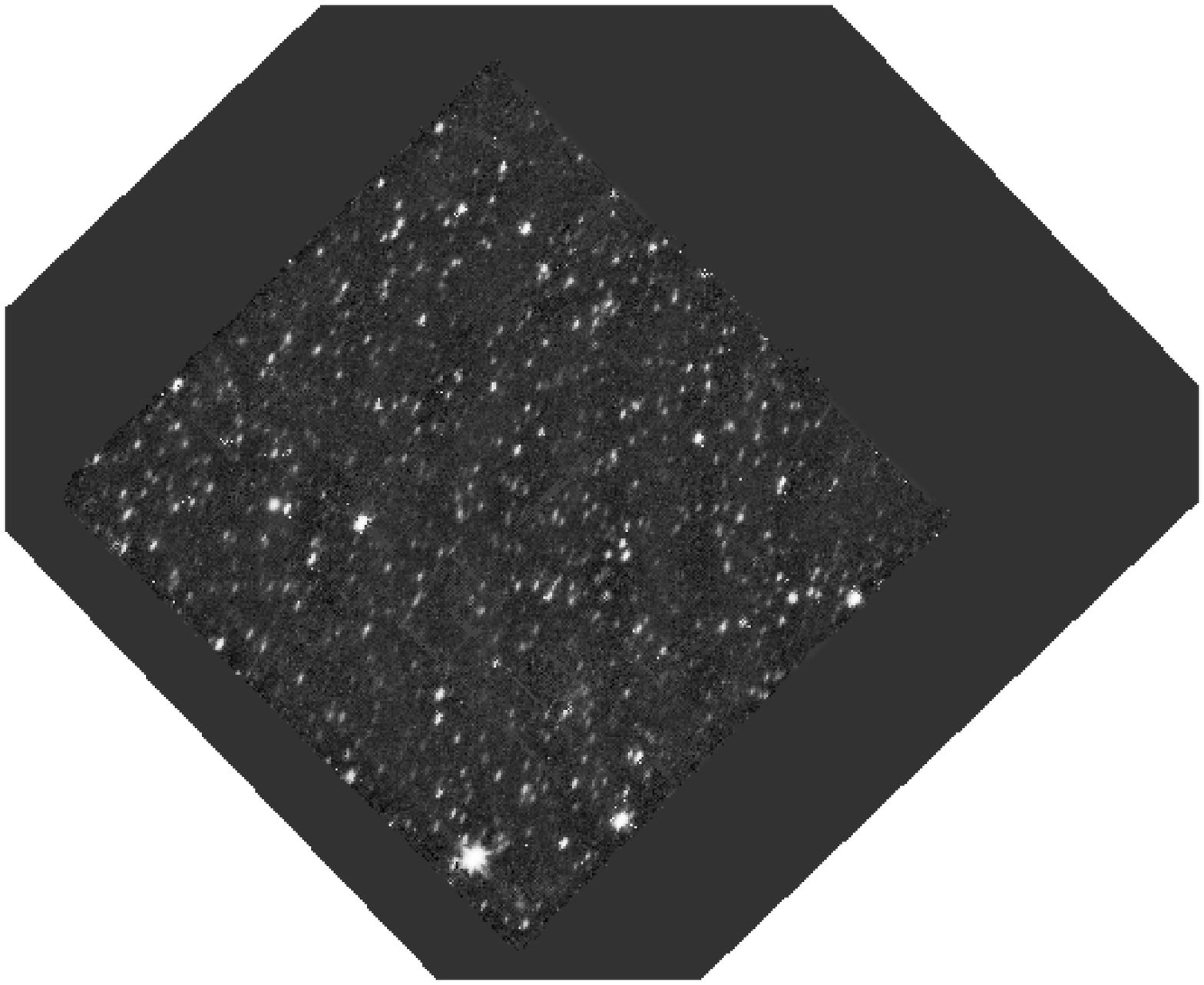,width=6cm}
}
\caption{{\it Top}: The final processed images produced by the {\it standard} IRC pipeline; ({\it left panel}) L18W band image, ({\it middle panel}) S11 band image, ({\it right panel}) N4 band image. {\it Bottom}: The final processed images produced by terminating the IRC pipeline before the distortion correction step and using the spacecraft jitter to provide dithering to correct for hot pixels described in the text; ({\it left panel}) L18W band image, ({\it middle panel}) S11 band image, ({\it right panel}) N4 band image. Note the absence of hot pixels (appearing as black spots in the {\it Top panels} in the standard processed images) particularly in the L18W and S11 band images. The banding and bad columns in the original images are also greatly improved. 
\label{fig:finalimages}}
\end{figure*}

The final co-added images in the {\it AKARI} L18W, S11 \& N4 bands are shown in the {\it bottom panels} of Figure \ref{fig:finalimages}. For comparison in the {\it top panels} of Figure \ref{fig:finalimages} we also show the final equivalent images produced by the {\it standard} IRC pipeline processing (i.e. for the original undithered observations). The hot pixels that remain after the standard processing are immediately apparent in the L18W and S11 band images. Note also the bad columns in the standard processed images that have been almost entirely corrected for in the new co-added images. 

\subsection{Astrometry and source extraction}\label{sec:sourceextraction}

Nominally the   standard IRC pipeline toolkit calculates astrometry by matching bright stars in the IRC image to reference objects using the 2MASS star catalogue. In reality, this method is successful for the IRC-NIR and MIR-S channels, however for observations taken with the IRC MR-L channel, the number of bright stars is usually insufficient to match to the 2MASS data. In this scenario,   usually data taken simultaneously in the MIR-S/NIR channels is used to project the correct astrometry onto the MIR-L images. However, in our case we have already interrupted the standard IRC pipeline toolkit before the astrometry stage, however FU-HYU by definition  has the advantage of ancillary multi-wavelength data in place a priori. Therefore, instead we register the images onto the {\it Spitzer} reference frame using the brighter 24$\mu$m {\it Spitzer} sources.

Source extraction is made using a noise-weighted matched filter following the method of  \cite{serjeant03}.
 For any signal image $I$ with associated,  noise image $N$, with point spread function $P$, and  a weight image given by $W=1/N^{2}$, the minimum $\chi^{2}$ estimator for a point source flux at any position in the images is given by;

\begin{equation}
S= \frac{((I \times W)\bigotimes P)}{W\bigotimes  P^{2}}
\end{equation}

where $\bigotimes$ denotes a convolution. The associated error on this quantity is given by;

\begin{equation}
\Delta S =\frac{1}{\sqrt{W\bigotimes  P^{2}}}
\end{equation}

The 11$\mu$m PSF was estimated directly from the image, using 101 bright point sources in the zerofootprint image (see {\it bottom panel} of figure \ref{fig:s11noise}). Inevitably, some of the point sources used to compile the PSF will have neighbouring objects. To alleviate this problem, we used an algorithm originally developed for estimating mean quasar fluxes with nearby companions (\cite{serjeanthatz09}). We assume our measurements $x_{i}$ with estimated noise levels $\sigma_{i}$ have an additional unknown noise contribution  $\sigma_{o}$ caused in this case by the RMS fluctuations from companion galaxies. The maximum-likelihood solution is then determined for the mean level $\mu$ and for  $\sigma_{o}$ simultaneously. When using this technique on all observations of any given pixel of the point spread function, it is found to be a less noisy estimator than for example, the median value. We repeated this technique for all pixels of the point spread function. While this procedure is less sensitive to outliers caused by single companions, it remains possible that some excess flux remains, nevertheless, if such a low-level excess is typical around our objects, it is appropriate and indeed necessary to include this in the point source filter. We neglect any possible variation in the PSF over the field of view at the present time. Since the L18W image does not suffer from the same small scale structure we can assume a Gaussian PSF with the standard instrument estimation of the FWHM  equal to 2.3 pixels.

\begin{figure*}
\centering
\centerline{
\psfig{ figure= 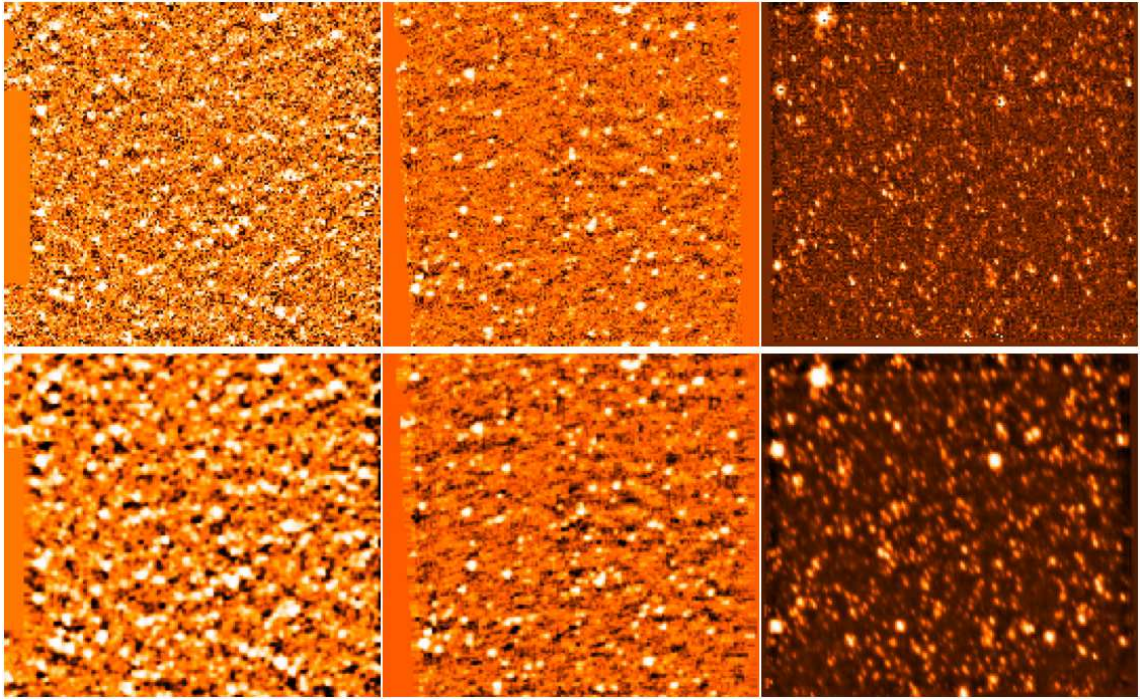,width=15cm}
}
\caption{Final GOODS-N signal-to-noise images in the L18W band ({\it left panels}),  S11 band ({\it middle panels}) \& in the N4 band ({\it right panels})  , before the point source filtering ({\it top}, scaled from -5$\sigma$ to 10$\sigma$) and after ({\it bottom}, scaled from -10$\sigma$ to 20$\sigma$).
\label{fig:snrimage}}
\end{figure*}

The best fit flux at any point in the image map is given by the value in the PSF convolved image at the extracted source position and the corresponding errors from the values in the noise map. Sources are extracted by thresholding the signal-to-noise ($\sigma=S/\Delta S$) image. In practice, objects that are separated at a high signal-to-noise threshold may be
connected at a lower threshold. We therefore thresholded the
signal-to-noise images from 3$\sigma$ to 10$\sigma$ in $\sigma$ steps, using the locations of the peak signal-to-noise as the source position, and selected only the unique sources from the resulting lists to produce our source catalogues.
 The final signal-to-noise images, before and after PSF convolution, are shown in Figure \ref{fig:snrimage}. 
 
 Note that although the standard {\it AKARI} IRC Instrument manual (\cite{lorente07}) provides conversion factors from the raw instrument ADU units to Jy, this conversion factor is based on aperture photometry of standard stars with a flux aperture and assumed sky annulus. This conversion factor is not appropriate for our source extraction algorithm and we have thus returned to the original calibration stars (see \cite{tanabe08}) and reprocessed the raw calibration data using exactly the same method as applied to the GOODS-N data in order to calculate the appropriate conversion factors for our photometry method.

In our final extracted source lists we find 233 sources at $>$3$\sigma$ in the L18W image with fluxes $>$100$\mu$Jy, and 242
$>$3$\sigma$ sources in the S11 image with fluxes $>$50$\mu$Jy. The L18W survey covers around 101.3 arcmin$^{2}$ with 1$\sigma$ noise levels below 100$\mu$Jy, and 98.3 arcmin$^{2}$ below 50$\mu$Jy. The S11 11$\mu$m survey covers 98.06 arcmin$^{2}$ with noise levels below 100$\mu$Jy, and 95.6 arcmin$^{2}$ below 50$\mu$Jy. The
median noise levels at 18$\mu$m and  11$\mu$m are 37.7$\mu$Jy and 23.0$\mu$Jy respectively. These
figures assume conversions of 1 ADU/s = 27.36$\mu$Jy at 18$\mu$m and 22.23$\mu$Jy at 11$\mu$m.
There is a possible absence of sources in the 11$\mu$m and 18$\mu$m images, tracing the bad
columns. It is possible that an over-zealous bad column correction has removed
astronomical flux. However this is also seen in  reduction using the standard IRC pipeline toolkit. The causes
(assuming it is an artefact) are still under investigation. In the N4 band we detect 340 sources down to a 3$\sigma$ flux limit of $\sim$8$\mu$Jy.

\section{Cross associations and catalogue creation}\label{sec:Association}

\subsection{Cross-associations with {\it Spitzer} IRAC \& MIPS data}\label{sec:spitzer}

As part of the FU-HYU Mission program,  \cite{negrello09} have cross-associated the {\it Spitzer} data for the  GOODS-N field with our {\it AKARI} catalogue. The  {\it Spitzer} Legacy Data Product release 3  are available in the public domain\footnote {http://data.spitzer.caltech.edu/popular/goods/Documents/goodsdataproducts.html} and consist of both the images in the four IRAC bands (3.6, 4.5, 5.8, 8.0$\mu$m) and the image and source catalogue from the MIPS 24$\mu$m band. Sources were extracted from the four IRAC images using SExtractor  (\cite{bertin96}) resulting in catalogues containing  5,792, 5,576, 2,328, 2,186 3$\sigma$ sources in the 3.6, 4.5, 5.8 \& 8.0$\mu$m bands respectively.  Sources with SExtractor  'STAR\_CLASS' equal to either 1 (cosmicray-like) or 0 (star-like) were then removed (See  \cite{negrello09}  for details). The MIPS source catalogue consists of 1,199 3$\sigma$ sources to a flux limit of  80 $\mu$Jy. However, the MIPS24 image obviously includes more sources than the catalogue and we have reprocessed the image using DAOPHOT to extract sources deeper than the 80 $\mu$Jy limit.

As a reference for our band merged catalogue we have started with the catalogue created from the deep IRAC 3.6$\mu$m band image and cross-correlated all catalogues with this. Sources between catalogues were matched over a radius of  $r = \sqrt{(\sigma_{3.6\mu m}^{2} + \sigma_{\lambda}^{2})}$, where $\sigma$ is twice the Gaussian rms width of the instrument beam (equivalent to the Full Width at Half Maximum, $FWHM/ (2\sqrt{2ln2})$) and $\lambda$ is the wavelength of the cross-match catalogue.
 Only 3.6$\mu$m sources with a single unique counterpart in at least one of the other catalogues were included in the band-merged catalogue. When no counterparts were found in a given catalogue  the flux and error were set to -1. Finally any pairs of objects sharing the same flux were removed from the final band-merged catalogue.

The final band merged catalogue contains a total of 4393 sources at 3.6$\mu$m. There are 4344, 2045 \& 1873 associations in the 4.5, 5.8 \& 8$\mu$m IRAC bands  respectively. In the MIPS 24$\mu$m band we find 1143 associations. In the {\it AKARI} bands we find  340, 181 \& 192 associations in the N4, S11 \& L18W bands respectively.

For the {\it AKARI} sources in the S11 band 13 11$\mu$m sources  and in the L18W band, 12 18$\mu$m sources have no counterpart at 3.6$\mu$m respectively and are therefore not included in the catalogue. On visual inspection we find that almost all of these sources either lie on bright sources or around the periphery of the  {\it AKARI} image making their reliability dubious.

In Table \ref{tab:catalogue} the sources with detections in all 3 {\it AKARI} bands are listed with fluxes in $\mu$Jy for the {\it Spitzer} IRAC 3.6, 4.5, 5.8, 8$\mu$m, IRS 16$\mu$m, MIPS24 bands, the {\it AKARI} IRC N4, S11, L18W bands and the {\it ISO} 15$\mu$m band. The full FU-HYU catalogue will be made available to the astronomical community in the future.

\begin{table*}
  \caption{Source catalogue in the FU-HYU GOODS-N survey for sources detected in all IRC bands with fluxes given for  the {\it Spitzer} IRAC 3.6, 4.5, 5.8, 8$\mu$m, IRS 16$\mu$m, MIPS24 bands, the {\it AKARI} IRC N4, S11, L18W bands and the {\it ISO} 15$\mu$m band.}
  \label{tab:catalogue}
  \begin{center}
    \begin{tabular}{lllllllllllll}
      \hline \hline
ID & RA & DEC & IRAC3.6 & IRAC4.5 & IRAC5.8 & IRAC8 & IRS16 & MIPS24 & N4 & S11 & L18W & ISO15 \\
     & (deg.) & (deg.) & ($\mu$Jy) & ($\mu$Jy) & ($\mu$Jy) & ($\mu$Jy) & ($\mu$Jy) & ($\mu$Jy) & ($\mu$Jy) & ($\mu$Jy) & ($\mu$Jy) & ($\mu$Jy) \\
     \hline
357 & 189.233 & 62.136 & 108.50 & 76.00 & 73.21 & 73.86 & 568.00 & 798.05 & 82.17 & 213.41 & 383.99 & ... \\
417 & 189.194 & 62.143 & 111.38 & 80.98 & 72.10 & 84.81 & ... & 990.90 & 104.39 & 115.08 & 599.85 & ... \\
447 & 189.138 & 62.143 & 102.71 & 90.25 & 91.81 & 109.02 & ... & 769.89 & 84.15 & 164.61 & 428.38 & ... \\
456 & 189.256 & 62.145 & 37.67 & 20.72 & 23.05 & 21.29 & 130.00 & 184.85 & 24.03 & 69.56 & 122.91 & ... \\
481 & 189.209 & 62.146 & 101.31 & 86.28 & 79.78 & 129.94 & 348.00 & 561.86 & 80.18 & 323.41 & 292.51 & ... \\
485 & 189.151 & 62.145 & 50.99 & 36.53 & 36.47 & 35.01 & ... & 189.86 & 36.91 & 179.11 & 148.63 & ... \\
500 & 189.167 & 62.145 & 135.39 & 86.98 & 82.80 & 426.77 & ... & 313.06 & 80.06 & 326.84 & 266.81 & ... \\
533 & 189.127 & 62.148 & 77.82 & 52.27 & 48.90 & 40.32 & ... & 380.92 & 48.77 & 99.75 & 254.62 & ... \\
884 & 189.320 & 62.169 & 75.16 & 66.17 & 54.47 & 132.94 & ... & 579.73 & 81.09 & 361.80 & 299.22 & ... \\
1082 & 189.121 & 62.180 & 98.14 & 88.33 & 76.03 & 70.62 & 465.00 & 720.49 & 84.08 & 75.10 & 433.77 & ... \\
1101 & 189.349 & 62.180 & 92.92 & 62.13 & 54.92 & 242.26 & 141.00 & 182.99 & 60.26 & 183.92 & 128.81 & ... \\
1261 & 189.089 & 62.186 & 136.10 & 102.12 & 80.44 & 68.78 & ... & ... & 123.53 & 83.09 & 290.68 & ... \\
1355 & 189.228 & 62.191 & 50.10 & 43.73 & 27.11 & 118.23 & ... & 166.28 & 35.55 & 180.29 & 103.31 & 42.00 \\
1381 & 189.330 & 62.192 & 74.11 & 52.31 & 48.52 & 50.71 & 213.00 & 201.20 & 54.04 & 171.39 & 200.44 & ... \\
1441 & 189.078 & 62.198 & 32.97 & 22.06 & 22.24 & 14.47 & ... & 155.75 & 27.19 & 60.66 & 99.02 & ... \\
1464 & 189.154 & 62.193 & 170.63 & 114.91 & 113.53 & 545.24 & ... & 788.55 & 94.52 & 370.46 & 376.82 & 300.00 \\
1471 & 189.257 & 62.196 & 60.04 & 70.31 & 69.05 & 54.47 & ... & 673.02 & 59.82 & 65.06 & 384.33 & 15.00 \\
1590 & 189.246 & 62.203 & 62.69 & 42.63 & 35.08 & 24.90 & ... & 264.50 & 45.69 & 85.11 & 179.07 & 157.00 \\
1680 & 189.316 & 62.200 & 300.51 & 192.04 & 126.56 & 101.12 & ... & 217.99 & 210.98 & 89.12 & 99.20 & ... \\
1698 & 189.090 & 62.208 & 47.05 & 42.06 & 31.48 & 72.24 & ... & 251.90 & 47.72 & 199.56 & 169.25 & ... \\
1716 & 189.153 & 62.204 & 126.41 & 83.66 & 68.90 & 50.37 & 300.00 & 395.21 & 88.33 & 101.26 & 260.65 & 202.00 \\
1760 & 189.066 & 62.210 & 70.31 & 64.61 & 41.47 & 155.21 & ... & 394.12 & 57.64 & 263.27 & 232.65 & ... \\
1771 & 189.144 & 62.204 & 316.28 & 244.52 & 196.02 & 338.39 & 853.00 & 1309.87 & 227.87 & 858.14 & 896.83 & 448.00 \\
1827 & 189.225 & 62.215 & 58.38 & 37.31 & 36.45 & 25.33 & 207.00 & 134.91 & 38.75 & 91.85 & 180.88 & 179.00 \\
1885 & 189.090 & 62.217 & 41.04 & 28.58 & 31.71 & 27.42 & ... & ... & 35.14 & 61.65 & 129.56 & ... \\
1907 & 189.081 & 62.215 & 129.55 & 105.91 & 100.16 & 167.43 & ... & 975.50 & 106.52 & 467.30 & 595.12 & ... \\
1985 & 189.140 & 62.222 & 49.92 & 32.84 & 34.63 & 23.08 & 320.00 & 319.17 & 40.65 & 68.49 & 192.08 & 122.00 \\
2109 & 189.096 & 62.230 & 39.30 & 35.63 & 30.37 & 69.09 & ... & 296.78 & 37.49 & 236.64 & 247.21 & ... \\
2253 & 189.074 & 62.236 & 52.69 & 57.54 & 56.07 & 46.33 & ... & 447.95 & 58.73 & 61.34 & 300.72 & ... \\
2345 & 189.041 & 62.240 & 63.87 & 40.00 & 42.13 & 26.40 & ... & 293.24 & 48.45 & 91.51 & 199.19 & ... \\
2355 & 189.058 & 62.238 & 59.06 & 48.61 & 36.09 & 213.25 & ... & 222.45 & 55.01 & 260.37 & 215.32 & ... \\
2412 & 189.201 & 62.241 & 86.26 & 60.13 & 54.77 & 386.56 & 283.00 & 463.48 & 68.03 & 363.90 & 290.78 & 307.00 \\
2483 & 189.148 & 62.240 & 71.64 & 99.93 & 163.45 & 282.56 & 615.00 & 1487.41 & 89.34 & 368.62 & 755.66 & 441.00 \\
2558 & 189.250 & 62.247 & 50.47 & 33.56 & 39.29 & 32.95 & ... & 453.35 & 37.32 & 165.57 & 310.12 & 295.00 \\
2769 & 189.096 & 62.257 & 46.39 & 51.08 & 72.85 & 130.06 & 335.00 & 519.55 & 50.95 & 159.34 & 347.93 & ... \\
2788 & 189.165 & 62.257 & 46.59 & 40.28 & 27.90 & 66.68 & 121.00 & 139.45 & 39.61 & 120.86 & 135.86 & ... \\
2885 & 189.193 & 62.258 & 84.54 & 56.75 & 54.15 & 42.26 & 433.00 & 524.84 & 59.16 & 119.42 & 350.03 & 418.00 \\
2955 & 189.132 & 62.268 & 44.94 & 29.99 & 30.94 & 23.77 & 245.00 & 286.83 & 32.54 & 86.56 & 238.93 & ... \\
2979 & 189.261 & 62.262 & 143.18 & 131.91 & 132.58 & 158.66 & ... & 482.65 & 125.25 & 265.34 & 297.76 & ... \\
3023 & 189.176 & 62.263 & 126.59 & 102.57 & 106.10 & 120.96 & ... & 829.52 & 129.85 & 190.16 & 457.31 & 459.00 \\
3073 & 189.252 & 62.271 & 41.08 & 29.99 & 32.23 & 28.41 & ... & 490.26 & 31.41 & 78.12 & 285.35 & ... \\
3107 & 189.145 & 62.275 & 58.86 & 40.10 & 38.46 & 34.78 & 368.00 & 458.81 & 41.26 & 104.18 & 299.68 & ... \\
3191 & 189.245 & 62.277 & 81.26 & 70.72 & 48.76 & 148.61 & ... & 282.12 & 64.82 & 272.17 & 260.71 & ... \\
3233 & 189.158 & 62.297 & 162.41 & 120.65 & 72.01 & 132.01 & ... & 385.86 & 108.53 & 294.75 & 313.63 & ... \\
3398 & 189.225 & 62.294 & 80.09 & 72.90 & 56.07 & 143.53 & ... & 349.00 & 72.39 & 345.16 & 318.29 & ... \\
3479 & 189.183 & 62.285 & 30.85 & 25.36 & 20.70 & 42.54 & ... & 210.30 & 19.17 & 95.30 & 114.45 & ... \\
3725 & 189.277 & 62.292 & 94.69 & 83.88 & 57.85 & 150.07 & ... & 303.69 & 76.49 & 285.99 & 242.33 & ... \\
    \hline
    \end{tabular}
  \end{center}
\end{table*}

\subsection{Cross-associations with other data}\label{sec:xcorrelate}

In addition to the {\it Spitzer} IRAC \& MIPS data, the GOODS-N field enjoys copious multi-wavelength coverage at other wavelengths. Using the same matching criteria described above we have added additional bands to our FU-HYU-GOODS-N catalogue. A 36 square arcminute area of the GOODS-N field was also imaged by {\it Spitzer}  at 16$\mu$m down to a flux level of $\sim$0.08mJy  by \cite{teplitz05} using the Infra-Red Spectrograph (IRS) blue peak-up filter, detecting around 153 sources. Cross-correlating with our data set we find a total of 83 sources in our catalogue.
The GOODS-N field was also imaged by {\it ISO} at 15$\mu$m, down to 0.2mJy detecting of the order of 100 objects (\cite{aussel99}). We find numerous close pairs with similar flux levels and after  selecting unique sources we find 59 associations with our catalogue.

The GOODS-N field has also been imaged at optical wavelengths in the U,B,V,R,I,z$^ \prime$ bands over the entire field and partially in H,K$^ \prime$ using both space-bourne Hubble Space Telescope / Advanced Camera for Surveys (HST/ACS) and the Subaru, KPNO \& Hawaii 2.2m ground based observatories (\cite{capak04}). we find approximately 3600 sources in the UBVRIz$^ \prime$ bands associated with sources in our catalogue. Specifiacally, at the UBVRI, z$^ \prime$, HK$^ \prime$ detection limits of 27.1, 26.9, 26.8, 26.6, 25.6, 25.4, 22.1, 22.1 AB magnitudes we associate 3536, 3609, 3617, 3586, 3617, 3612, 3380, 3380 sources in our catalogue.

In the ultra-violet, the GOODS field has been observed as part of the surveys made by the Galaxy Evolution Explorer (GALEX, \cite{martin05}) in two bands (232nm and 154nm) referred to as the near-uv (nuv) and far-UV (fuv). For cross-associations with our FU-HYU catalogue we use the data of \cite{morrissey07} with fluxes derived from the data reduction pipeline of  \cite{burgarella07}. In total we find 829 associations with our catalogue in both the GALEX bands.

\smallskip
\section{Results}\label{sec:results}

\subsection{Final catalogue}\label{sec:finalcatalogue}
The final catalogue contains data for 4393 sources over 19 bands from mid-infrared to UV wavelengths and their associated errors. However, in the final product, we find only 4 sources with data in all 19 bands. Concentrating on our near to mid-infrared sources only we find a total of 7 sources with fluxes in all 10 infrared bands (i.e. {\it Spitzer} IRAC 3.6, 4.5, 5.8, 8$\mu$m, IRS 16$\mu$m, MIPS 24$\mu$m, {\it AKARI} IRC N4, S11, L18W \& {\it ISO} 15$\mu$m bands). Two of the 3/7 sources without data in all 19 bands have no detection in the GALEX bands and the other of the 3/7 sources has no detection in either the GALEX bands or optical wavelengths. 

The total number of sources with complete infrared coverage increases to 17 sources if the constraint on the {\it ISO} detection is relaxed. Removing both the relatively smaller 16$\mu$m {\it Spitzer}-IRS and 15$\mu$m {\it ISO} samples we retrieve a sample of around 50 sources. Whilst relaxing the constraint on the {\it AKARI} N4 band we are left with a sample of 100 sources with complete coverage of their infrared spectrum from 3-24$\mu$m (See Figure \ref{fig:bandcoverage}). We refer to this hereafter as the  the {\it total mid-infrared (MIR) flux} (the total flux in the {\it Spitzer} IRAC 3.6, 4.5, 5.8, 8$\mu$m, {\it AKARI} S11, L18W \& {\it Spitzer} MIPS 24$\mu$m bands).

\begin{figure}
\centering
\centerline{
\psfig{ figure=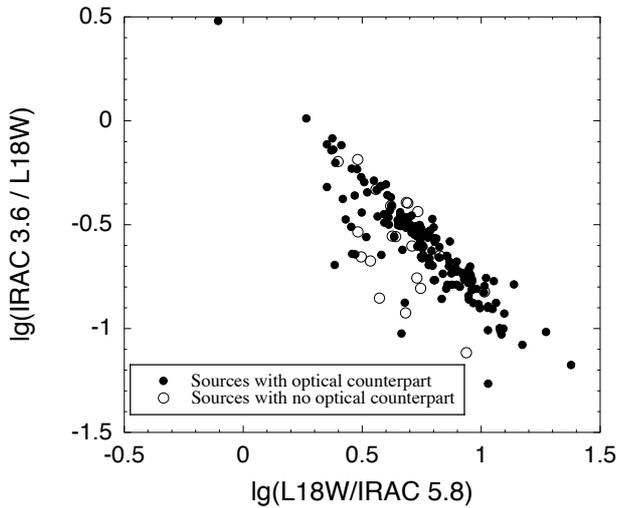,width=8cm}
}
\caption{Infrared colour-colour diagram for sources n the FU-HYU-GOODS-N field with and without optical counterparts. Colours are IRAC 3.6$\mu$m / IRC L18W \&  IRC L18W / IRAC 5.8$\mu$m. There is no discernible difference in the colour-colour space populated by the sources without optical counterparts.
\label{fig:optID}}
\end{figure}

There is also a sample of 27 sources with no optical counterpart in any of the optical bands with {\it Spitzer} IRAC, MIPS and {\it AKARI} MIR-L identifications.  To investigate the significance of this population we plot in Figure ~\ref{fig:optID} the infrared colours of the sources without optical counterparts along with the sources with optical counterparts. It is found that the two populations occupy the same colour-colour space with no discernible differences.

\subsection{Sources with full infrared coverage}\label{sec:seds}
\begin{figure*}
\centering
\centerline{
\psfig{figure=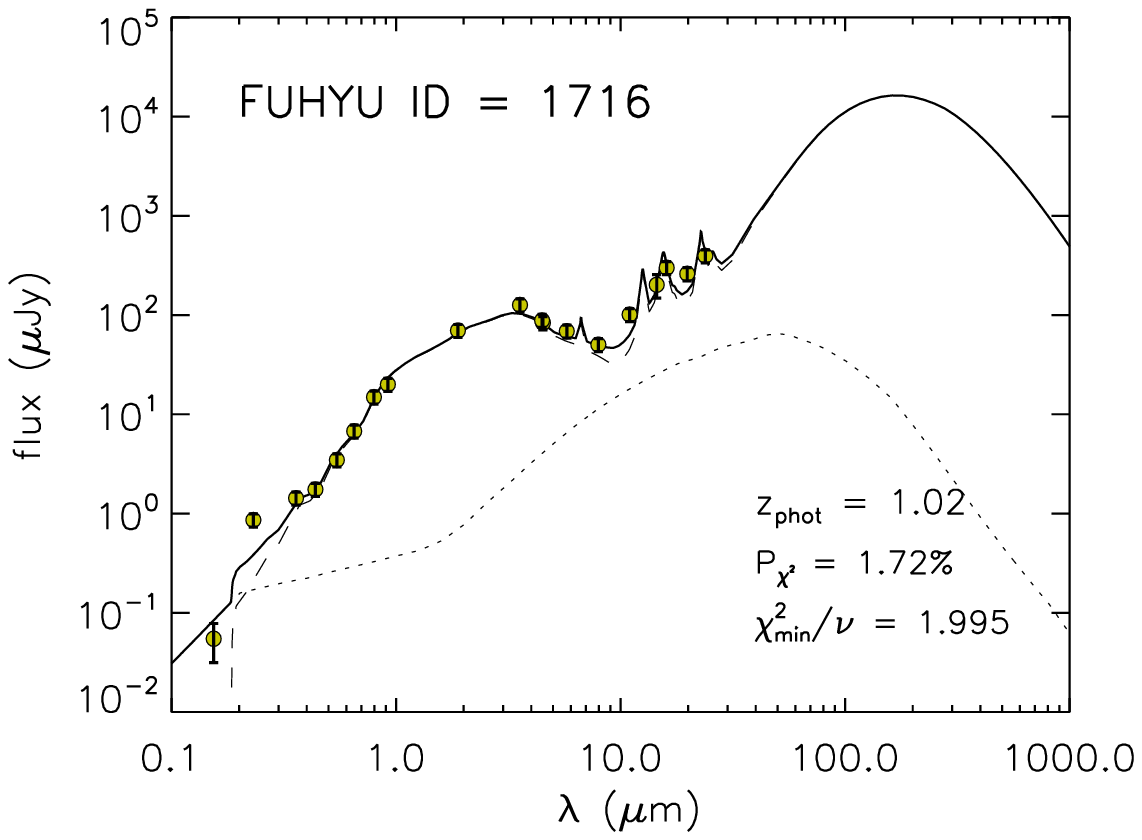,  width=5cm}
\psfig{figure=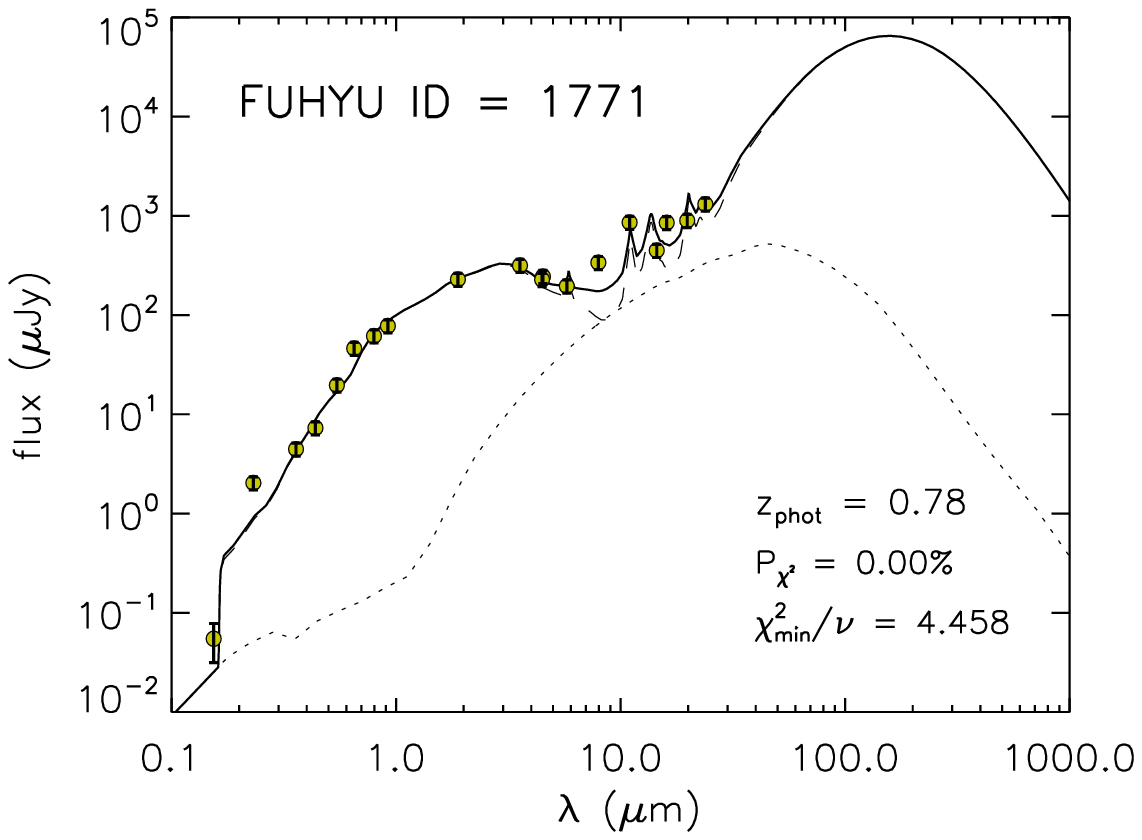, width=5cm}
\psfig{figure=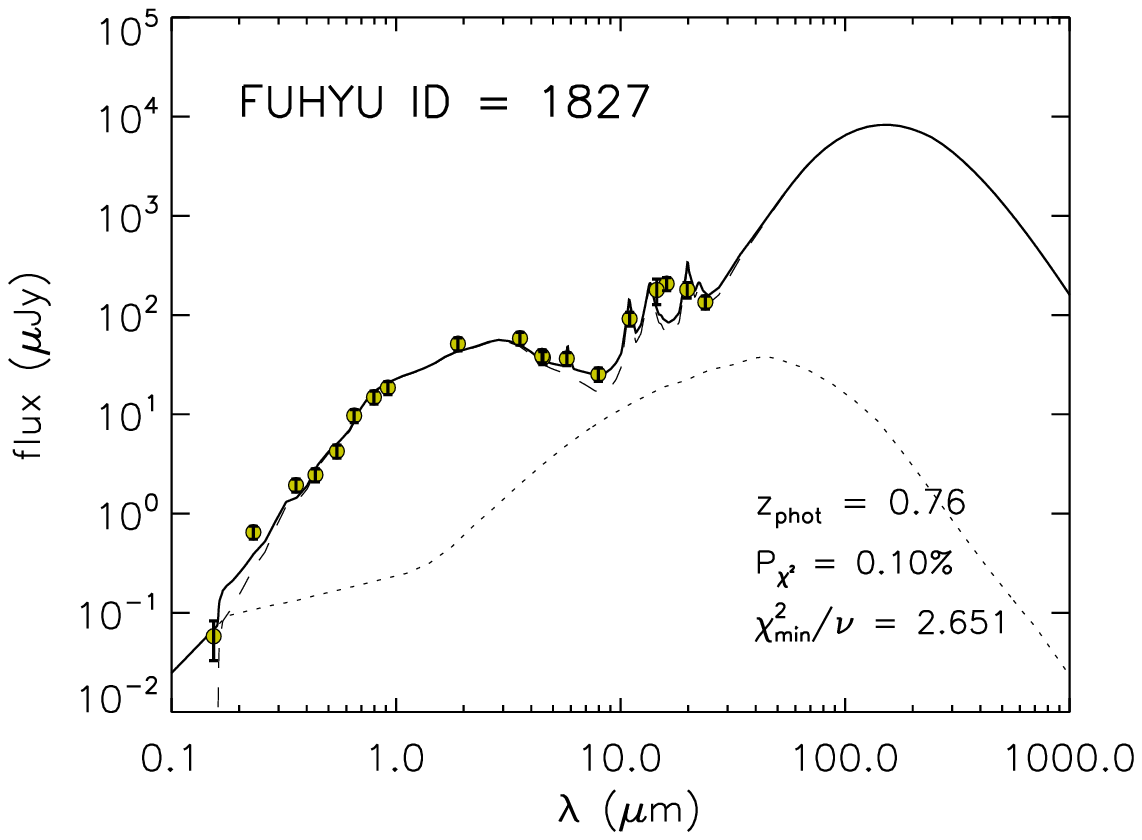, width=5cm}
\psfig{figure=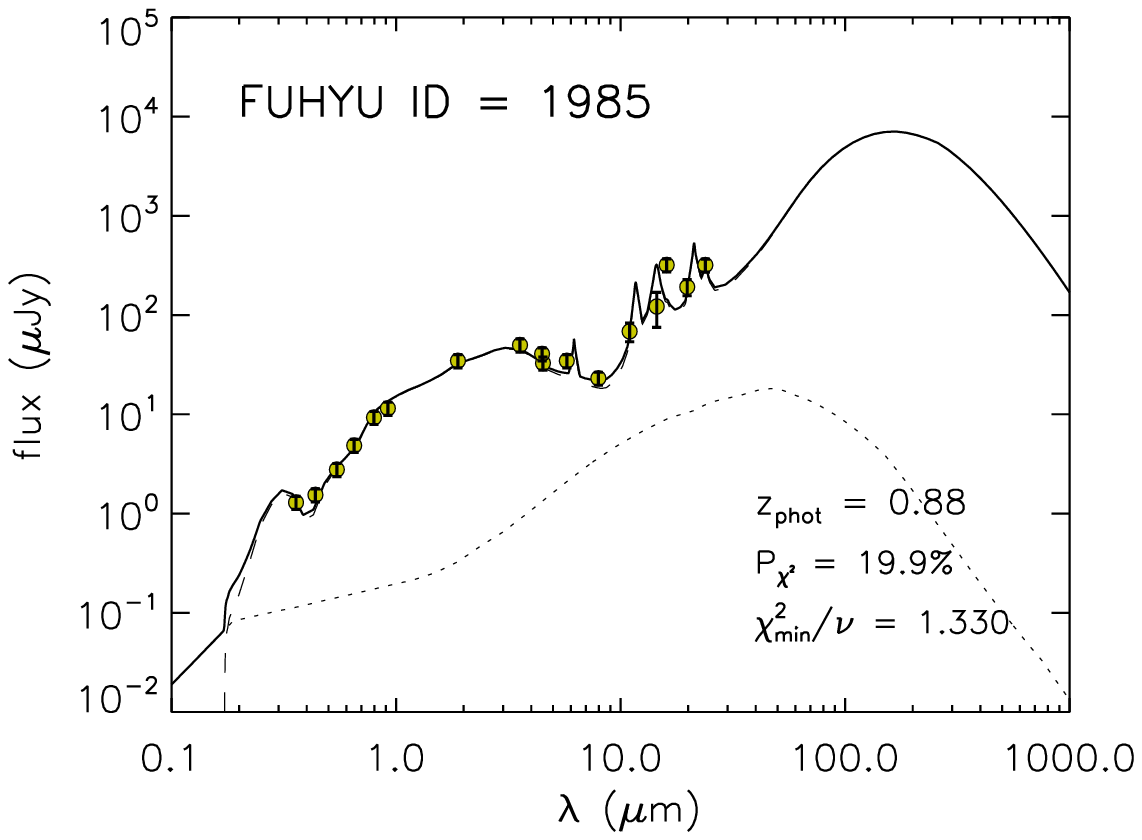, width=5cm}
}
\vspace{0.2cm}
\centerline{
\psfig{figure=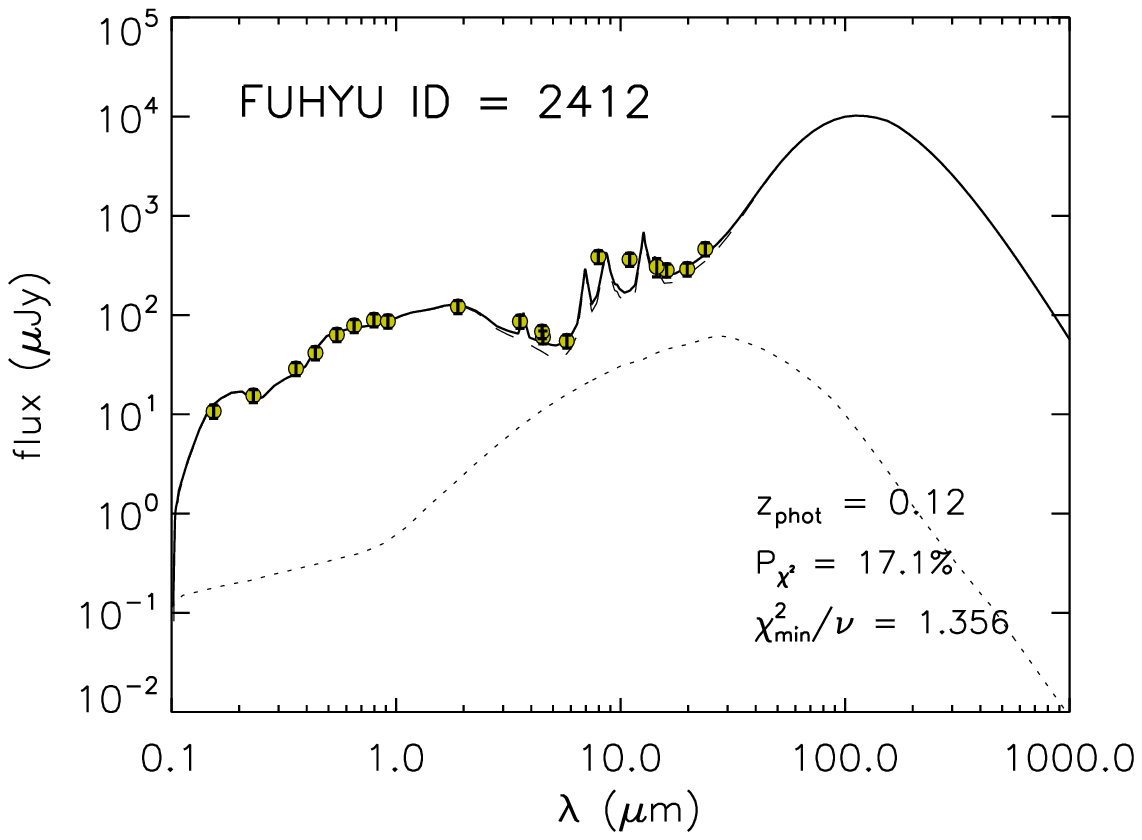, width=5cm}
\psfig{figure=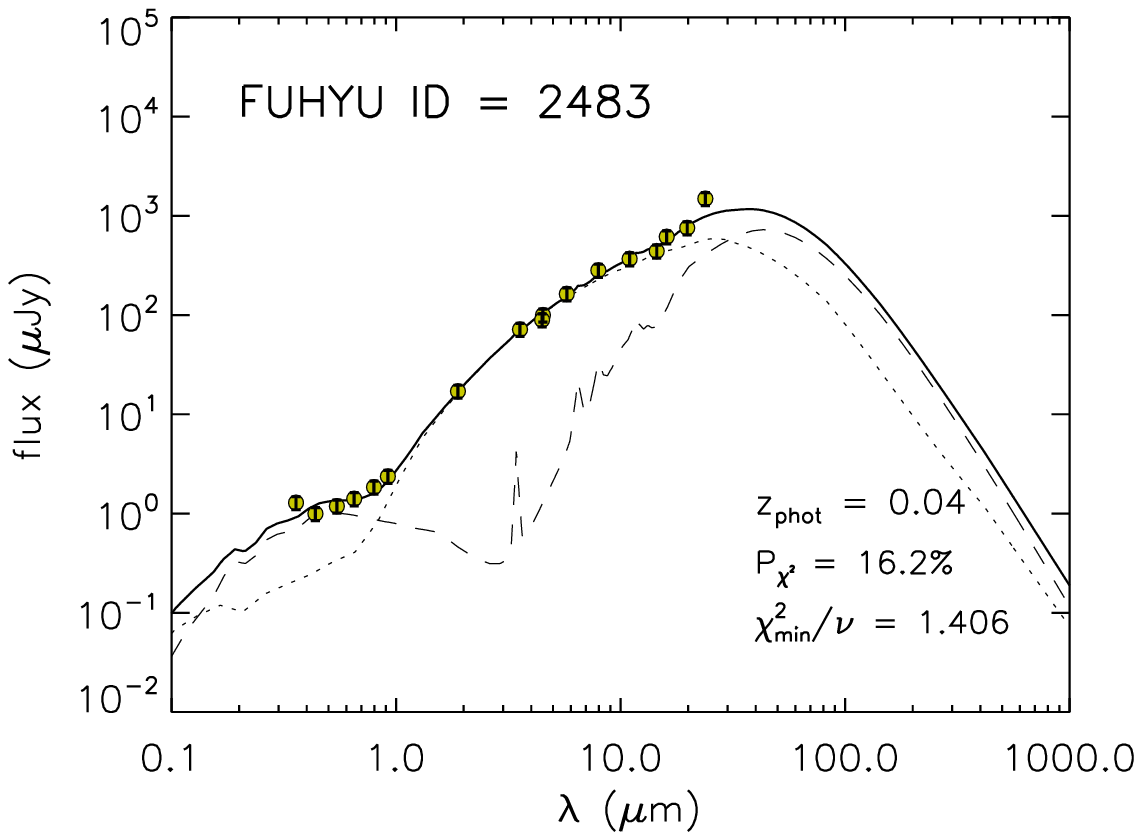,  width=5cm}
\psfig{figure=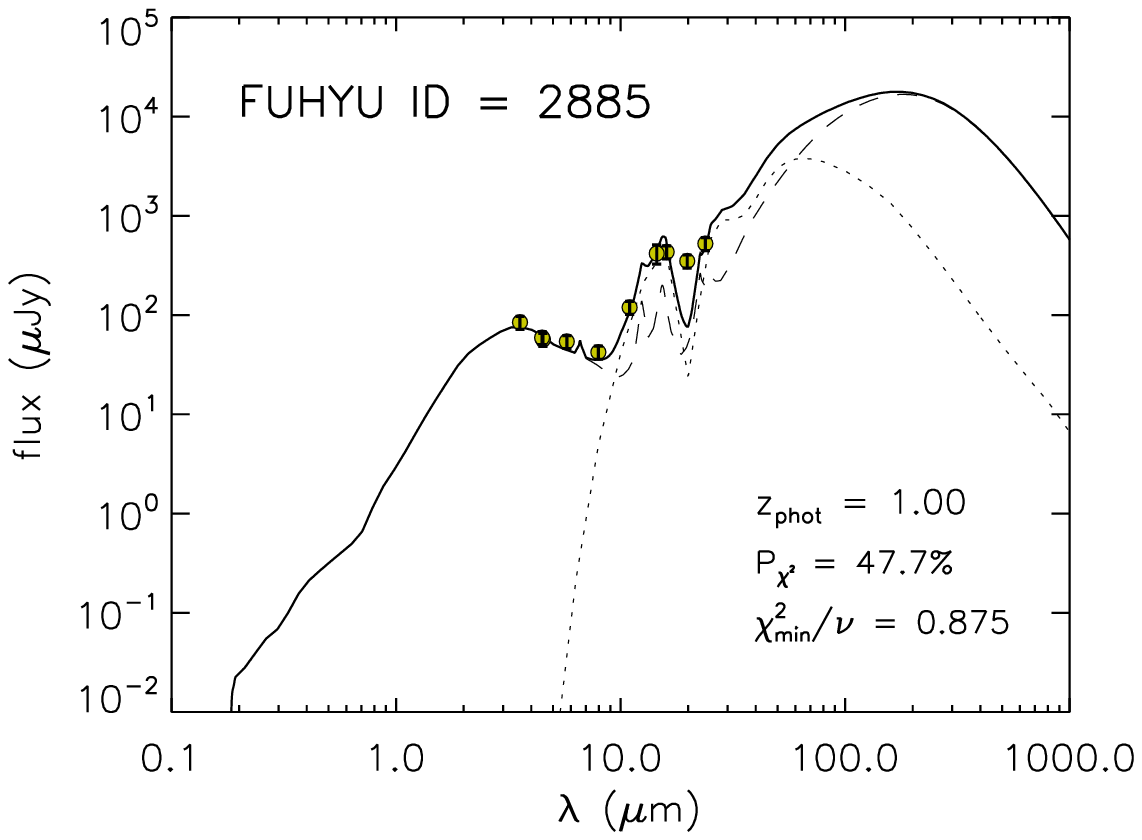,  width=5cm}
}
\caption{Photometric data for the seven sources in the FU-HYU GOODS-N field with fluxes in all 10 infrared bands (i.e. {\it Spitzer} IRAC 3.6, 4.5, 5.8, 8$\mu$m, IRS 16$\mu$m, MIPS 24$\mu$m, {\it AKARI} IRC N4, S11, L18W \& {\it ISO} 15$\mu$m bands).  Also shown are the model fits for these sources using the photo-metric redshift code of \cite{negrello09}. The {\it dashed} and  {\it dotted lines} are the fitted starburst and AGN components from the spectral libraries of  \cite{takagi03} \& \cite{efstathiou95} respectively and the {\it thick solid line} is the composite fit. Photometric redshift fitting parameters are listed in each panel and tabulated in Table \ref{tab:allbandfits} following \cite{negrello09}.
\label{fig:sedplots}}
\end{figure*}

The spectral energy distributions (SED) of the 7 sources with complete infrared coverage  in all 10 {\it AKARI},  {\it Spitzer} \&  {\it ISO} infrared bands  are shown in Figure \ref{fig:sedplots} and listed below. In  Figure \ref{fig:sedplots} the best spectral fit and photometric redshift from a combination of starburst and AGN templates using the algorithm of \cite{negrello09} is shown.  The starburst templates are taken from the libraries of \cite{takagi03} which use three types of dust model to model the extinction, referred to as the Milky Way (MW), Large Magellanic Cloud (LMC), and Small Magellanic Cloud (SMC) dust models
and are a function of the metallicity. Model parameters are the star formation timescale (100Myr), metallicity (assumed to be 0.1), the compactness of the starburst region ($\Theta$) and the starburst age. The variation in the spectra is explained by the difference in the age (ranging from 0.01-6Gyr) and the compactness of the starburst region (ranging from 0.3 -- 5.0 which can be considered as a measure of the optical depth with values of $\Theta$ of 0.3 and 5.0 corresponding to high and low optical depths respectively).  The AGN templates are taken from the libraries of \cite{efstathiou95} who assume a tapered disc geometry for the AGN with the the torus height increasing with distance from the central source, tapering to a constant value at large distances. The tapered
disc is assumed to have a $1/r$ density distribution and a ratio between the outer and the inner radius of the torus of  20. The spectra of the torus depends only upon the viewing angle of the torus, $\theta_{\rm view}$, with $\theta_{\rm view}$=0$^{\circ}$, 90$^{\circ}$ corresponding to an edge on and face on torus respectively.
The model parameters used for the fits, and the resulting redshift and  luminosity contributed by the starburst and AGN components respectively are tabulated in Table \ref{tab:allbandfits}.  All the quoted errors represent the 99 per cent confidence interval.

\begin{description}
\item[{\bf FU-HYU ID=G1716}] 
This source has a 24$\mu$m flux of 395$\pm$8$\mu$Jy and an {\it ISO} 15$\mu$m flux of  202$\pm$58$\mu$Jy (HDF PM3 8 in the source list of \cite{aussel99}). The best spectral fit is a PAH dominated starburst (i.e ULIRG with L$\sim$10$^{12}L_{\sun}$) at redshift 1.02.  
\item[{\bf FU-HYU ID=G1771}] 
This is a bright 24$\mu$m source with flux of 1310$\pm$12$\mu$Jy. The  {\it ISO} 15$\mu$m flux is  448$\pm$68$\mu$Jy (HDF PM3 2 in the source list of \cite{aussel99}). The best spectral fit is a PAH starburst galaxy (i.e ULIRG with L=10$^{12.3}L_{\sun}$)  with underlying mid-infrared AGN at redshift 0.78.
\item[{\bf FU-HYU ID=G1827}] 
This source has a relatively faint 24$\mu$m flux of 134.91$\pm$6$\mu$Jy. The  {\it ISO} 15$\mu$m flux is  179$\pm$60$\mu$Jy (HDF PM3 33 in the source list of \cite{aussel99}). The best spectral fit  is a PAH dominated luminous infrared galaxy (LIRG,  L=10$^{11.4}L_{\sun}$)  at redshift 0.76.
\item[{\bf FU-HYU ID=G1985}] 
This source has a 24$\mu$m flux of 319.7$\pm$12$\mu$Jy. The  {\it ISO} 15$\mu$m flux is  122$\pm$54$\mu$Jy (HDF PS3 3 in the source list of \cite{aussel99}). The best spectral fit  is a  pure PAH dominated luminous infrared galaxy (LIRG,  L=10$^{11.5}L_{\sun}$)   at redshift 0.88 with strong silicate absorption.
\item[{\bf FU-HYU ID=G2412}] 
This source has a 24$\mu$m flux of 463$\pm$7$\mu$Jy. The  {\it ISO} 15$\mu$m flux is  307$\pm$62$\mu$Jy (HDF PM3 24 in the source list of \cite{aussel99}). The best spectral fit is a pure PAH dominated moderate starburst (L=10$^{10.1}L_{\sun}$)  at redshift 0.12.
\item[{\bf FU-HYU ID=G2483}] 
This source is the brightest infrared source in our sample of complete coverage with a power law spectrum and  24$\mu$m flux of 1487$\pm$11$\mu$Jy. The  {\it ISO} 15$\mu$m flux is  441$\pm$43$\mu$Jy (HDF PM3 5 in the source list of \cite{aussel99}).The best spectral fit from \cite{negrello09} is an AGN dominated source with a featureless mid-infrared pectrum at redshift 0.04. Although the model fitting predicts and roughly equal partition between the starburst and AGN luminosities, the uncertainty on the starburst luminosity is large. This source has no detections in the  GALEX UV bands.
\item[{\bf FU-HYU ID=G2885}]
This source has a 24$\mu$m flux of 524$\pm$8$\mu$Jy. The  {\it ISO} 15$\mu$m flux is  418$\pm$91$\mu$Jy (HDF PM3 21 in the source list of \cite{aussel99}). The best spectral fit from \cite{negrello09} is a composite starburst \& AGN at redshift 1.0 and would be classed as a ULIRG with L=10$^{12}L_{\sun}$ . This source has no detections in either the UBVRIz$^ \prime$HK$^ \prime$ optical bands or the GALEX UV bands.
\end{description}

\begin{table*}
  \vspace{0.0cm}
  \begin{center}
    \small
    \caption{Best fit results for the 7 objetcs in Figure \ref{fig:sedplots}.}
  \label{tab:allbandfits}
    \vspace{-0.2cm}
    \begin{tabular}{lrrrrrlrrrrr}
      \hline
      \hline
      ID & $z_{\rm phot}$ & $\chi^{2}_{\rm min}/\nu$ & $P_{\chi^2}$ & $\nu$ & ext. & Age & $\Theta$ & $\theta_{\rm view}$ & $\log[L_{sb}]$ & $\log[L_{agn}]$ & $\log[L_{tot}]$ \\
         &              &                         &             &       &      & (Myr) &       & (deg)             &  ($L_{\odot}$) &  ($L_{\odot}$) &  ($L_{\odot}$)  \\
      \hline
1716 & 1.02$_{-0.27}^{+0.12}$ & 1.99 & 0.01 & 13 & LMC & 500$_{-100}^{+100}$ & 1.4$_{-0.1}^{+0.6}$ & 47$_{-3}^{+42}$ & 12.0$_{-0.11}^{+0.16}$ & 10.4$_{-0.51}^{+0.17}$ & 12.0$_{-0.11}^{+0.16}$ \\
1771 & 0.78$_{-0.11}^{+0.93}$ & 4.45 & 0.00 & 13 & SMC & 400$_{-370}^{+200}$ & 1.6$_{-0.4}^{+1.4}$ & 42$_{-3}^{+3}$ & 12.3$_{-0.31}^{+0.93}$ & 10.9$_{-0.02}^{+1.09}$ & 12.3$_{-0.27}^{+0.92}$ \\
1827 & 0.76$_{-0.14}^{+0.07}$ & 2.65 & 0.00 & 13 & LMC & 500$_{-100}^{+100}$ & 1.6$_{-0.1}^{+0.1}$ & 47$_{-3}^{+10}$ & 11.4$_{-0.23}^{+0.12}$ & 9.92$_{-0.18}^{+0.07}$ & 11.4$_{-0.24}^{+0.11}$ \\
1985 & 0.88$_{-0.15}^{+0.27}$ & 1.33 & 0.19 & 11 & MW & 500.$_{-470.}^{+100.}$ & 2.0$_{-1.3}^{+0.4}$ & 50$_{-42}^{+3}$ & 11.5$_{-0.06}^{+0.47}$ & 9.87$_{-0.29}^{+1.41}$ & 11.5$_{-0.06}^{+0.48}$ \\
2412 & 0.12$_{-0.06}^{+0.06}$ & 1.35 & 0.17 & 13 & MW & 600$_{-100}^{+0}$ & 3.0$_{-0.1}^{+0.1}$ & 47$_{-3}^{+3}$ & 10.0$_{-0.53}^{+0.53}$ & 8.52$_{-1.04}^{+0.38}$ & 10.1$_{-0.65}^{+0.41}$ \\
2483 & 0.04$_{-0.03}^{+2.53}$ & 1.40 & 0.16 & 11 & LMC & 10$_{-0}^{+590}$ & 1.2$_{-0.9}^{+1.8}$ & 45$_{-3}^{+3}$ & 9.67$_{-1.77}^{+3.38}$ & 9.24$_{-1.73}^{+3.44}$ & 10.7$_{-2.65}^{+2.52}$ \\
2885 & 1.00$_{-0.38}^{+0.48}$ & 0.87 & 0.47 & 4 & SMC & 600$_{-590}^{+100}$ & 0.7$_{-0.4}^{+4.3}$ & 8$_{-3}^{+30}$ & 11.9$_{-0.17}^{+0.51}$ & 11.5$_{-1.71}^{+0.02}$ & 12.0$_{-0.28}^{+0.40}$ \\
      \hline
\multicolumn{12}{l}{  $z_{\rm phot}$: Best fit photometric redshift }\\
\multicolumn{12}{l}{ $\chi^{2}_{\rm min}/\nu$: reduced minimum $\chi^2$ for  $\nu$ degrees of freedom }\\
\multicolumn{12}{l}{ $P_{\chi^2}$: probability associated to the minimum $\chi^2$  }\\
\multicolumn{12}{l}{ ext.: extinction curve }\\
\multicolumn{12}{l}{ Age: age of the starburst }\\
\multicolumn{12}{l}{ $\Theta$: starburst compactness factor }\\
\multicolumn{12}{l}{ $\theta_{\rm view}$: viewing angle of the AGN torus }\\
\multicolumn{12}{l}{ $\log[L_{sb}]$:  Luminosity contributed by the starburst component }\\
\multicolumn{12}{l}{ $\log[L_{agn}]$: Luminosity contributed by the AGN component }\\
\multicolumn{12}{l}{  $\log[L_{tot}]$: Total Luminosity }\\
    \end{tabular}
  \end{center}
\end{table*}

\smallskip
\subsection{Source fluxes and colours as a function of the total mid-infrared emission}\label{sec:totalmir}

\begin{figure*}
\centering
\centerline{
\psfig{figure=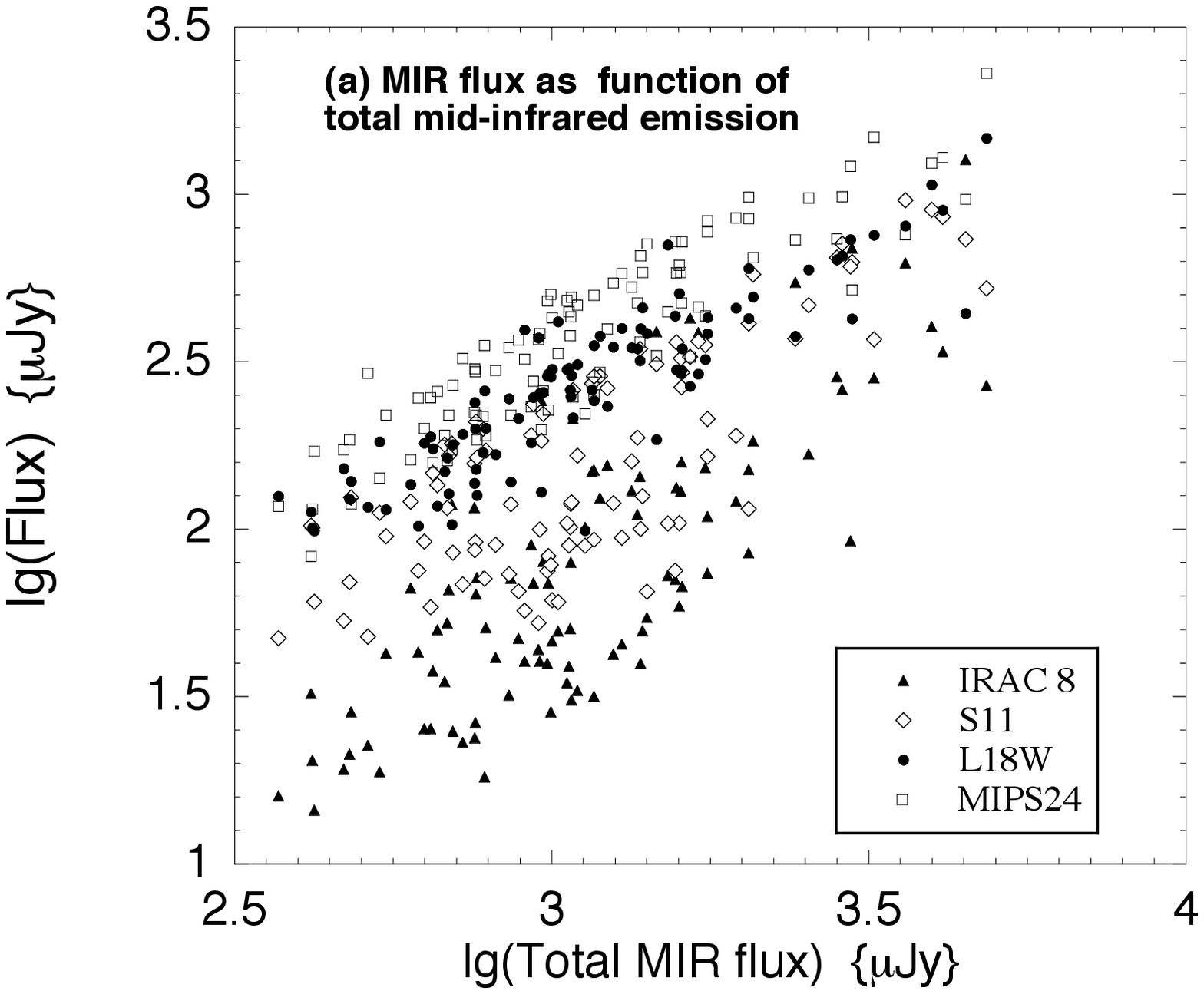,width=9cm}
\psfig{figure=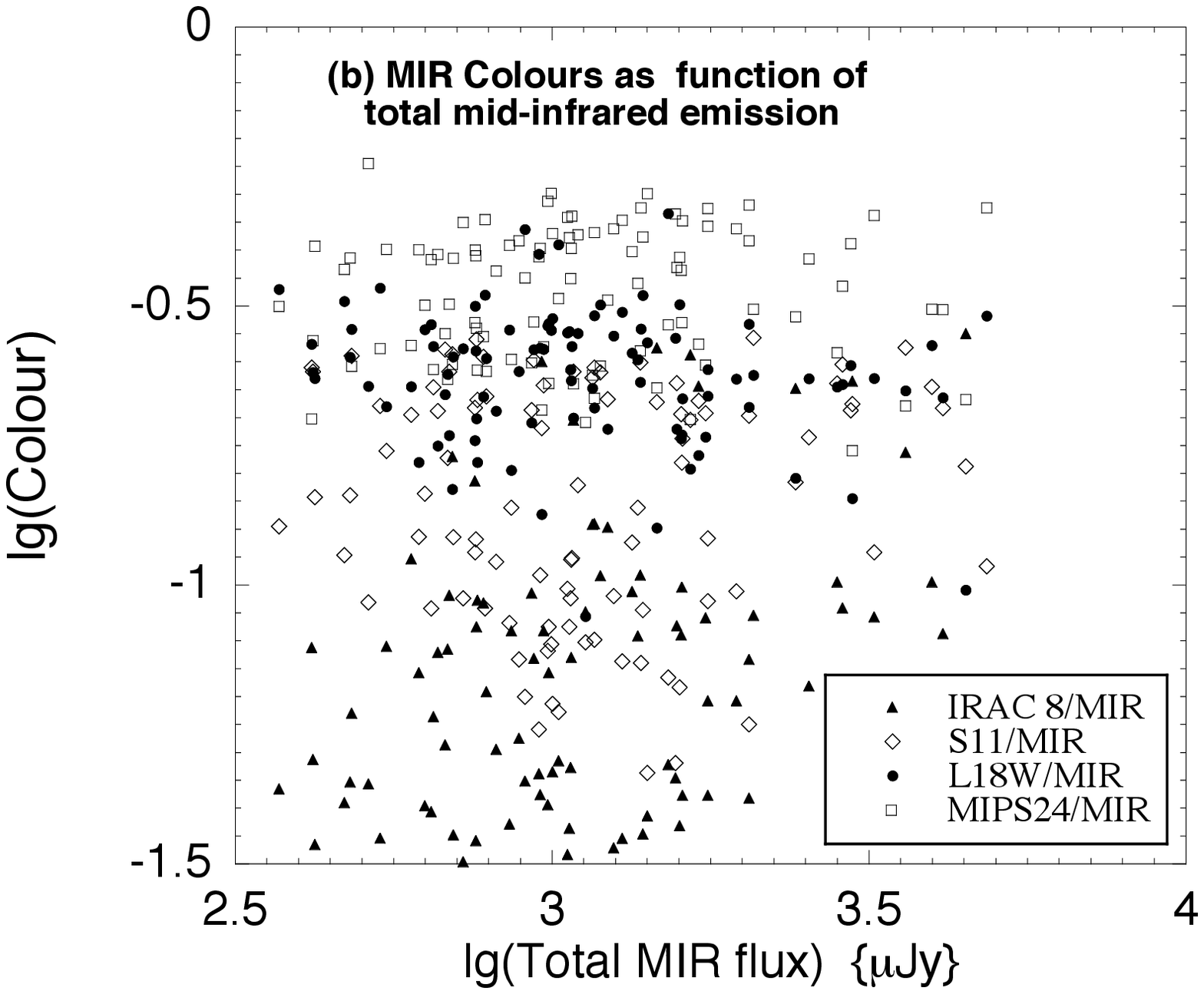,width=9cm}
}
\caption{ {\bf (a)}: Mid-infrared source fluxes in  the  {\it Spitzer} \&  {\it AKARI} bands as a function of the total mid-infrared flux (the total flux in the {\it Spitzer} IRAC 3.6, 4.5, 5.8, 8$\mu$m, {\it AKARI} S11, L18W \& {\it Spitzer} MIPS 24$\mu$m bands) for the sources in the FU-HYU-GOODS-N field. {\bf (b)}: Mid-infrared colours  (mid-IR flux/Total MIR flux) as a function of the total mid-infrared flux for the sources in the FU-HYU-GOODS-N field.
\label{fig:mirir}}
\end{figure*}

\begin{figure*}
\centering
\centerline{
\psfig{figure=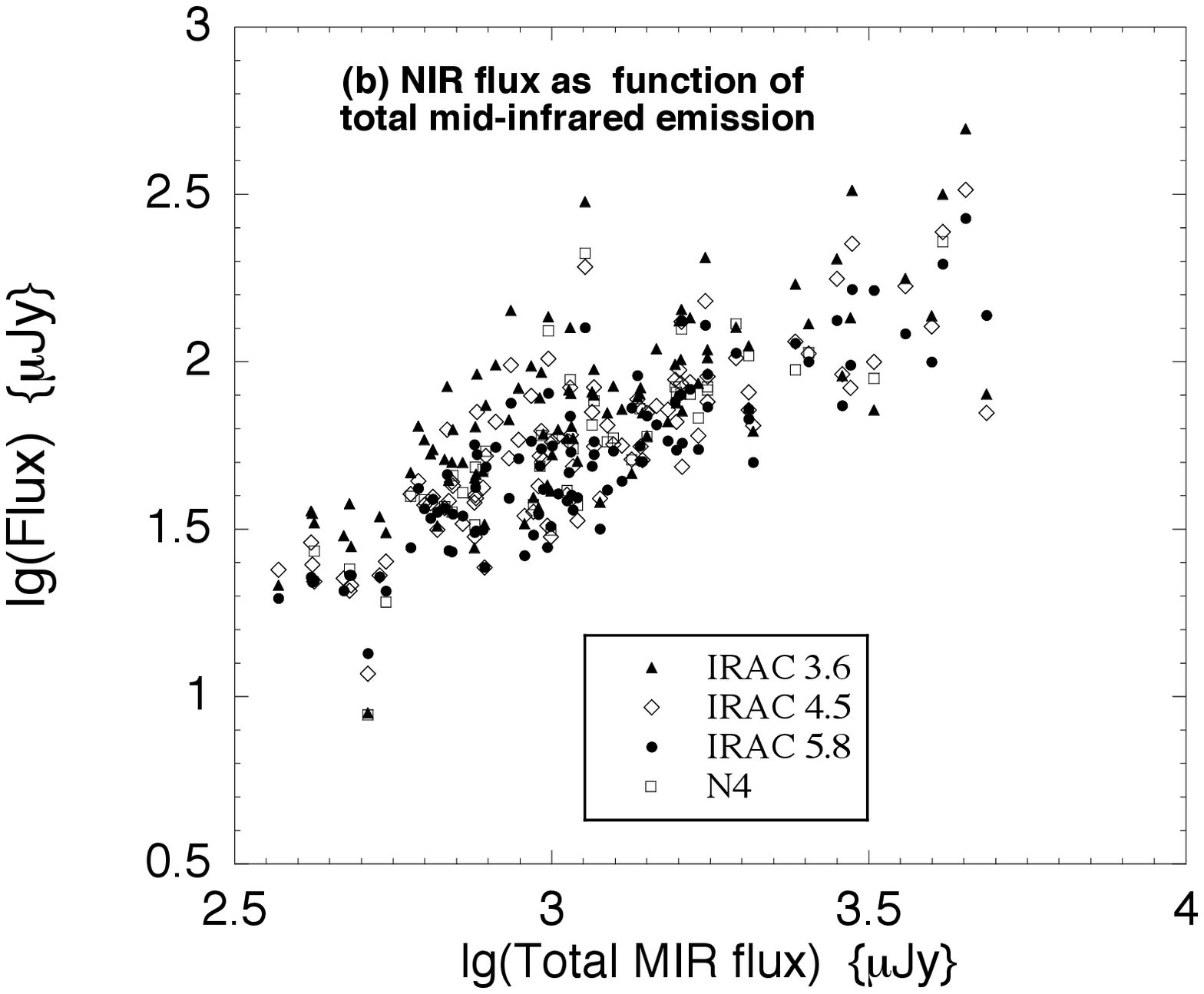,width=9cm}
\psfig{figure=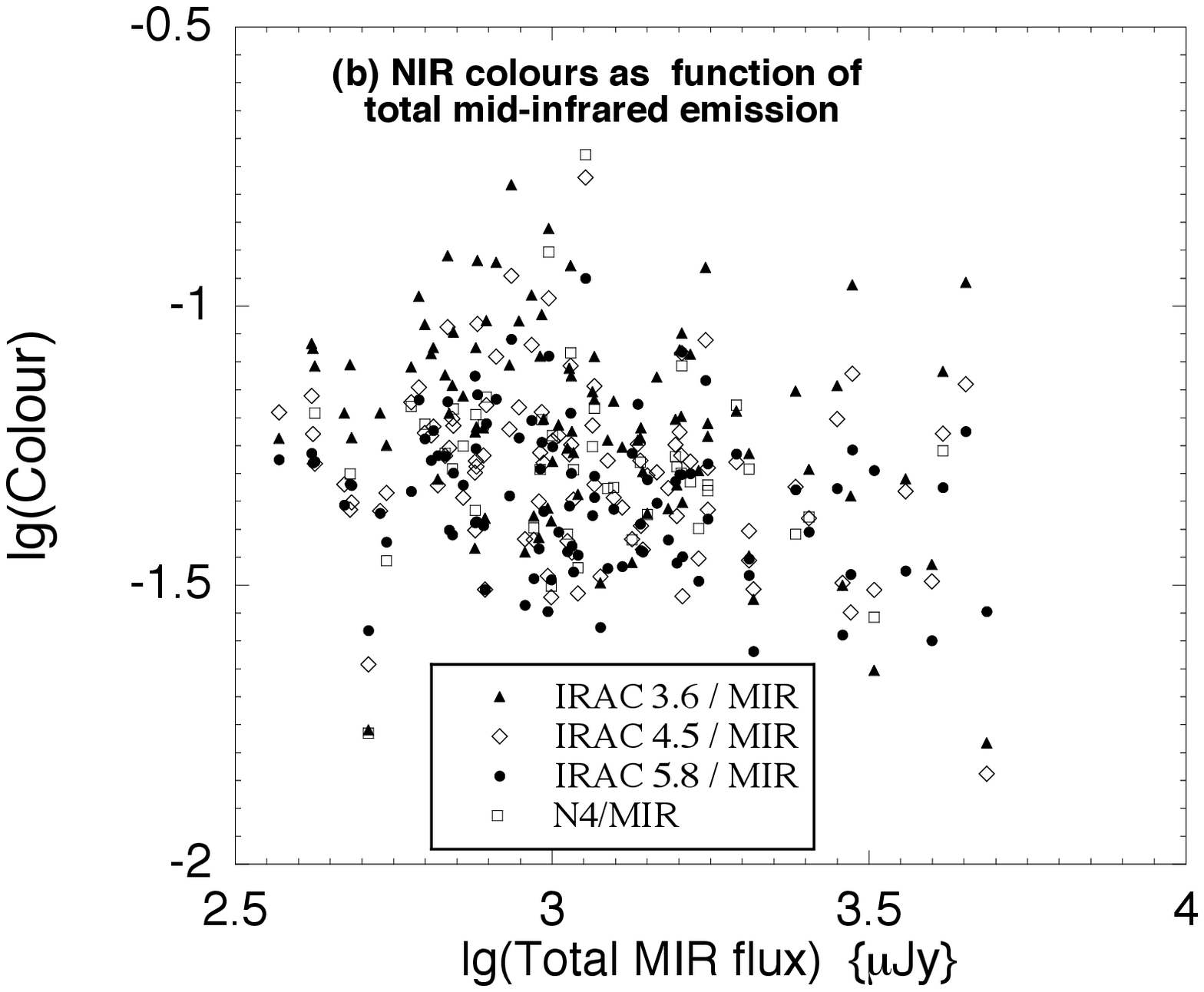,width=9cm}
}
\caption{{\bf (a)}: Near-infrared source fluxes in  the  {\it Spitzer} \&  {\it AKARI} bands as a function of the total mid-infrared flux (the total flux in the {\it Spitzer} IRAC 3.6, 4.5, 5.8, 8$\mu$m, {\it AKARI} S11, L18W \& {\it Spitzer} MIPS 24$\mu$m bands) for the sources in the FU-HYU-GOODS-N field.fluxes, {\it \bf (b)}: Near-infrared colours (near-IR flux/Total MIR flux)  as a function of the total mid-infrared flux for the sources in the FU-HYU-GOODS-N field.
\label{fig:nirir}}
\end{figure*}

Given the comprehensive multi-wavelength mid-infrared coverage provided by the combined  {\it AKARI} \&  {\it Spitzer} bands,  we define the quantity referred to as the total mid-infrared flux as the total flux in the {\it Spitzer} IRAC 3.6, 4.5, 5.8, 8$\mu$m, {\it AKARI} S11, L18W \& {\it Spitzer} MIPS 24$\mu$m bands for the sources in the FU-HYU-GOODS-N field. This quantity is representative  of the total emission from the mid-infrared spectra, dominated by the PAH emission features in our sources. In Figure \ref{fig:mirir} this total mid-infrared flux is plotted as a function of the mid-infrared fluxes ({\it panel (a)}) and colours ({\it panel (b)}) in the   {\it Spitzer}-IRAC 8$\mu$m \&  -MIPS 24$\mu$m bands and the  {\it AKARI}-IRC S11, L18W bands respectively. We observe a tight correlation between the total mid-infrared flux and the longer wavelength MIPS 24$\mu$m and IRC L18W bands increasing as a function of increasing total MIR flux but a larger scatter in the shorter wavelength bands revealing that the longer mid-infrared bands are better indicators of the total mid-infrared flux of these sources. The colours exhibit an almost constant value for the the case of L18W/MIR \& MIPS24/MIR as a function of total MIR flux where as the IRAC8/MIR, S11/MIR colours show a slight redder trend to high values of the total mid-infrared flux.
The corresponding  near-infrared flux is plotted in Figure \ref{fig:nirir} as a function of the mid-infrared fluxes ({\it panel (a)}) and colours ({\it panel (b)}) for the   {\it Spitzer}-IRAC 3.6,4.5, 5.8$\mu$m \&   {\it AKARI}-IRC N4 bands. In general there is less dispersion in both the fluxes and the colours (for all IRAC and the IRC-N4 bands) as a function of total mid-infrared flux since the near-infrared spectra of these sources are less affected by the emission features. In general there is a trend of increasing near-infrared flux as a function of total mid-infrared flux but bluer colours in the colour-total MIR flux plane.

\begin{figure*}
\centering
\centerline{
\psfig{figure=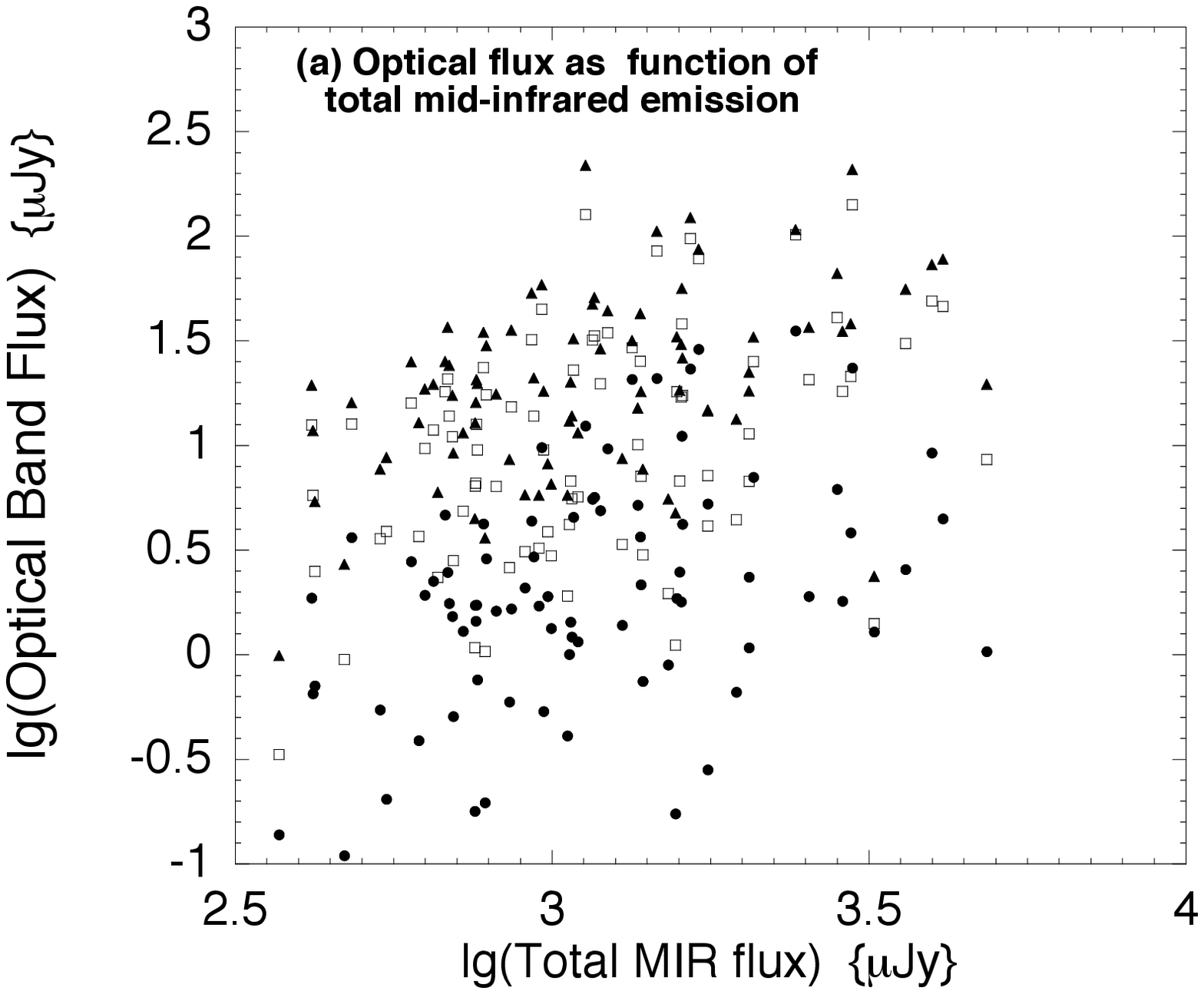,width=9cm}
\psfig{figure=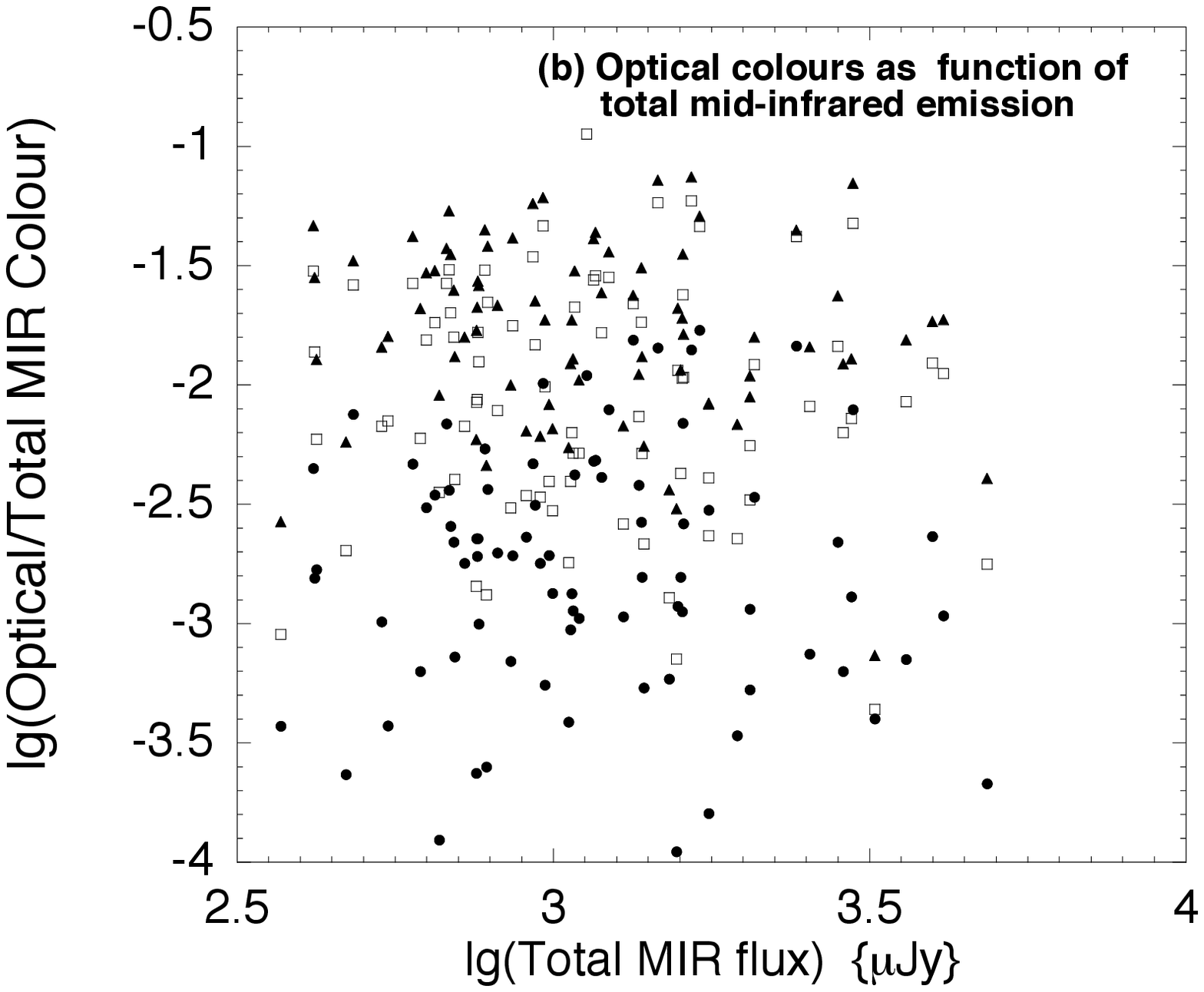,width=9cm}
}
\caption{{\bf (a)}: Optical source fluxes ({\it circles} - U-band, {\it squares} - R-band \& {\it triangles} - z$^ \prime$-band) as a function of the total mid-infrared flux (the total flux in the {\it Spitzer} IRAC 3.6, 4.5, 5.8, 8$\mu$m, {\it AKARI} S11, L18W \& {\it Spitzer} MIPS 24$\mu$m bands) for the sources in the FU-HYU-GOODS-N field. {\bf (b)}: Optical colours (optical flux/Total MIR flux) as a function of the total mid-infrared flux for the sources in the FU-HYU-GOODS-N field.
\label{fig:miropt}}
\end{figure*}

\begin{figure*}
\centering
\centerline{
\psfig{figure=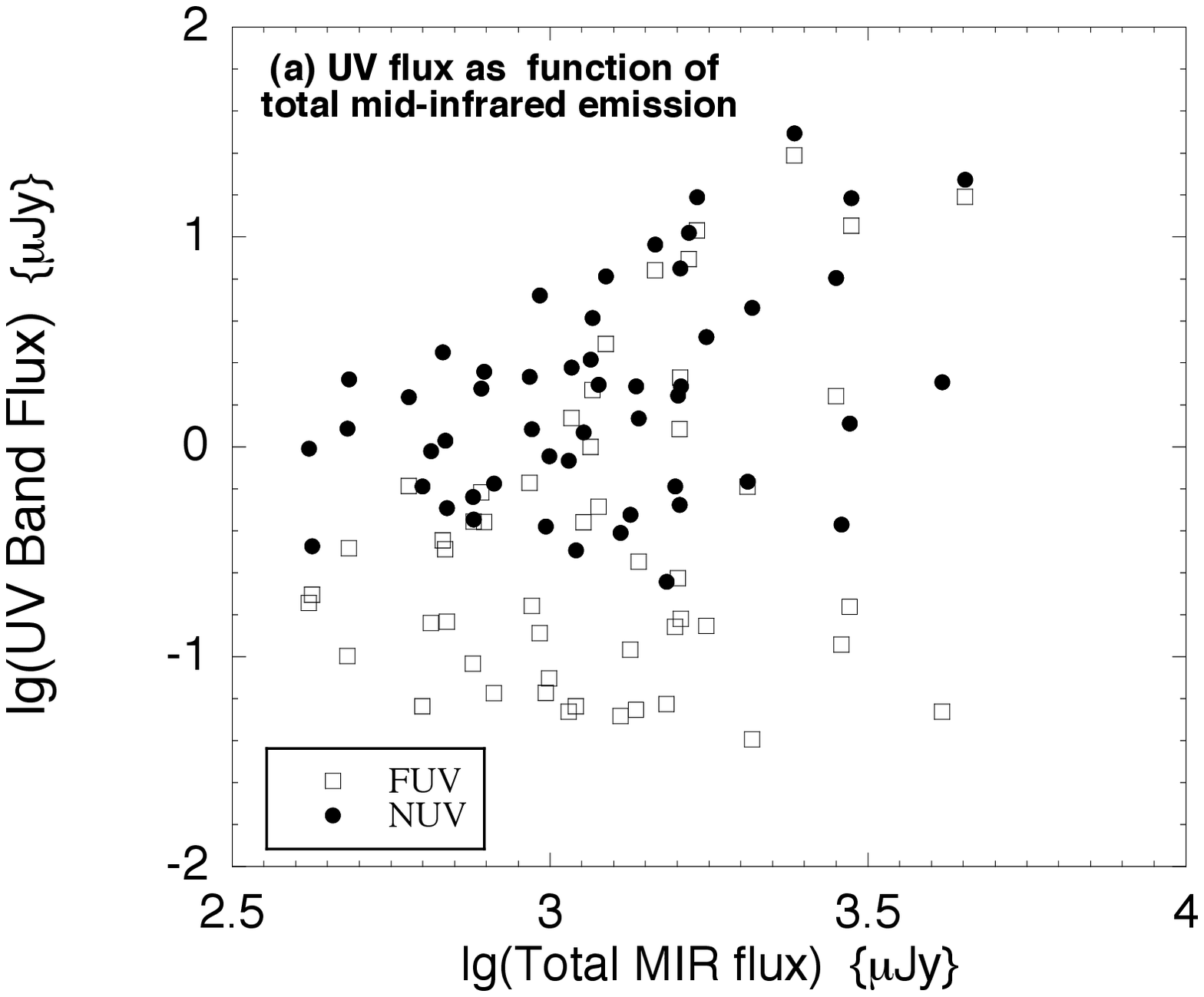,width=9cm}
\psfig{figure=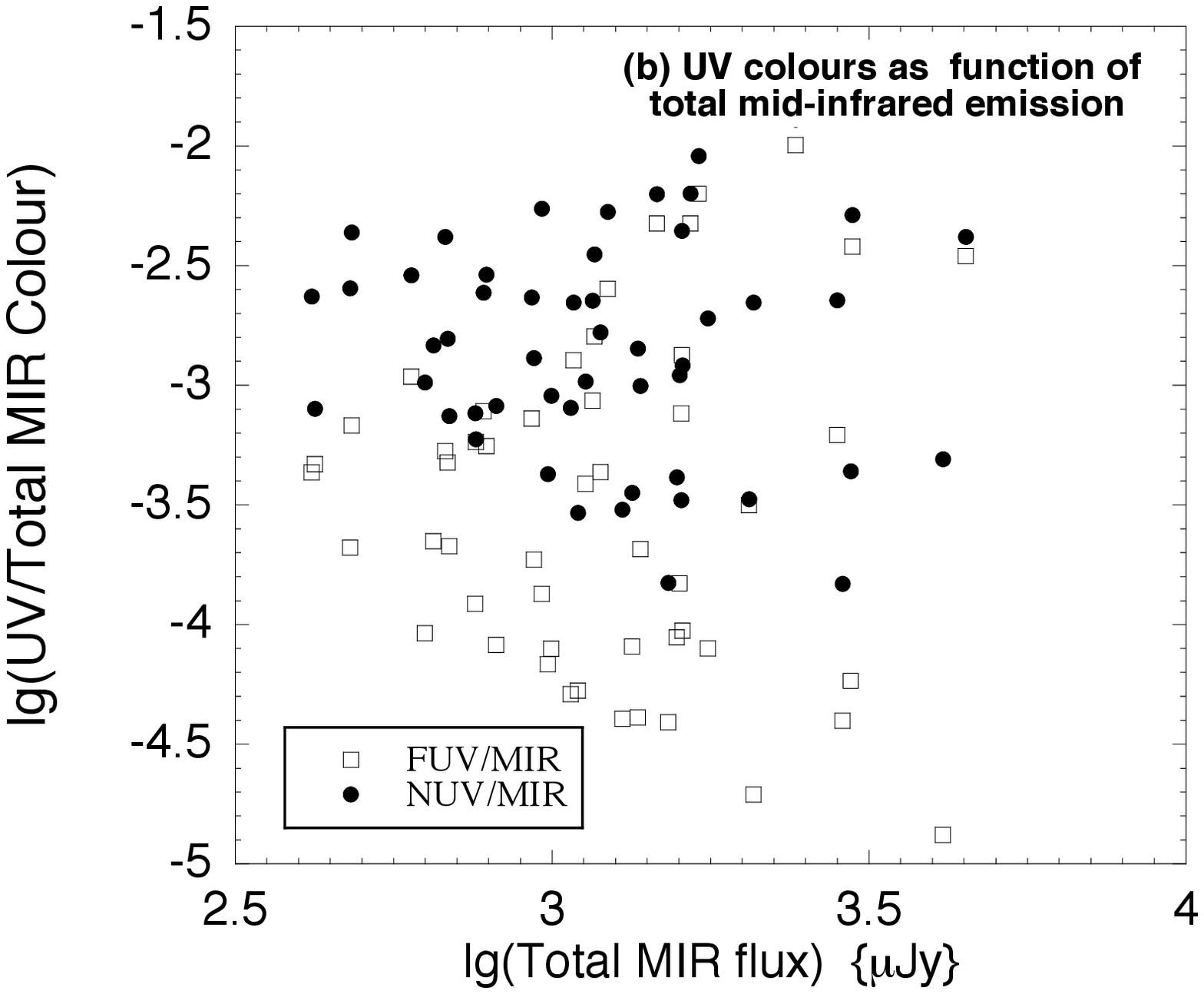,width=9cm}
}
\caption{{ \bf (a)}:  {\it GALEX} UV band source fluxes as a function of the total mid-infrared flux (the total flux in the {\it Spitzer} IRAC 3.6, 4.5, 5.8, 8$\mu$m, {\it AKARI} S11, L18W \& {\it Spitzer} MIPS 24$\mu$m bands) for the sources in the FU-HYU-GOODS-N field. {\bf (b)}:  {\it GALEX} UV band colours (UV flux/Total MIR flux) as a function of the total mid-infrared flux for the sources in the FU-HYU-GOODS-N field.
\label{fig:miruv}}
\end{figure*}

In Figure \ref{fig:miropt} the total mid-infrared flux is plotted against the optical fluxes ({\it panel (a)}) and colours ({\it panel (b)}) for the sources in our catalogue. For clarity we include only 3 bands (U, R, z$^ \prime$). A wide dispersion in optical fluxes is seen over the entire range of total MIR flux implying that the two are not particularly correlated in general. There is a trend to increasing optical flux as a function of increasing total MIR emission although as expected, it is not as pronounced as in the infrared band case.  Similarly, there appears little or no correlation of optical/MIR colour as a function of total MIR flux, in particular the U-band/MIR colour is distributed over 2 orders of magnitude at all values of the total MIR flux. Figure \ref{fig:miruv} shows the corresponding flux  ({\it panel (a)}) and colour  ({\it panel (b)}) for the GALEX UV bands. Similarly to the U-band plots, no significant correlation is seen in either the flux distribution or colours as a function of total MIR flux, although the near-UV (NUV) emission shows less dispersion than the far-UV (FUV) emission.

\smallskip
\subsection{The nature of the galaxy populations in the FU-HYU-GOODS sample}\label{sec:colourcolour}

\begin{figure*}
\centering
\centerline{
\psfig{figure=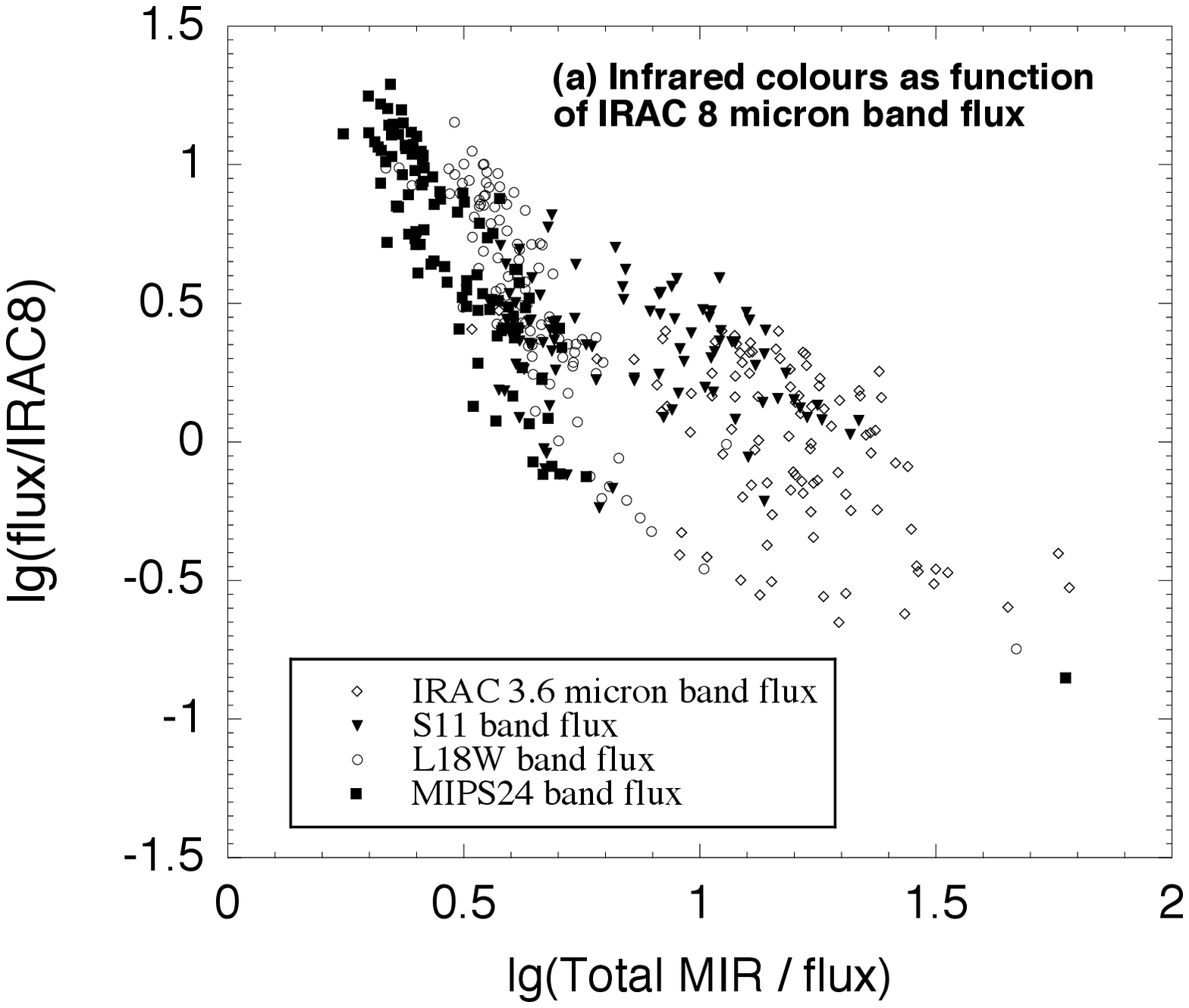,width=9cm}
\psfig{figure=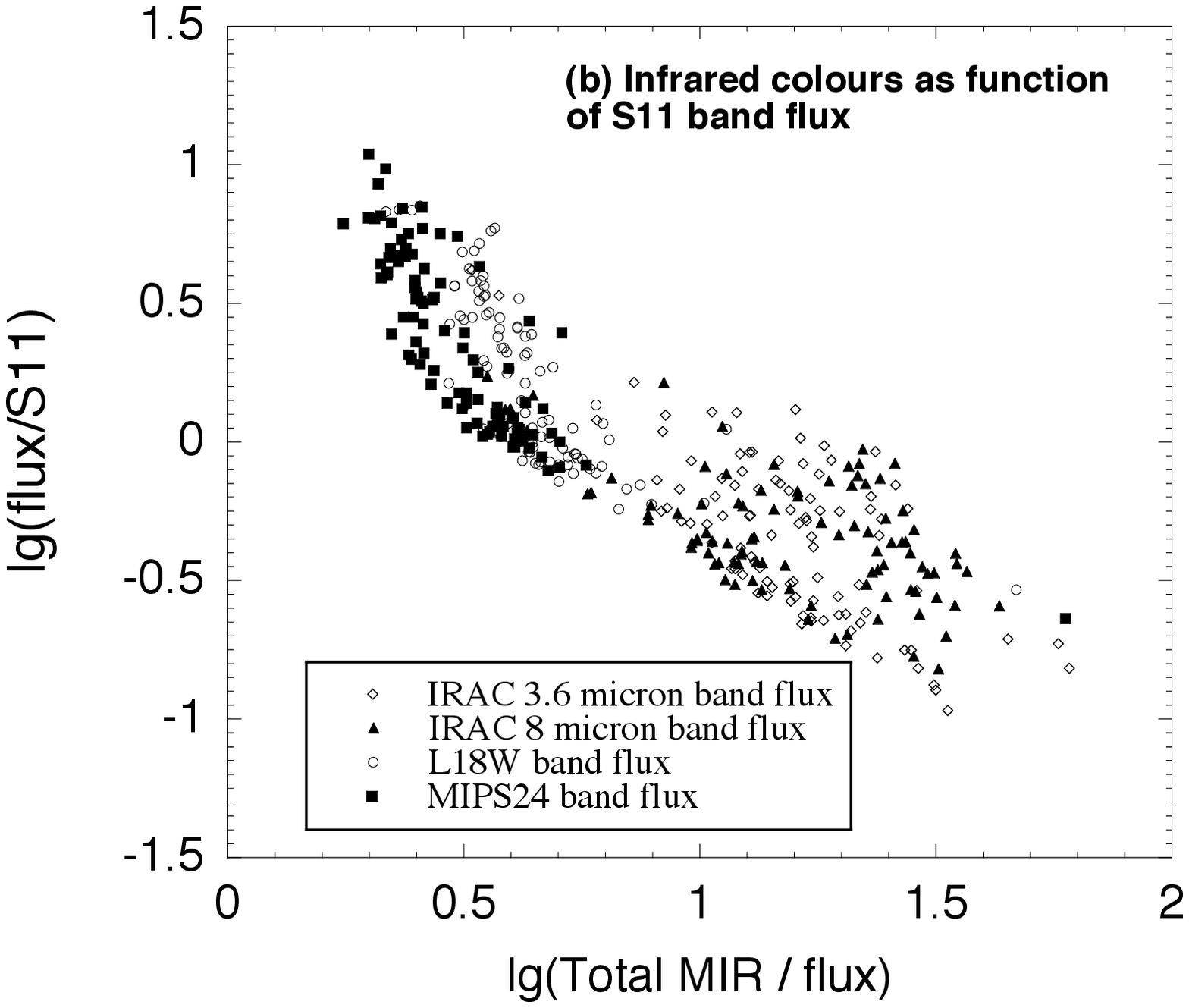,width=9cm}
}
\caption{Colour-Colour diagrams for  sources in the FU-HYU-GOODS-N field as a function of the total mid-infrared flux (the total flux in the {\it Spitzer} IRAC 3.6, 4.5, 5.8, 8$\mu$m, {\it AKARI} S11, L18W \& {\it Spitzer} MIPS 24$\mu$m bands). {\bf (a)}: as a function of the {\it Spitzer} IRAC 8$\mu$m band.  {\bf (b)}: as a function of the {\it AKARI} IRC S11 band.
\label{fig:nircolcol}}
\end{figure*}

\begin{figure*}
\centering
\centerline{
\psfig{figure=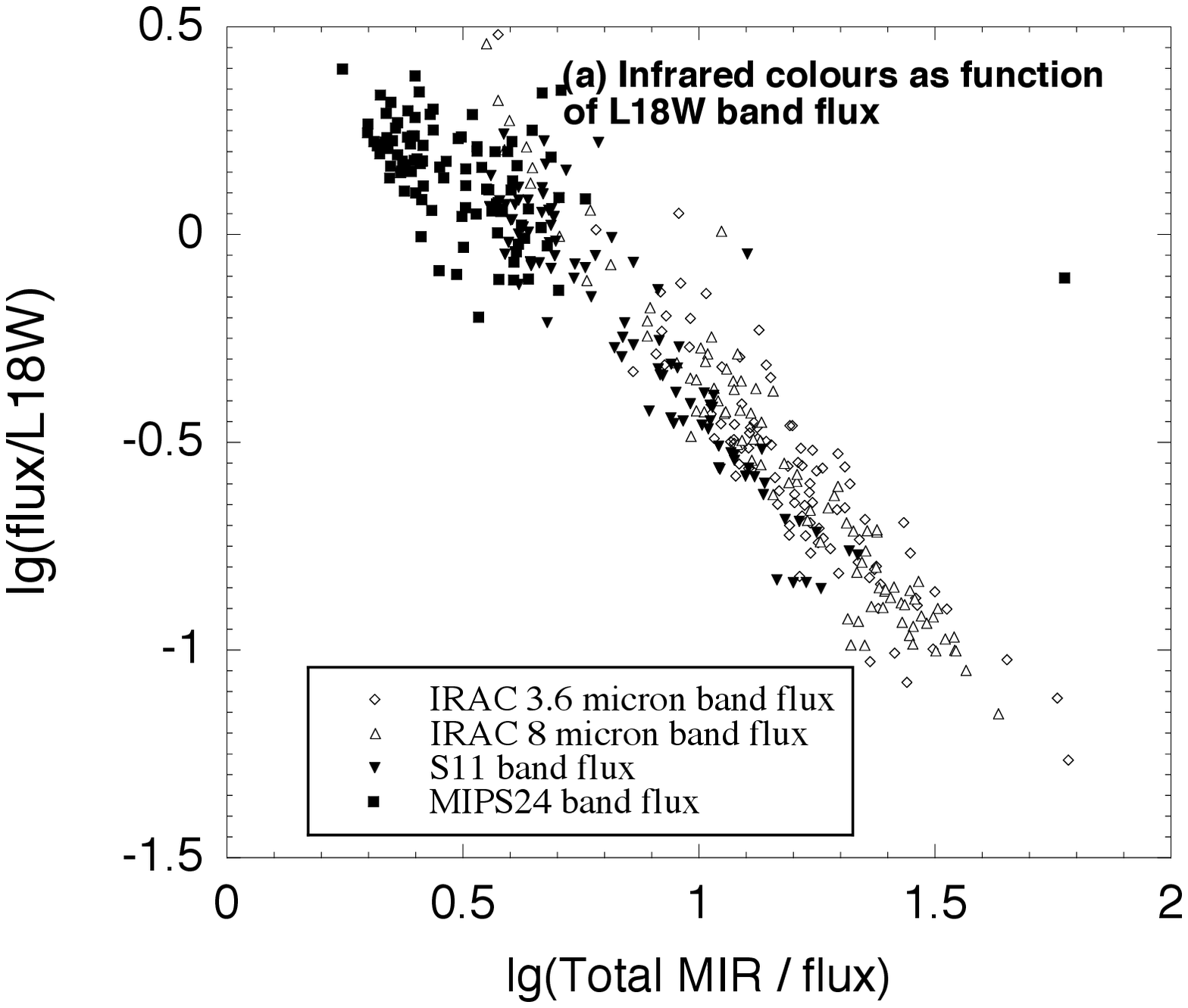,width=9cm}
\psfig{figure=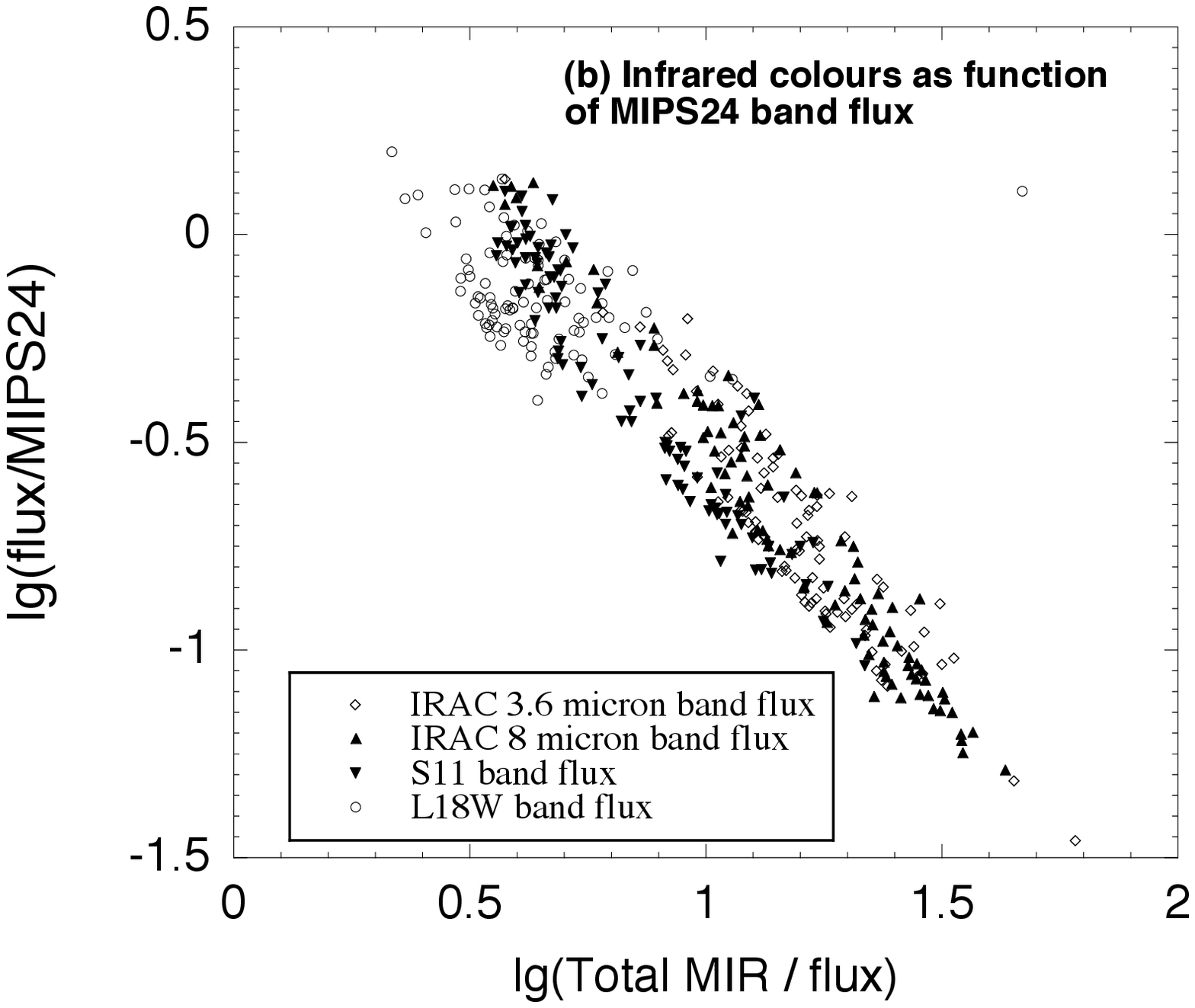,width=9cm}
}
\caption{Colour-Colour diagrams for  sources in the FU-HYU-GOODS-N field as a function of the total mid-infrared flux (the total flux in the {\it Spitzer} IRAC 3.6, 4.5, 5.8, 8$\mu$m, {\it AKARI} S11, L18W \& {\it Spitzer} MIPS 24$\mu$m bands). {\bf (a)}: as a function of the {\it AKARI} IRC L18W band.  {\bf (b)}: as a function of the {\it Spitzer} MIPS 24$\mu$m band.
\label{fig:mircolcol}}
\end{figure*}

In Figures \ref{fig:nircolcol} \&  \ref{fig:mircolcol} we plot a selection of colour-colour distributions from the FU-HYU-GOODS sample covering the wavelength range from 3.6-24$\mu$m in the  {\it Spitzer}-IRAC 3.6, 8$\mu$m bands, the  {\it AKARI}-IRC S11, L18W bands and the {\it Spitzer}-MIPS 24$\mu$m band.
Figures ~\ref{fig:nircolcol}{\it (a)} \& {\it (b)} plot the above fluxes as a function of {\it flux}/IRAC8 -- Total MIR/ {\it flux} colour and {\it flux}/IRC S11 -- Total MIR/ {\it flux} colour respectively. A distinct separation is seen between the longer L18W \& MIPS24 band colours (which show very similar colours and trends) and the shorter wavelength IRAC3.6 \& IRC S11 band colours. The anomalous  L18W \& MIPS24 band point at lg(MIR/L18W) \& lg(MIR/MIPS24) colour $\sim$1.7 is a bright local galaxy. Note that although the IRAC band colours occupy similar colour-colour space, the  {\it AKARI}-IRC S11/IRAC8 colours in Figure \ref{fig:nircolcol}{\it (a)} extend to  redder values (bluer values on the Total MIR/S11 axis) implying an excess in the S11 band emission not seen in the shorter wavelength IRAC bands. 

\begin{figure}
\centering
\centerline{
\psfig{ figure=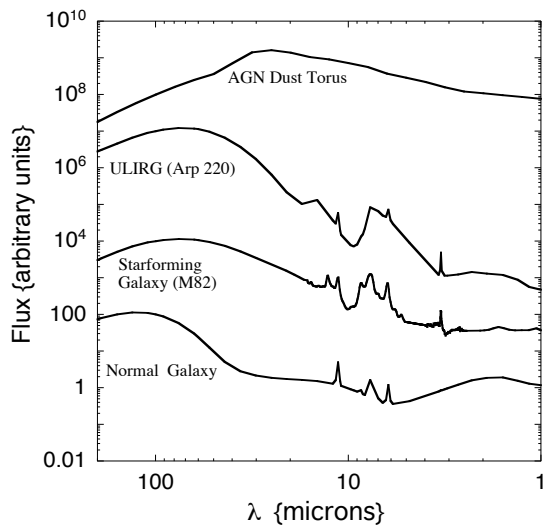,width=7cm}
}
\caption{Model spectral templates adopted for the colour tracks plotted in Figures \ref{fig:sedcolcol} \& \ref{fig:silicatebreak} and described in the text.
\label{fig:modelseds}}
\end{figure}  

In Figures ~\ref{fig:mircolcol}{\it (a)} \& {\it (b)} the above fluxes are plotted as a function of the longer wavelength {\it flux}/MIPS24 -- Total MIR/ {\it flux} colour and {\it flux}/IRC L18W -- Total MIR/ {\it flux} colour respectively. Similarly to Figure \ref{fig:nircolcol}, the same separation in colour-colour space is seen for the longer (MIPS24, IRC L18W) wavebands from the shorter wavelength IRAC bands. The tight clustering of the MIPS24/L18W -- Total MIR/MIPS24 in Figure \ref{fig:mircolcol}{\it (a)} and the L18W/MIPS24 -- Total MIR/L18W in Figure \ref{fig:mircolcol}{\it (b)} emphasis the fact that the longer wavelength bands provide an accurate measure of the Total MIR emission. In contrast the shorter wavelength bands show a wider dispersion as a function of colour and indeed in some cases exhibit a bi-modal distribution. Note, the apparent excess in the S11 band flux seen in Figure \ref{fig:nircolcol}{\it (a)} manifests itself by producing two distinct populations in the S11/MIPS24 -- Total MIR/S11 and S11/L18W -- Total MIR/S11 planes, one population with similar colours to the IRAC band colours and another population with colours more akin to the longer wavelength (MIPS24, L18W) band colours.

\begin{figure*}
\centering
\centerline{
\psfig{ figure=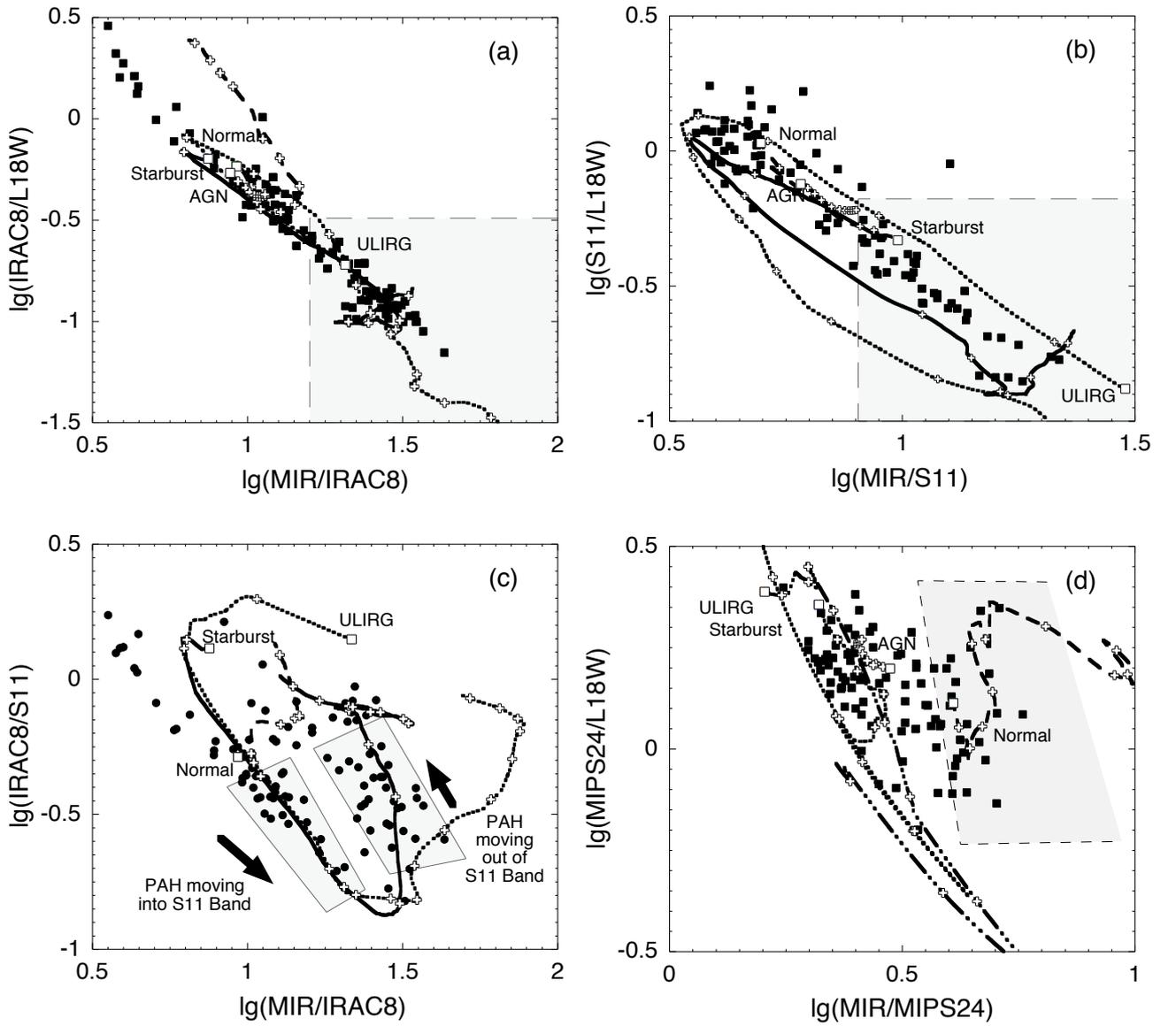,width=17cm}
}
\caption{Infrared colour-colour diagrams for sources in the FU-HYU-GOODS-N field  as a function of the {\it AKARI} L18W (or S11) band and the total mid-infrared flux (the total flux in the {\it Spitzer} IRAC 3.6, 4.5, 5.8, 8$\mu$m, {\it AKARI} S11, L18W \& {\it Spitzer} MIPS 24$\mu$m bands). {\bf (a)}: as a function of the {\it Spitzer} IRAC 8$\mu$m band.  {\bf (b)}:  as a function of the {\it AKARI} IRC S11 band. {\bf (c)}: as a function of the{\it Spitzer} IRAC 8$\mu$m band (to {\it AKARI} IRC S11 band).  {\bf (d)}: as a function of the {\it Spitzer} MIPS 24$\mu$m band. Model predictions for colour-colour tracks are also shown for normal galaxy ({\it dashed}), star-forming galaxy ({\it dotted}), ULIRG ({\it triple dot-dash}) and AGN ({\it solid}) spectral templates. The large squares mark the zero redshift template and the crosses along the tracks are spaced at equal redshift intervals of $\Delta$z=0.2.
\label{fig:sedcolcol}}
\end{figure*}

In Figure \ref{fig:sedcolcol} a selection of colour-colour distributions are individually extracted from Figures \ref{fig:nircolcol} \&  \ref{fig:mircolcol}. Overlaid on each individual colour-colour distribution are the galaxy template spectral energy distributions for a normal quiescent galaxy, a star-forming galaxy, an ultraluminous infrared galaxy and an AGN. 
 The normal spectral template is taken from the SED libraries of  \cite{efstathiou03} which assume an input radiation field based on  \cite{bruzual93} and assumed interstellar dust field. There are two free parameters,  the ratio of the radiation field  to the local solar neighborhood, $\chi$, and the visual extinction $A_{V}$. Our adopted template corresponds to  $\chi$=5 , $A_{V}$=0.5. The star-forming galaxy and ultraluminous infrared galaxy templates are taken from the libraries of \cite{efstathiou00}. The templates assume a Salpeter initial mass function (IMF) from 0.1-125M$_{\sun}$ with an $e$-folding time of 20Myr.  The evolution of the stellar population within giant molecular clouds follows the stellar synthesis population models of  \cite{bruzual93}. The individual starburst models are defined by two free parameters: the age of the starburst in Myr; $T_{AGE}$ and the initial optical depth ($\tau_{V}$) of the molecular clouds. We select two specific templates that represent the archetypical starforming galaxy M82 and ULIRG Arp 220 which correspond to model parameters of $T_{AGE}$=10 \& 26 Myr and $\tau_{V}$=50 \& 150 respectively. For M82 we replace the model SED with the 18-5$\mu$m region measured by the ISOCAM Circular Variable Filter and the true optical SED (\cite{forster03}). 
Finally our AGN SED is derived from the dust torus models of \cite{efstathiou95} with the near-infrared-optical spectrum following \cite{king03}. The dust torus model assumes an opening angle of 45$^\circ$, ratio of outer to inner torus radius of 20 and a viewing angle of 10$^{\circ}$ . The adopted model spectral templates are shown in Figure \ref{fig:modelseds} .

The SED tracks in Figure ~\ref{fig:sedcolcol} are marked by a {\it large square} at the redshift =0 position and have crosses for every 0.2 steps in redshift thereafter extending to a redshift of 2.
Figure \ref{fig:sedcolcol}{\it (a)} shows the IRAC 8$\mu$m/L18W against the Total MIR flux/ IRAC 8$\mu$m colours.  The shaded area bounded by lg(IRAC8/L18W)$<$-0.5,  lg(MIR/IRAC8)$>$1.2 denotes sources that are expected to be higher redshift (z$>$0.7) star-forming galaxies. These colours can be compared with similar work carried out by  \cite{brand06} using IRAC8, MIPS24 band {\it Spitzer} data. Our colours agree well with the corresponding segregation of low and high redshift star-forming population to redder and bluer colours respectively either side of the locus of AGN colours  lg(IRAC8/L18W)$<$-0.3. 
 In Figure \ref{fig:sedcolcol}{\it (b)} the IRC S11 / L18W colour is plotted against the total MIR /  IRC S11 band colour. A similar trend can be seen in these colours as Figure \ref{fig:sedcolcol}{\it (a)} and the colours produce an effective means of segregating a star-forming population from the normal galaxy population at  high (lg(MIR flux / S11 flux)$>$0.1, z$>$1) redshifts. Note that this result can explain the bimodal distribution seen in Figure\ref{fig:mircolcol} for the S11/L18W, S11/MIPS colours, as a population of lower and higher redshift source populations. 
The multi-band coverage of the mid-infrared spectra of our sources can also be used to provide powerful diagnostics of the PAH emission in our galaxy spectra using colour-colour information. In Figure \ref{fig:sedcolcol}{\it (c)} the IRAC 8$\mu$m / IRC S11 band colour is plotted against the total MIR /  IRAC 8$\mu$m colour. The two highlighted regions on the plot show the passage of prominent (predominantly 7.7$\mu$m) PAH features through the {\it AKARI} S11 band. From redshifts of $\sim$0.4 --0.6 the  7.7$\mu$m feature enters the S11 band and a sharp decrease in the IRAC 8$\mu$m / IRC S11 colour is observed in the SEDs of starforming (starburst + ULIRG) galaxies. At redshifts $>$ 0.8--1 the feature is redshifted out of the S11 band and a gradual climb back up the  IRAC 8$\mu$m / IRC S11 colour  track is seen in these sources. Note that for clarity, the AGN track is not shown on this plot although the tracks are constrained to lg(IRAC8/S11)$>$-0.15 \& lg(MIR/IRAC8)$<$1. 
Finally, in Figure  \ref{fig:sedcolcol}{\it (d)} the MIPS24/L18W colour is plotted against the total MIR/MIPS24 band colour. Interestingly, it is found that following our SED templates, the quiescent normal galaxies can be well separated from the evolving galaxy population by their  MIR/MIPS24 colour alone. This could be indicative of the 24$\mu$m flux as an excellent tracer of star formation in galaxies where the 24$\mu$m emission in normal quiescent sources would be systematically lower with a greater dispersion. Note that \cite{bavouzet08} have indicated that the 24$\mu$m flux is a good indicator of the mid-infrared and following this the bolometric infrared luminosity of star forming galaxies  (STFG) where as \cite{dale07} have shown for a limited sample of local quiescent galaxies a larger dispersion in the mid- to far-infrared flux ratios. In addition, for the star-forming populations in Figure \ref{fig:sedcolcol}{\it (d)} a blue tail to the MIPS24/L18W colours is observed. This is sparsely populated and almost certainly due to the passage of the 9.7 silicate absorption feature through the MIPS24 band (see Section \ref{sec:silicatebreak}).

\smallskip
\subsection{Silicate Break galaxies in the GOODS-N field}\label{sec:silicatebreak}

\begin{figure*}
\centering
\centerline{
\psfig{figure=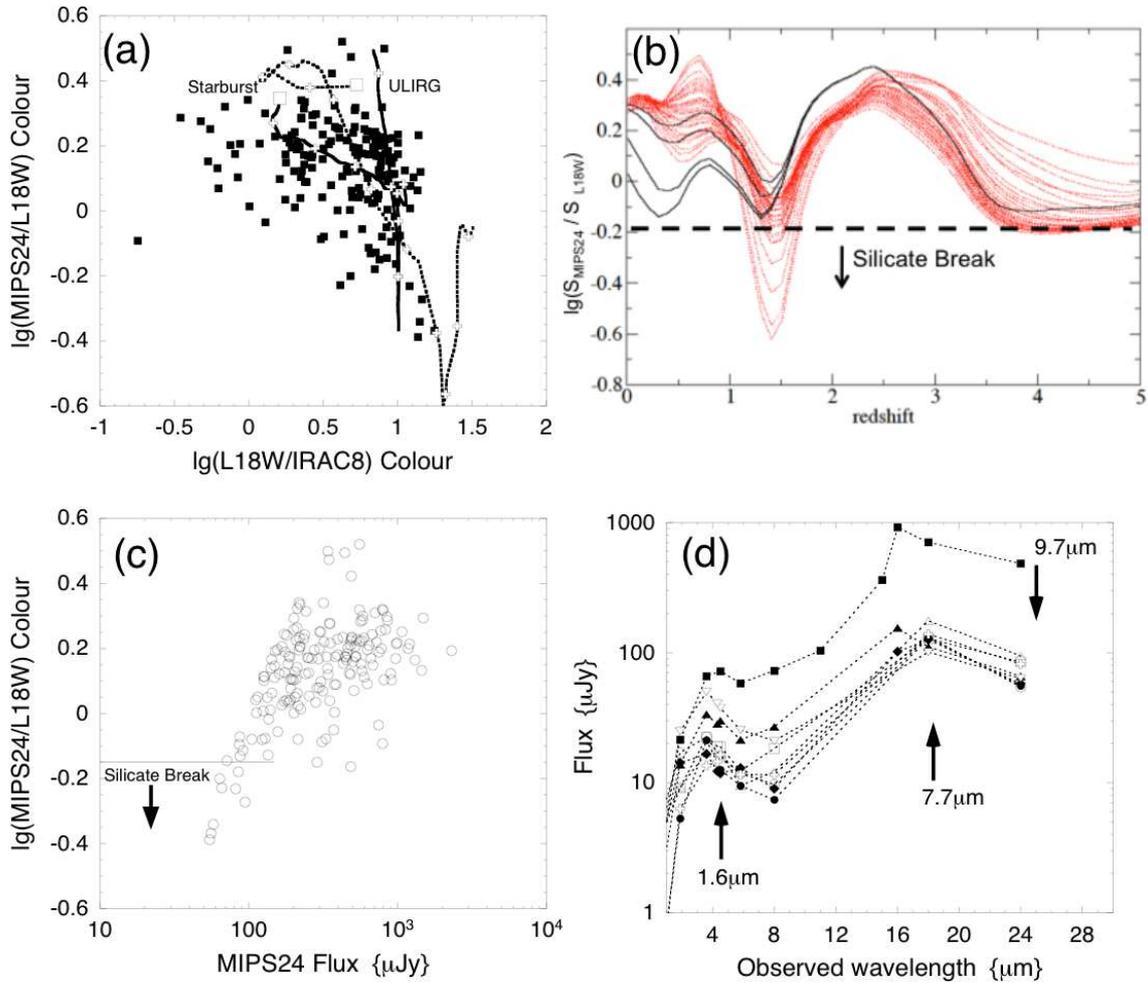,width=15cm}
}
\caption{Selection of potential Silicate Break Galaxies in the FU-HYU sample.  {\bf (a)}   The  MIPS24/L18W plotted against the  L18W/IRAC8 colour. Overlaid are the SEDs of a  moderate starburst and an ultra-luminous galaxy.  The SED tracks  are marked by a {\it large square} at the redshift =0 position and have crosses a step of 0.2 in redshift thereafter extending to a redshift of 2. The passage of the silicate absorption feature through the mid-infrared bands can be clearly seen in the starburst and ULIRG templates and map a distinct tail  towards lower MIPS24/L18W colours in the colour-colour plane. {\bf (b)} MIPS24/L18W  mid-infrared colour as a function of redshift for a variety of star-forming galaxies adapted from \cite{takagi05}. {\it Dotted lines} are the starburst models of \cite{takagi03}, {\it solid lines} are comparative quiescent galaxy spectral  templates from \cite{dale01}. The silicate break selection criteria is set to be $\sim$lg(MIPS24/L18W)$<$-0.15. {\bf (c)}  The diagnostic lg(MIPS24/L18W) colour against the  MIPS 24$\mu$m band flux. The solid horizontal line represents the Silicate Break detection criteria of lg(MIPS24/L18W)$<$-0.15 below which around nine potential candidates are selected.  {\bf (d)} Mid-infrared photometry in all available bands for the nine candidate Silicate-Break galaxies in the FU-HYU GOODS-N field. Several emission features are highlighted in the observed frame of the galaxies implying that all (except one) sources are between redshifts 1.2 -- 1.6.
\label{fig:silicatebreak}}
\end{figure*}

\begin{figure*}
\centering
\centerline{
\psfig{figure=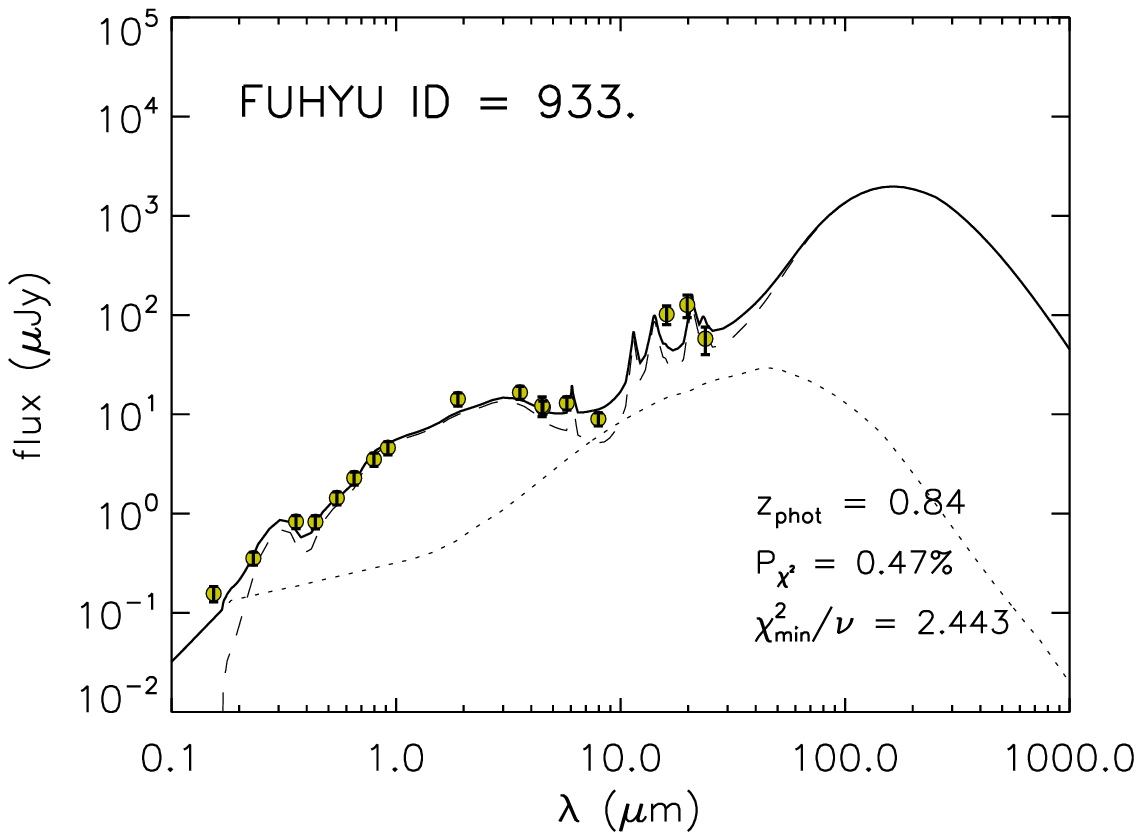,  width=6cm}
\psfig{figure=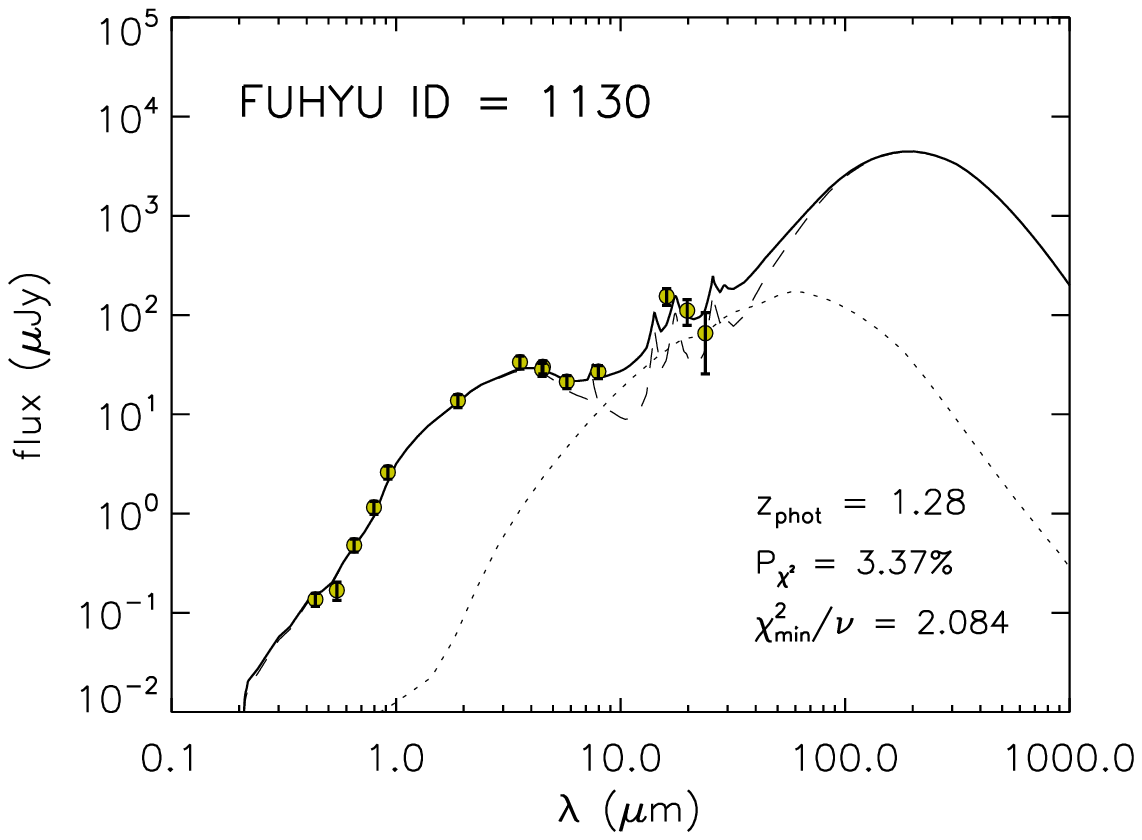, width=6cm}
\psfig{figure=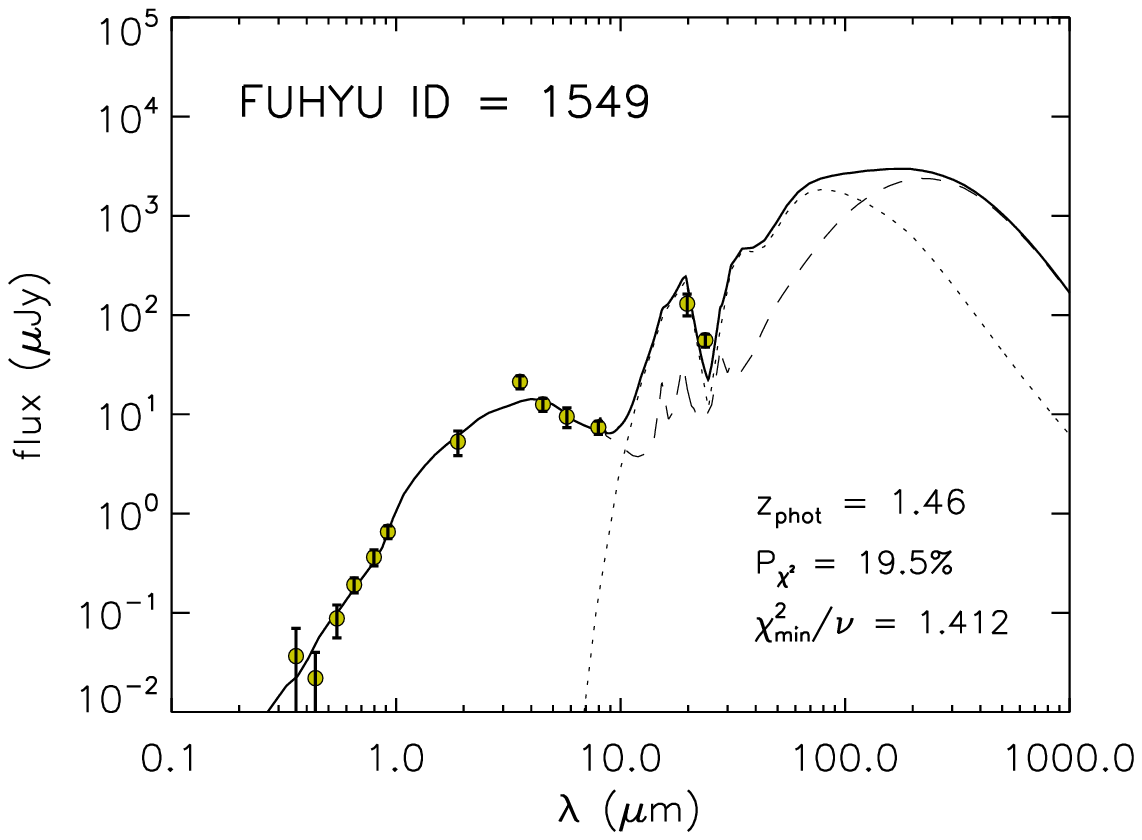, width=6cm}
}
\vspace{0.2cm}
\centerline{
\psfig{figure=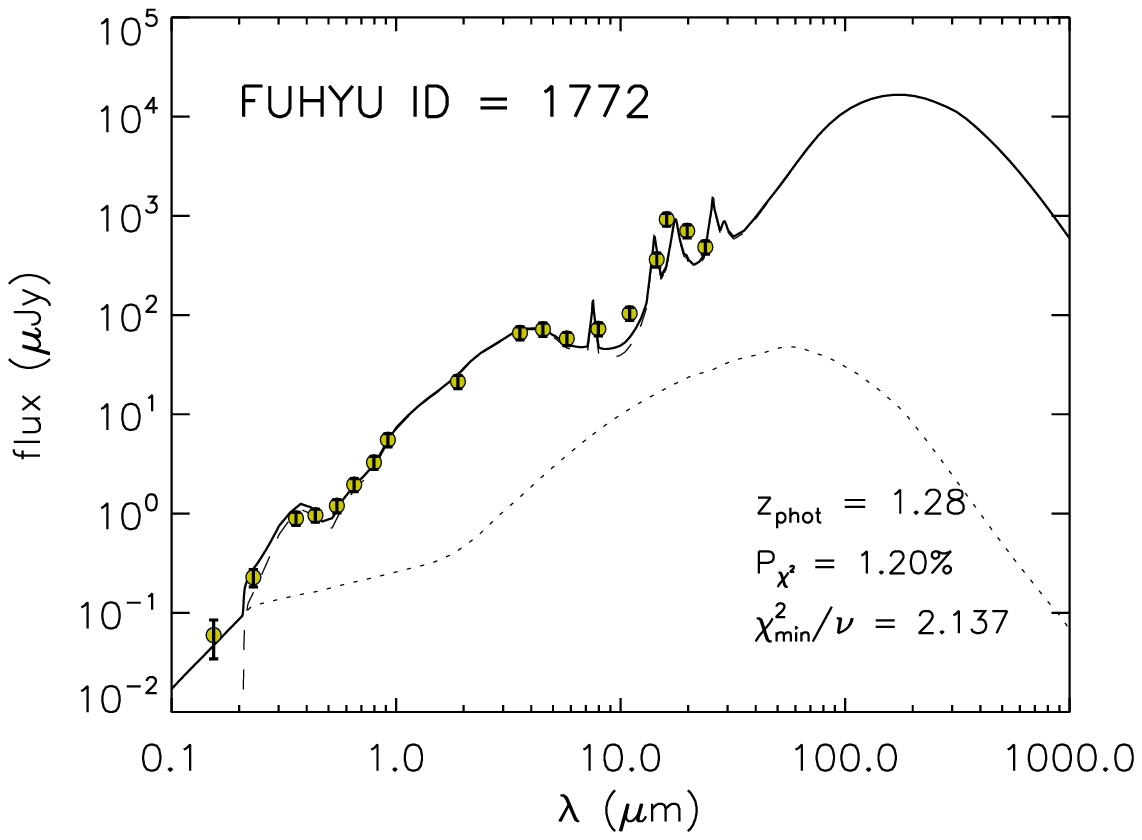, width=6cm}
\psfig{figure=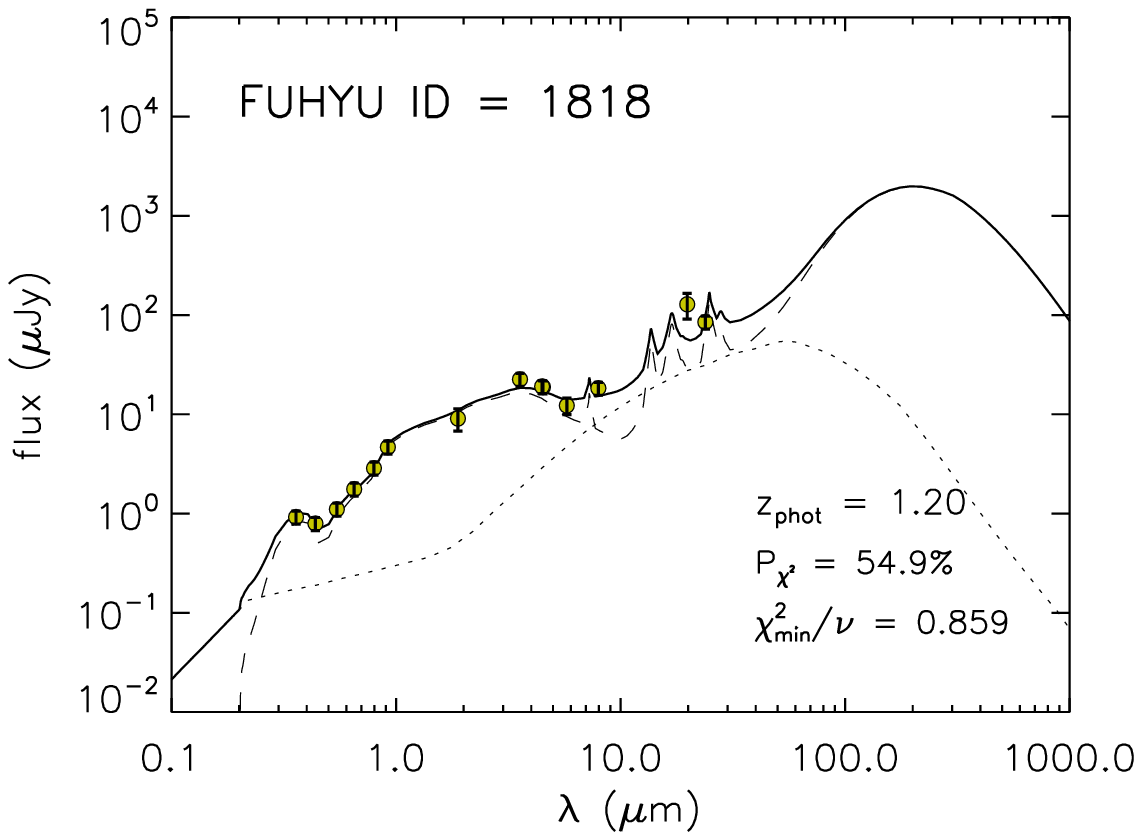,  width=6cm}
\psfig{figure=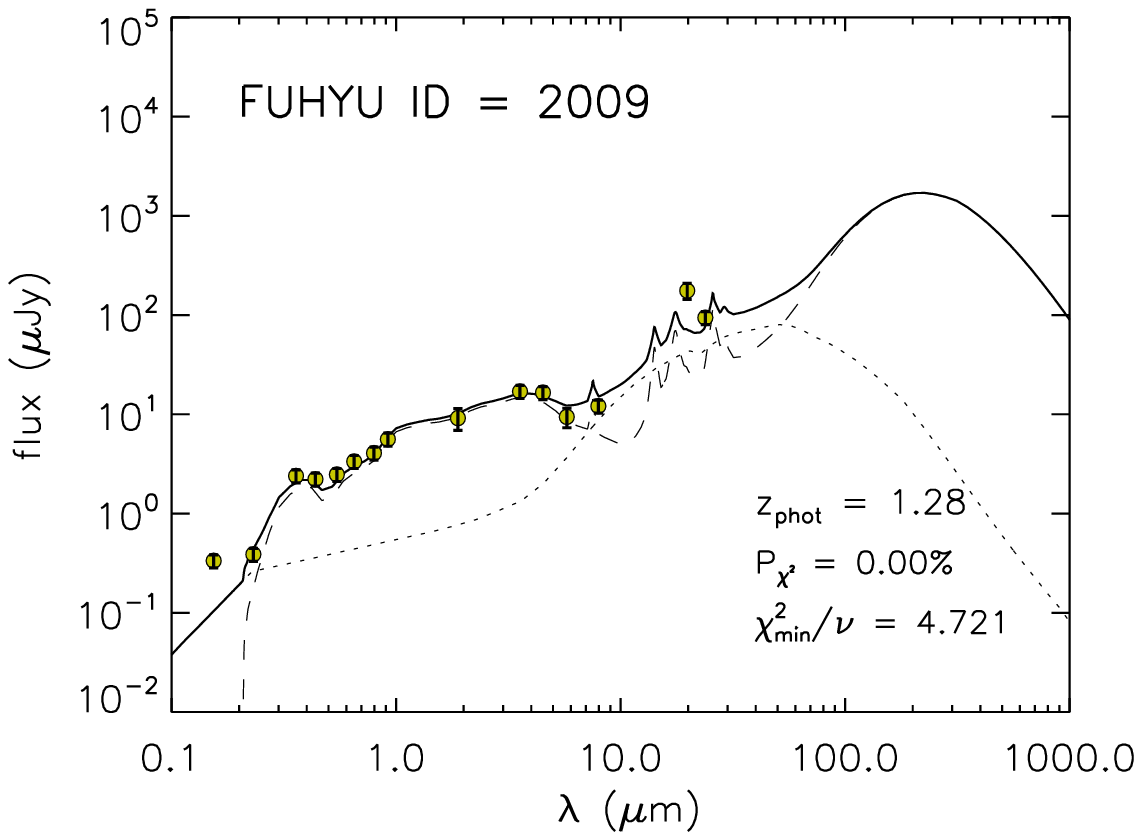,  width=6cm}
}
\vspace{0.2cm}
\centerline{
\psfig{figure=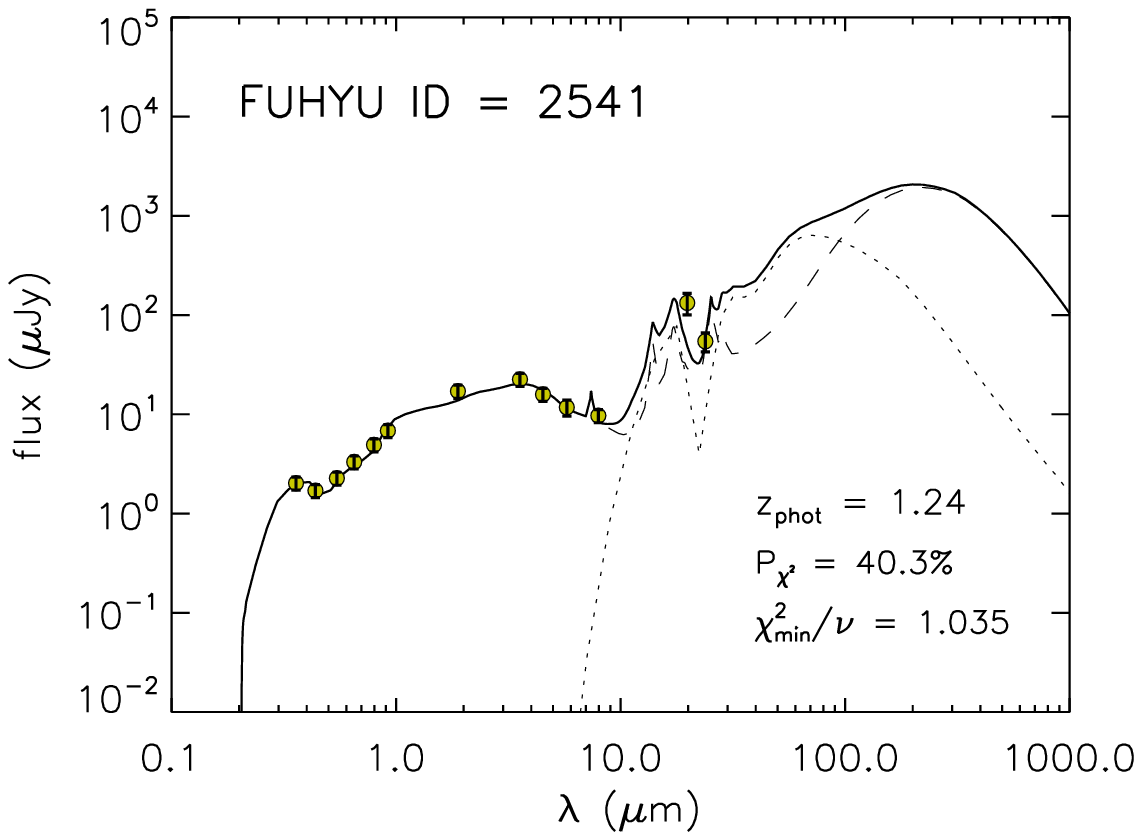,  width=6cm}
\psfig{figure=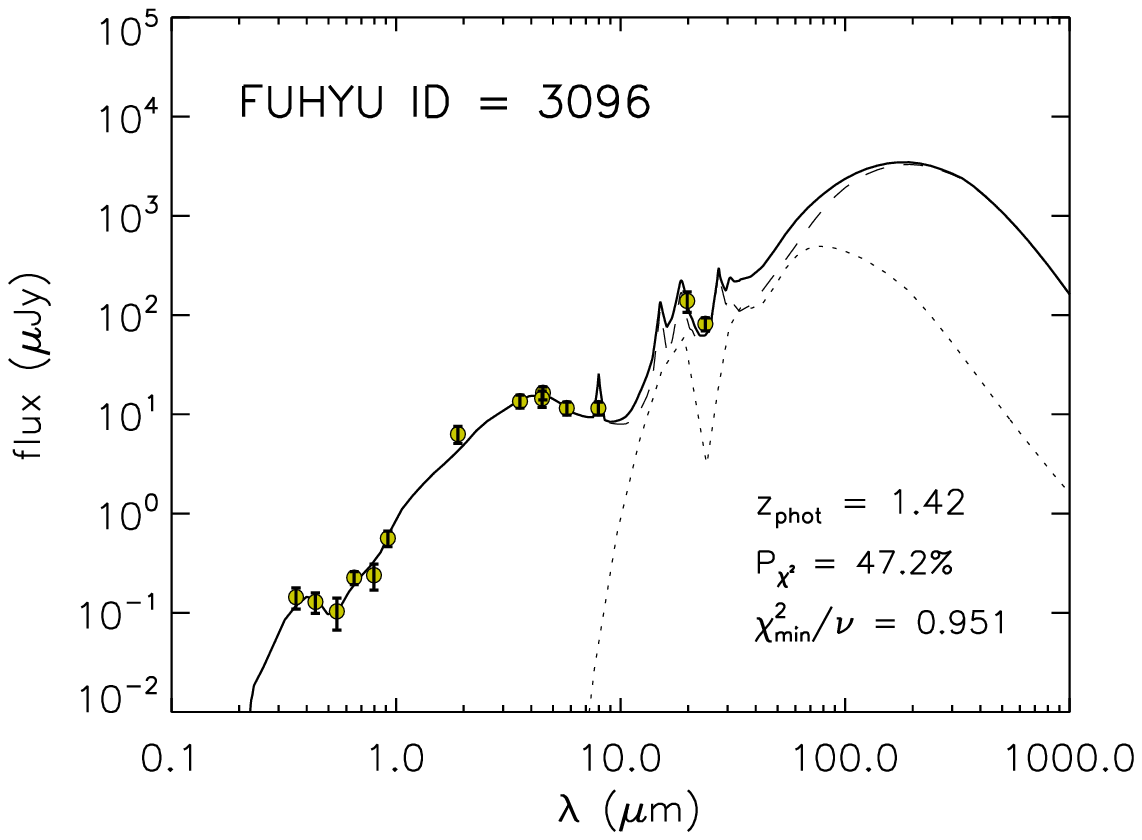,  width=6cm}
\psfig{figure=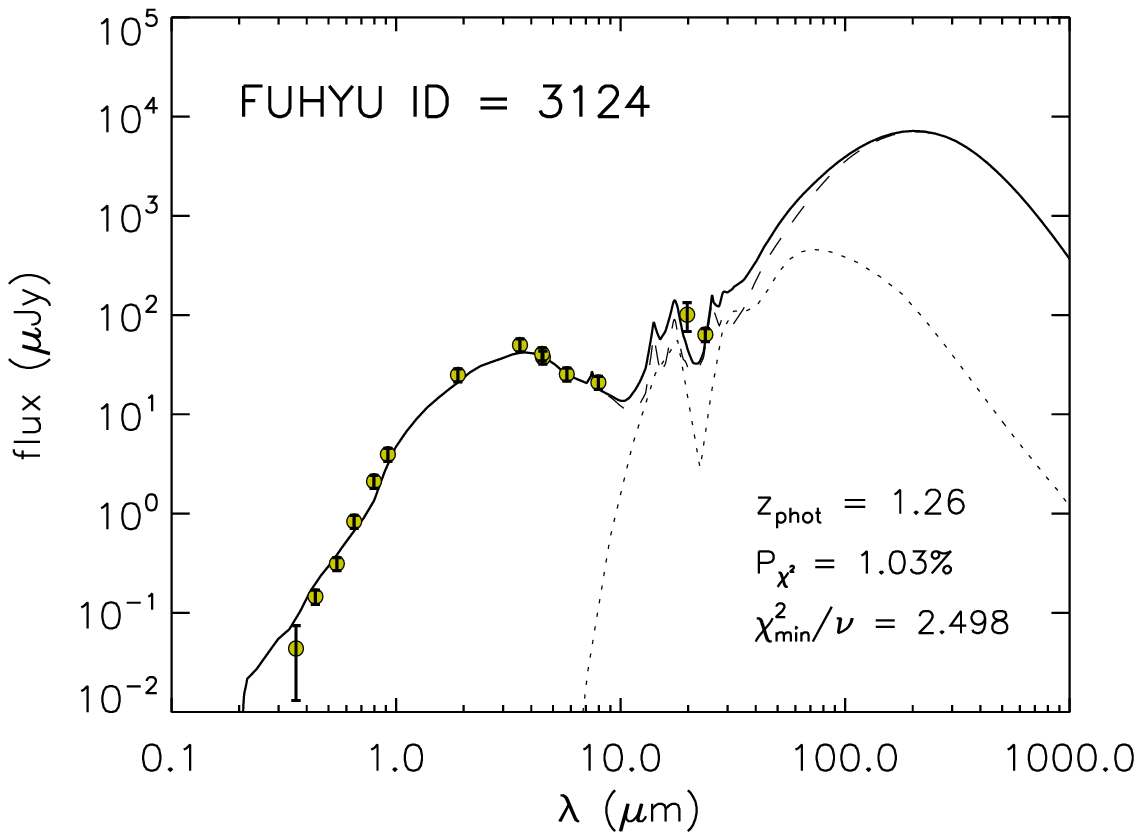,  width=6cm}
}
\caption{Model fits for the candidate Silicate Break galaxies using the photo-metric redshift code of \cite{negrello09}. The {\it dashed} and  {\it dotted lines} are the fitted starburst and AGN components from the spectral libraries of  \cite{takagi03} \& \cite{efstathiou95} respectively and the {\it solid line} is the composite fit. Photometric redshift fitting parameters are listed in each panel and tabulated in Table \ref{tab:silcatebreakfits} following \cite{negrello09}.
\label{fig:silicatebreakSED}}
\end{figure*}

The FU-HYU colour-colour diagrams can also be used to search for interesting populations on the basis of unique colours or colour separations. As an example, we use the colour information in our catalogue to segregate high absorption dusty galaxies from the general population. \cite{takagi05} postulated that dusty, heavily obscured starbursts could be detected via their deep silicate absorption feature at 9.7$\mu$m using a "{\it drop-out}" technique. In this technique, a sharp dip is observed in the mid-infrared colours of these sources as the silicate absorption feature is redshifted through one of the bands.  \cite{takagi05} coined galaxies selected in this way "{\it Silicate Break}" galaxies and such galaxies were subsequently identified by  \cite{charmandaris04} using the IRS peak up imager on {\it Spitzer}. 
\cite{takagi05} performed simulations for the detection of potential Silicate Break galaxies for a combination of {\it AKARI} \& {\it Spitzer} filters concluding that such sources would exhibit colour drops due to a deficit in their MIPS 24$\mu$m flux between redshifts of 1.2 -- 1.6.
In Figure  \ref{fig:silicatebreak}{\it (a)} the  MIPS24/L18W colour is plotted against the  L18W/IRAC8 colour with  the galaxy template spectral energy distributions for a moderate starburst and ULIRG overlaid. The passage of the silicate absorption feature through the mid-infrared bands can be clearly seen in the starburst and ULIRG templates and map a distinct tail  towards lower MIPS24/L18W colours from -0.15$>$ lg(MIPS24/L18W)$>$-0.4  in the colour-colour plane. In Figure  \ref{fig:silicatebreak}{\it (a)}, we adapt the simulations of  \cite{takagi05} for the {\it AKARI} L18W band plotting the MIPS 24/L18W colour as a function of redshift for an ensemble of template star-forming galaxies from the libraries of \cite{takagi03} (also included are comparative quiescent galaxy spectral  templates from \cite{dale01}). Enforcing a selection criteria of lg(MIPS 24/L18W) colours $<-0.15$ should preferentially select galaxies in the redshift range of 1.2 -- 1.6.
 In Figure  \ref{fig:silicatebreak}{\it (c)} the lg(MIPS24/L18W) colour is plotted against the MIPS 24$\mu$m band flux for the FU-HYU sources. The threshold for the Silicate Break detection is marked as a solid line and it can be seen that there are approximately 9 potential candidates in our sample with faint MIPS  24$\mu$m fluxes and  lg(MIPS24/L18W)$<$-0.15. Extracting all available photometry for the 9 candidate Silicate Break galaxies allows the creation of photometric spectral energy distributions shown in Figure \ref{fig:silicatebreak}{\it (d)}. The presence of the strong emission (1.6$\mu$m bump and 7.7$\mu$m PAH) and absorption (9.7$\mu$m silicate) features can be clearly seen in the SEDs. 
We have used the photometric redshift code of  \cite{negrello09}, to fit model spectral energy distributions to the 9 {\it Silicate-Break} galaxy candidates in Figue \ref{fig:silicatebreakSED}.   The model parameters used for the fits, and the resulting redshift and  luminosity contributed by the starburst and AGN components respectively are tabulated in Table \ref{tab:silcatebreakfits} where all the quoted errors represent the 99 per cent confidence interval.  In total 8/9 of our candidates are well fitted by SEDs with silicate absorption in the redshift range 1.2 -- 1.6, i.e. the expected range sensitive to the Silicate-Break drop out method. The erroneous candidates, ID933  appears to have a photometric redshift of z=0.84 assigned although it seems plausible that even in this case the  7.7$\mu$m PAH feature is being mistaken for the 11.2$\mu$m PAH feature and that in fact this galaxy too could lie in the preferred redshift range.. 
The brightest source from Figure \ref{fig:silicatebreakSED} (ID1772) is in fact the ISO-HDF source HDF PM3 3 (ISOHDF3 J123634+621238, \cite{goldscmidt97}) at a redshift $\sim$1.28. The galaxy is a dusty ultra-luminous merging sub-millimetre source with strong emission features and a strong silicate absorption feature in its mid-infrared spectrum (\cite{frayer08}, \cite{pope08}). The photometric redshift is correct to within 10$\%$, highlighting the successful identification of dusty galaxies in the silicate-break redshift range.

\begin{table*}
  \vspace{0.0cm}
  \begin{center}
    \small
    \caption{Best fit results for the 7 objetcs in Figure \ref{fig:silicatebreakSED} for candidate Silicate Break galaxies.}
    \label{tab:silcatebreakfits}
    \vspace{-0.2cm}
    \begin{tabular}{lrrrrrlrrrrr}
      \hline
      \hline
      ID & $z_{\rm phot}$ & $\chi^{2}_{\rm min}/\nu$ & $P_{\chi^2}$ & $\nu$ & ext. & Age & $\Theta$ & $\theta_{\rm view}$ & $\log[L_{sb}]$ & $\log[L_{agn}]$ & $\log[L_{tot}]$ \\
         &              &                         &             &       &      & (Myr) &       & (deg)             &  ($L_{\odot}$) &  ($L_{\odot}$) &  ($L_{\odot}$)  \\
      \hline
933 & 0.84$_{-0.17}^{+0.17}$ & 2.44 & 0.00 & 11 & MW & 500$_{-10}^{+10}$ & 2.2$_{-1.9}^{+2.8}$ & 50$_{-50}^{+40}$ & 10.9$_{-0.12}^{+0.14}$ & 10.1$_{-0.36}^{+0.19}$ & 10.9$_{-0.14}^{+0.15}$ \\
1130 & 1.28$_{-0.35}^{+0.17}$ & 2.08 & 0.03 & 8 & LMC & 600$_{-10}^{+10}$ & 1.0$_{-0.3}^{+2.8}$ & 40$_{-32}^{+3}$ & 11.5$_{-0.24}^{+0.22}$ & 10.4$_{-0.00}^{+0.89}$ & 11.6$_{-0.23}^{+0.29}$ \\
1549 & 1.46$_{-0.60}^{+0.46}$ & 1.41 & 0.19 & 7 & SMC & 600$_{-100}^{+0}$ & 1.0$_{-0.5}^{+0.8}$ & 8$_{-8}^{+83}$ & 11.4$_{-0.15}^{+0.16}$ & 11.5$_{-2.09}^{+0.00}$ & 11.7$_{-0.33}^{+0.00}$ \\
1772 & 1.28$_{-0.11}^{+0.13}$ & 2.13 & 0.01 & 12 & MW & 300$_{-10}^{+100}$ & 2.0$_{-0.1}^{+0.1}$ & 47$_{-47}^{+43}$ & 12.2$_{-0.14}^{+0.12}$ & 10.4$_{-0.12}^{+1.36}$ & 12.2$_{-0.13}^{+0.22}$ \\
1818 & 1.20$_{-0.27}^{+0.30}$ & 0.85 & 0.54 & 8 & MW & 600$_{-300}^{+0}$ & 2.0$_{-1.6}^{+1.0}$ & 47$_{-3}^{+42}$ & 11.2$_{-0.16}^{+0.42}$ & 10.5$_{-0.14}^{+0.09}$ & 11.3$_{-0.15}^{+0.44}$ \\
2009 & 1.28$_{-0.75}^{+0.94}$ & 4.72 & 0.00 & 9 & MW & 500$_{-300}^{+100}$ & 3.0$_{-1.6}^{+2.0}$ & 90$_{-90}^{+0}$ & 11.3$_{-0.72}^{+0.20}$ & 10.7$_{-0.83}^{+0.97}$ & 11.4$_{-0.71}^{+0.50}$ \\
2541 & 1.24$_{-0.21}^{+0.18}$ & 1.03 & 0.40 & 7 & MW & 600$_{-100}^{+0}$ & 2.6$_{-0.1}^{+0.1}$ & 8$_{-3}^{+40}$ & 11.4$_{-0.21}^{+0.07}$ & 11.2$_{-1.60}^{+0.14}$ & 11.5$_{-0.29}^{+0.21}$ \\
3096 & 1.42$_{-0.42}^{+0.25}$ & 0.95 & 0.47 & 8 & MW & 400$_{-100}^{+200}$ & 1.6$_{-1.1}^{+0.2}$ & 8$_{-3}^{+42}$ & 11.6$_{-0.33}^{+0.09}$ & 10.9$_{-1.18}^{+0.39}$ & 11.7$_{-0.41}^{+0.06}$ \\
3124 & 1.26$_{-0.19}^{+0.85}$ & 2.49 & 0.01 & 8 & SMC & 600$_{-100}^{+0}$ & 1.0$_{-0.1}^{+0.4}$ & 7$_{-7}^{+32}$ & 11.8$_{-0.14}^{+0.28}$ & 9.71$_{-0.00}^{+1.73}$ & 12.0$_{-0.33}^{+0.13}$ \\
      \hline
\multicolumn{12}{l}{  $z_{\rm phot}$: Best fit photometric redshift }\\
\multicolumn{12}{l}{ $\chi^{2}_{\rm min}/\nu$: reduced minimum $\chi^2$ for  $\nu$ degrees of freedom }\\
\multicolumn{12}{l}{ $P_{\chi^2}$: probability associated to the minimum $\chi^2$  }\\
\multicolumn{12}{l}{ ext.: extinction curve }\\
\multicolumn{12}{l}{ Age: age of the starburst }\\
\multicolumn{12}{l}{ $\Theta$: starburst compactness factor }\\
\multicolumn{12}{l}{ $\theta_{\rm view}$: viewing angle of the AGN torus }\\
\multicolumn{12}{l}{ $\log[L_{sb}]$:  Luminosity contributed by the starburst component }\\
\multicolumn{12}{l}{ $\log[L_{agn}]$: Luminosity contributed by the AGN component }\\
\multicolumn{12}{l}{  $\log[L_{tot}]$: Total Luminosity }\\
    \end{tabular}
  \end{center}
\end{table*}

\smallskip
\section{Summary}\label{sec:summary}

We have presented the initial results from the {\it AKARI} FU-HYU Mission Program in the GOODS-N field  describing the data reduction process including the additional processing steps required to analyze the FU-HYU data due to the omission of dithering cycles when the observations were originally taken. Combining all the available data in the GOODS-N field, a final FU-HYU catalogue has been produced containing more than 4393 sources with almost 200 sources detected in the {\it AKARI} bands. Using the combination of the {\it AKARI} and {\it Spitzer} multi-wavelength coverage a total mid-infrared flux has been defined to be representative of the total mid-infrared emission of the sources and a measure of the total luminosity of the galaxies (c.f. \cite{elbaz02}). This total MIR flux is tightly correlated with the longer wavelength mid-infrared (IRC L18W, MIPS 24) bands but less correlated with the shorter IRAC and IRC bands. The implication is that the {\it Spitzer} MIPS24 or {\it AKARI} L18W band are representative of the total mid-infrared luminosity of these galaxies in agreement with the results of \cite{bavouzet08} who have indicated that the 24$\mu$m flux is a good indicator of the mid and bolometric infrared luminosity of star forming galaxies.

The mid-infrared colours have been used to track the passage of the PAH emission features through the observation bands and to segregate the star-forming population at z$\sim$1 which have higher total MIR / single band colours than the more quiescent population. In particular we have shown that an excess in emission in the {\it AKARI} S11 band is indicative of a moderate redshift population.

Using the  {\it AKARI} IRC L18W to {\it Spitzer} MIPS 24 band colour as an example diagnostic, we have shown that it is possible to segregate specific populations in the colour colour plane, thus we have used the "{\it Silicate-Break}" technique to extract extinct, dusty galaxies from our FU-HYU sample. This population is sensitive to the passage of the 9.7$\mu$m absorption feature through the MIPS 24 band in the redshift range $\sim$1.2 -- 1.6 and this technique has been successful in identifying 8 possible candidates in the GOODS-N field. Models fits of spectra using a photometric redshift code have indeed confirmed that the sources lie in the redshift range expected for Silicate-Break galaxies confirming that the silicate break selection method can provide a powerful means to detect dusty ULIRGs at moderate redshift.

\begin{acknowledgements}
The authors would like to thank Denis Burgarella for kindly supplying the GALEX data for the GOODS-N region and the referee, whose comments improved the clarity of this work. W-SJ and HML are supported by Korea Astronomy and Space Science Institute. HML was supported by National Research Foundation of Kore (NRF) grant No. 2006-341-C00018.\\
The {\it AKARI} Project is an infrared mission of the Japan Space Exploration Agency (JAXA) Institute of Space and Astronautical Science (ISAS), and is carried out with the participation of mainly the following institutes; Nagoya University, The University of Tokyo, National Astronomical Observatory Japan, The European Space Agency (ESA), Imperial College London, University of Sussex, The Open University (UK), University of Groningen / SRON (The Netherlands), Seoul National University (Korea). The far-infrared detectors were developed under collaboration with The National Institute of Information and Communications Technology.
\end{acknowledgements}



\label{lastpage}

\end{document}